\documentclass[physrev,reprint,aps]{revtex4-2} 
\usepackage[utf8]{inputenc}

\usepackage{graphicx}
\usepackage{gensymb} 
\usepackage{textgreek} 
\usepackage{siunitx} 
\usepackage{makecell} 
\usepackage{textcomp} 
\usepackage{miller} 
\usepackage{lineno}

\usepackage[table]{xcolor} 
\definecolor{lightgrey}{RGB}{238,239,240}
\usepackage{colortbl}

\usepackage[acronym, nonumberlist, nomain]{glossaries} 
\makeglossaries

\usepackage[shortlabels]{enumitem} 
\usepackage[normalem]{ulem} 
\usepackage{xfrac} 
\usepackage[version=4]{mhchem}

\makeatletter
\def\l@subsubsection#1#2{}
\makeatother

\usepackage{booktabs}
\usepackage{multirow} 
\usepackage{tabularx} 
\usepackage[caption=false]{subfig} 
\usepackage{changepage}

\usepackage{array} 
\newcolumntype{P}[1]{>{\centering\arraybackslash}p{#1}} 
\newcolumntype{M}[1]{>{\centering\arraybackslash}m{#1}} 
\newcolumntype{L}[1]{>{\raggedright\arraybackslash}m{#1}} 

\usepackage[hidelinks]{hyperref} 

\def\atp{\,\emph{at.}\%}
\def\um{\,\textmu m}
\def\etal{~\emph{et al.}}
\def\bg{\gls{bg}}
\def\mobun{\,cm$^2$/Vs}

\graphicspath{ {images/} } 

\begin{document}

\newacronym{adas}{ADAS}{advanced driver-assistance system}
\newacronym{das}{DAS}{driver-assistance system}
\newacronym{abs}{ABS}{anti-lock braking system}
\newacronym{radar}{radar}{radio detection and ranging}
\newacronym{lidar}{LiDAR}{light detection and ranging}
\newacronym{tof}{ToF}{time-of-flight}
\newacronym{cmos}{CMOS}{complementary metal-oxide-semiconductor}
\newacronym{nir}{NIR}{near-infrared}
\newacronym{swir}{SWIR}{short-wave infrared}
\newacronym{spad}{SPAD}{single-photon avalanche photodiode}
\newacronym{sipm}{SiPM}{Si photo-multiplier}
\newacronym{pvd}{PVD}{physical vapor deposition}
\newacronym{pd}{PD}{photodiode}
\newacronym{apd}{APD}{avalanche photodiode}

\newacronym{snr}{SNR}{signal-to-noise ratio}
\newacronym{mbe}{MBE}{molecular beam epitaxy}
\newacronym{cvd}{CVD}{chemical vapor deposition}
\newacronym{fcc}{FCC}{face-centered cubic}
\newacronym{pqpr}{PQPR}{passive quenching, passive recharge}
\newacronym{aqar}{AQAR}{active quenching, active recharge}
\newacronym{spde}{SPDE}{single-photon detection efficiency}
\newacronym{nep}{NEP}{noise-equivalent power}
\newacronym{dcr}{DCR}{dark count rate}
\newacronym{btbt}{BTBT}{band-to-band tunneling}
\newacronym{tat}{TAT}{trap-assisted tunneling}
\newacronym{srh}{SRH}{Shockley-Read-Hall}
\newacronym{vb}{VB}{valence band}
\newacronym{cb}{CB}{conduction band}
\newacronym{sacm}{SACM}{separate absorption-charge-multiplication}
\newacronym{sagcm}{SAGCM}{separate absorption-grading-charge-multiplication}
\newacronym{ic}{IC}{integrated circuit}
\newacronym{fwhm}{FWHM}{full width at half maximum}
\newacronym{irf}{IRF}{instrument response function}

\newacronym{vbr}{$V_{br}$}{breakdown voltage}
\newacronym{vop}{$V_{op}$}{operating voltage}
\newacronym{vex}{$V_{ex}$}{excess bias}
\newacronym{vpt}{$V_{pt}$}{punch-through voltage}
\newacronym{rsc}{$R_{sc}$}{space-charge resistance}
\newacronym{abar}{$\bar{\alpha}$}{mean number of ionization events per free carrier}
\newacronym{eg}{$E_g$}{bandgap energy}
\newacronym{tcoeff}{$\gamma$}{temperature coefficient of breakdown voltage}

\newacronym{fet}{FET}{field-effect transistor}
\newacronym{mosfet}{MOSFET}{metal-oxide-semiconductor field-effect transistor}
\newacronym{uhv}{UHV}{ultra-high vacuum}
\newacronym{apcvd}{AP-CVD}{atmospheric-pressure \acrlong{cvd}}
\newacronym{lpcvd}{LP-CVD}{low-pressure \acrlong{cvd}}
\newacronym{rpcvd}{RP-CVD}{reduced-pressure \acrlong{cvd}}
\newacronym{led}{LED}{light emitting diode}
\newacronym{bg}{BG}{bandgap}
\newacronym{sg}{SG}{space group}
\newacronym{mqw}{MQW}{multi quantum well}
\newacronym{mwir}{MWIR}{medium-wave infrared}
\newacronym{lwir}{LWIR}{long-wave infrared}
\newacronym{spe}{SPE}{solid-phase epitaxy}
\newacronym{lpe}{LPE}{liquid-phase epitaxy}
\newacronym{fla}{FLA}{flash-lamp annealing}
\newacronym{exafs}{EXAFS}{extended x-ray absorption fine structure}
\newacronym{apt}{APT}{atom probe tomography}
\newacronym{cer}{CER}{contactless electroreflectance}
\newacronym{vase}{VASE}{variable angle spectroscopic ellipsometry}
\newacronym{dft}{DFT}{density-functional theory}
\newacronym{pr}{PR}{photoreflectance}
\newacronym{dsr}{DSR}{degree of strain relaxation}
\newacronym{pl}{PL}{photoluminescence}
\newacronym{ot}{OT}{optical transmittance}
\newacronym{sro}{SRO}{short-range order}
\newacronym{lh}{LH}{light-hole}
\newacronym{hh}{HH}{heavy-hole}
\newacronym{tem}{TEM}{transmission electron microscopy}
\newacronym{sem}{SEM}{scanning electron microscopy}
\newacronym{xtem}{XTEM}{cross-section \acrlong{tem}}
\newacronym{md}{MD}{misfit dislocation}
\newacronym{td}{TD}{threading dislocation}
\newacronym{xrd}{XRD}{X-ray diffraction}
\newacronym{rta}{RTA}{rapid thermal annealing}
\newacronym{tdd}{TDD}{\acrlong{td} density}
\newacronym{gr}{GR}{growth rate}
\newacronym{sims}{SIMS}{secondary-ion mass spectroscopy}
\newacronym{rf}{RF}{radiofrequency}
\newacronym{dc}{DC}{direct current}
\newacronym{pdc}{pDC}{pulsed \acrlong{dc}}
\newacronym{hipi}{HiPI}{high-power impulse}
\newacronym{rc}{RC}{rocking curve}
\newacronym{lt}{LT}{low-temperature}
\newacronym{ht}{HT}{high-temperature}
\newacronym{tec}{TEC}{thermal expansion coefficient}
\newacronym{sf}{SF}{stacking fault}
\newacronym{cw}{CW}{continuous-wave}
\newacronym{ms}{MS}{magnetron sputtering}
\newacronym{pb}{P-B}{People and Bean's}
\newacronym{mb}{M-B}{Matthews and Blakeslee}
\newacronym{rbs}{RBS}{Rutherford backscattering spectrometry}
\newacronym{pas}{PAS}{positron annihilation spectroscopy}
\newacronym{gsmbe}{GS-MBE}{gas-source \acrlong{mbe}}
\newacronym{vdp}{VdP}{Van der Pauw}
\newacronym{cv}{CV}{capacitance-voltage}
\newacronym{iv}{IV}{current-voltage}
\newacronym{ecv}{e-CV}{electrochemical \acrlong{cv}}
\newacronym{ligt}{LITG}{light-induced transient grating}
\newacronym{soi}{SOI}{silicon-on-insulator}
\newacronym{goi}{GOI}{germanium-on-insulator}
\newacronym{pda}{PDA}{post-deposition annealing}
\newacronym{dlts}{DLTS}{deep-level transient spectroscopy}
\newacronym{spc}{SPC}{solid-phase crystallization}
\newacronym{pc}{PC}{photoconductivity}
\newacronym{ftir}{FTIR}{Fourier-transform infrared}
\newacronym{ed}{ED}{extended defect}
\newacronym{el}{EL}{electroluminesce}
\newacronym{trpls}{TRPLS}{time-resolved \acrlong{pl} spectroscopy}
\newacronym{trdr}{TRDR}{time-resolved differential reflectivity}
\newacronym{mwpcd}{\textmu W-PCD}{microwave photoconductance decay}
\newacronym{icpcd}{IC-PCD}{inductive-coupled photoconductance decay}
\newacronym{ir}{IR}{infrared}
\newacronym{pecvd}{PECVD}{plasma-enchanced \acrlong{cvd}}

\newacronym{undop}{$n_{un}$}{unintentional doping concentration}
\newacronym{mup}{$\mu_p$}{hole mobility}
\newacronym{rrms}{$R_{rms}$}{root mean square roughness}
\newacronym{tcr}{$t_{cr}$}{critical thickness of strain relaxation}
\newacronym{bvec}{\textbf{b}}{Burger's vector}
\newacronym{r}{$R$}{resistance}
\newacronym{tf}{$t_f$}{film thickness}

\newacronym{rsm}{RSM}{reciprocal space mapping}
\newacronym{afm}{AFM}{atomic force microscopy}
\newacronym{ml}{ML}{monolayer}
\newacronym{stem}{STEM}{scanning \acrlong{tem}}
\newacronym{hrtem}{HRTEM}{high-resolution \acrlong{tem}}
\newacronym{haadf}{HAADF}{high-angle annular dark field}
\newacronym{fft}{FFT}{fast Fourier transform}
\newacronym{za}{ZA}{zone axis}
\newacronym{edx}{EDX}{energy-dispersive x-ray spectroscopy}
\newacronym{bf}{BF}{bright field}
\newacronym{df}{DF}{dark field}
\newacronym{fib}{FIB}{focused ion beam}
\newacronym{vge}{vGe}{Ge virtual substrate}
\newacronym{sre}{SRE}{spontaneous-relaxation-enhancement}
\newacronym{su}{SU}{sputter-up}
\newacronym{sd}{SD}{sputter-down}
\newacronym{optp}{OPTP}{optical pump-THz probe spectroscopy}

\newacronym{t}{$T$}{temperature}
\newacronym{qe}{QE}{quasi-equilibrium}

\newacronym{f}{$F$}{electric field}
\newacronym{drie}{DRIE}{deep reactive-ion etching}
\newacronym{rca}{RCA}{Radio Corporation of America}
\newacronym{ibe}{IBE}{ion beam etching}
\newacronym{ar}{AR}{anti-reflection}

	
	\title[]{GeSn Defects and their Impact on Optoelectronic Properties: A Review}
	
	\author{Andrea Giunto}
    \affiliation{ 
		Laboratory of Semiconductor Materials, Institute of Materials, School of Engineering, École Polytechnique Fédérale de Lausanne, Lausanne, Switzerland
	}
	\author{Anna Fontcuberta i Morral}
	\email{anna.fontcuberta-morral@epfl.ch}
	\affiliation{ 
		Laboratory of Semiconductor Materials, Institute of Materials, School of Engineering, École Polytechnique Fédérale de Lausanne, Lausanne, Switzerland
	}%
    \affiliation{
  Institute of Physics, School of Basic Sciences, École Polytechnique Fédérale de Lausanne, Lausanne, Switzerland
	}%
	
	
	
	
	
	\date{\today}
	
	\begin{abstract}
	GeSn has emerged as a promising semiconductor with optoelectronic functionality in the mid-infrared, with the potential of replacing expensive III-V technology for monolithic on-chip Si photonics.
    Multiple challenges to achieve optoelectronic-grade GeSn have been successfully solved in the last decade. We stand today on the brink of a potential revolution in which GeSn could be used in many optoelectronic applications such as Light Detection and Ranging (LiDARs) devices and lasers.
    However, the limited understanding and control of material defects represents today a bottleneck in the performance of GeSn-based devices, hindering their commercialisation.
    Point and linear defects in GeSn have a strong impact on its electronic properties, namely unintentional doping concentration, carrier lifetime and mobility, which ultimately determine the performance of optoelectronic devices.
    In this review, after introducing the state-of-the-art of the fabrication and properties of GeSn, we provide a comprehensive overview of the current understanding of GeSn defects and their influence on the material (opto)electronic properties.
    Throughout the manuscript, we highlight the critical points that are still to solve.
    By bringing together the different fabrication techniques available and characterizations realized we provide a wholistic view on the field of GeSn and provide elements on how it could move forward.
	\end{abstract}

	\maketitle

\tableofcontents

\section{Introduction}
GeSn is nowadays recognized as a promising absorber and emitter that can be integrated monolithically on the Si platform, operating in the \gls{nir} and \gls{swir} wavelengths~\cite{Moutanabbir2021}.
The addition of Sn to the Ge lattice induces a red-shift in the material \gls{bg}, extending the absorption cut-off wavelength towards the infrared.
In addition, it has been shown that above 7-9\atp{} Sn~\cite{Ghetmiri2014,Gallagher2014,Wirths2015}, the GeSn alloy acquires a direct \gls{bg},  enabling its use as active material in light-emitting devices.
Ge-rich GeSn alloys have been demonstrated in a plethora of optoelectronic devices including photodetectors~\cite{Atalla2022,TalamasSimola2021,Tran2019,Zhang2018}, lasers~\cite{Wirths2015,Kim2022,Ojo2022,Marzban2023} and \glspl{led}~\cite{Cardoux2022,Huang2019,Chang2017}.
Furthermore, the high theoretical mobility of GeSn~\cite{Sau2007,Mukhopadhyay2021} has motivated its development \glspl{fet}~\cite{Liu2020,Huang2017,Wang2018}. The possibility of monolithic integration on Si platforms has also pushed the investigation of GeSn in silicon photonics and for on-chip thermoelectric applications~\cite{Portavoce2020,Spirito2021}.
To date, despite more than 15 years of intensive research in the field, there exists no commercial device based on GeSn. In fact, there are numerous challenges hindering the rise of this material for the next-generation (opto)electronics. 

In this review, we provide an overview of the withstanding challenges in the field of GeSn for (opto)electronic devices. Specifically, we illustrate the defects arising during material processing, and we discuss the current, albeit limited, understanding of their impact on the GeSn electrical properties.
While our primary focus is on the GeSn alloy, we also address and explore studies on Ge as they pertain to understanding GeSn.
In Sec.~\ref{sec:GeSn_HistPersp}, we provide a concise review of the historical development of GeSn as an optoelectronic grade semiconductor. This is accompanied by a description of the GeSn physical properties and the challenges associated with the material, respectively in Secs.~\ref{sec:GeSn_PhysicProp} and \ref{sec:GeSn_Challenges}.
In Sec.~\ref{sec:GeSn_Epitaxy}, we investigate the general efforts in GeSn epitaxy, as well as the challenges associated to increasing the Sn content in the alloy. Additionally, we provide a detailed review of magnetron sputtering epitaxy of Ge and GeSn, complementing existing literature reviews that focus on growth of these materials by \gls{cvd} and \gls{mbe}.
In Sec.~\ref{sec:GeSn-Ge_StrainRelax}, we discuss the defects arising during epitaxial growth of Ge and GeSn, considering both point defects and extended defects such as dislocations and stacking faults.
In Sec.~\ref{sec:GeSn_ElecProp}, we undertake a comprehensive examination of the optoelectronic characterization of Ge and GeSn films  reported in the literature. This consists in compiling measurement data into detailed tables and condensing key findings to establish a coherent understanding of the subject.
Finally, in Sec~\ref{sec:Conclusion}, we conclude with a brief discussion of the open questions in the field.

\section{Historical Perspectives}
\label{sec:GeSn_HistPersp}
\subsubsection*{Early days of c-GeSn}
In 1982, C.H.L. Goodman first proposed crystalline GeSn (cGeSn) as a group-IV direct-\gls{bg} material, hypothesizing alloy properties governed by Vegard's law between Ge and the diamond cubic phase of Sn (i.e., \textalpha Sn)~\cite{Goodman1982}. High mobility values were predicted due to the absence of polar scattering, typical in III-V and II-VI compounds. The Ge-Sn solid solution was thus suggested as alternative to III-V and II-VI materials for high-mobility \glspl{fet} and infrared photodetectors in the \gls{swir} (1.5\um-3\um), \gls{mwir} (3\um-8\um), and \gls{lwir} wavelengths (8\um-15\um).
Challenges in experimental realization of this material were expected due to the low solubility limit of 1\atp{} of Sn in Ge (and vice-versa)~\cite{Trumbore1956,Olesinski1984}, and the lack of a lattice-matched substrate to stabilize metastable GeSn phases.
Metastable, micro-crystalline GeSn with more than 20\atp{} Sn was demonstrated one year later, thanks to the use of out-of-equilibrium synthesis~\cite{Oguz1983}. Ge$_{0.78}$Sn$_{0.22}$ was crystallized from amorphous Ge$_{0.70}$Sn$_{0.30}$ by means of a UV pulsed laser.
The first epitaxial metastable GeSn film appeared in 1987, with growth up to 8\atp{} Sn on a Ge(001) substrate by sputtering~\cite{Shah1987}. The growth temperature was maintained at 150\degree C to prevent Sn segregation due to the reduced solubility in Ge.

\begin{figure*}[htb]
	\centering
	\includegraphics[width=\textwidth]{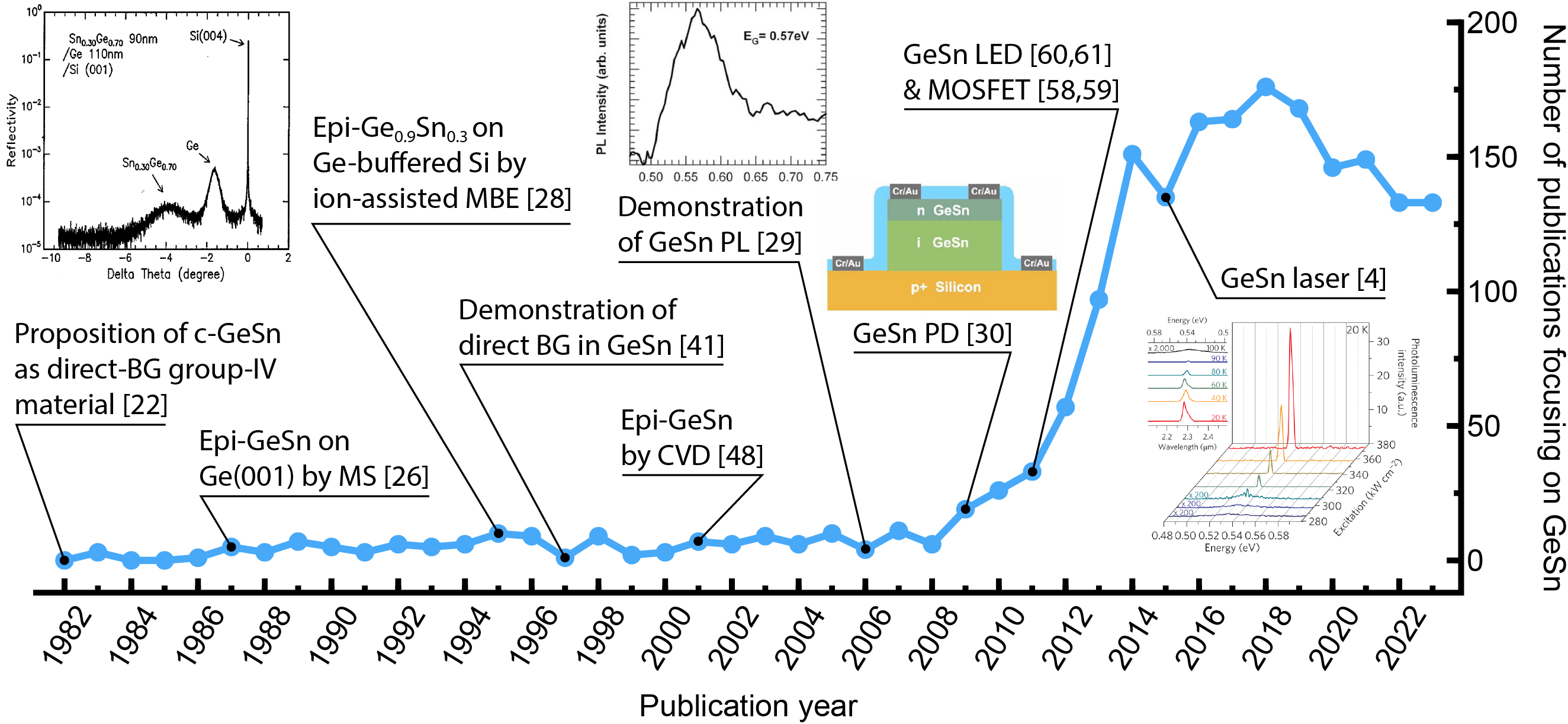}
	\caption{Plot of number of publications per year focusing on GeSn, with selected milestones achieved in the field.	Publication data was selected using the web tool \emph{Scopus Analytics}~\cite{Elsevier}, searching with the ``OR'' operator among the following keywords in the title or abstract: ``GeSn'', ``SnGe'', ``Ge1-xSnx'', ``SnxGe1-x'', ``Ge 1-x Sn x'', ``Sn x Ge 1-x'', accessed in March 2024. Figures in the insets are reprinted with permission from the corresponding authors: respectively, from left to right, He\etal~\cite{He1995a}, \copyright\,1995 \emph{Elsevier}; Soref\etal~\cite{Soref2006}, \copyright\,2006 \emph{Springer Nature}; Mathews\etal~\cite{Mathews2009}, \copyright\,2009 \emph{AIP Publishing}; Wirths\etal~\cite{Wirths2015}, \copyright\,2015 \emph{Springer Nature}.
}
	\label{fig:GeSnHistory}
\end{figure*}

\subsubsection*{First monocrystalline GeSn films}
Slow, but notable progress in out-of-equilibrium synthesis processes was obtained in the 20 years following the demonstration of epitaxial GeSn on Ge(001).
In 1989, Pukite\etal~\cite{Pukite1989} employed the \gls{mbe} method to push the Sn content to 30\atp{} in polycrystalline GeSn films on Si(100) at 170\degree C.
Despite the successful sputtering growth of GeSn epitaxial films, \gls{mbe} became the growth method of choice in the 1990s, hence seeing considerable improvements in the following years. 
After a few more studies reporting polycrystalline growth~\cite{Gossmann1990,Harwit1990}, monocrystalline \gls{mbe}-grown GeSn was obtained in 1992 on Ge(001), but only up to thicknesses of 2\,nm. Thicker layers could not be obtained due to Sn segregation and the low growth temperatures employed~\cite{Wegscheider1992}.
Low growth temperatures ($T<200$\degree C) reduce the adatom mobility, causing kinetic roughening and strain-induced roughening~\cite{Desjardins1999}, limiting the epitaxial thickness to tens of nm for Sn fractions higher than 10\atp~\cite{Gurdal1995,Gurdal1998,Bratland2005}.
Issues with Sn segregation and low-temperature growth were overcome in 1995 with the introduction of ion-assisted (Ar$^+$) \gls{mbe}~\cite{He1995}. Mimicking the analogous effect in sputtering, it was found that slight Ar bombardment induced collisional mixing of Sn adatoms with the film, increasing Sn incorporation and limiting its segregation~\cite{He1995a}. With this method, He\etal~\cite{He1995a} obtained 20-nm-thick monocrystalline GeSn films on Ge-buffered Si(001) substrates with Sn contents up to 34\atp{}. The same research group realized the first pseudomorphic GeSn film on Ge(001) in 2000, using standard thermal \gls{mbe}~\cite{Ragan2000}.

\subsubsection*{Demonstration of GeSn direct bandgap behavior}
In 1997, He\etal~\cite{He1997} experimentally observed direct-\gls{bg} behavior in GeSn for the first time. The \gls{bg} crossing from indirect to direct character was  found to occur for a Sn content of around 11\atp. This constitutes a Sn alloy fraction considerably lower than the cross-over composition expected at the time.
The first theoretical studies  of the GeSn alloy band structure using tight-binding models predicted in fact an indirect-to-direct \gls{bg} transition around 20\atp{} Sn~\cite{Jenkins1987}, similar to what expected from Vegard's law.
This discrepancy between theory and experiments was understood in the 2000s with first-principle computations~\cite{Chibane2003} and confirmed with subsequent experimental observation~\cite{DCosta2006}. It was found that the behavior of the GeSn band energies follows a positive deviation from Vegard's law, described with the use of a bowing parameter ($b$):
\begin{equation}
	\label{eq:EnergyBGbowing}
	E_{GeSn}=(1-x)E_{Ge}+xE_{\alpha Sn}-b(1-x)x
\end{equation}
where $E$ is the \gls{bg} energy of the respective materials~\cite{Yin2008}.
Numerous measurements and computations of $b$ have been reported in the literature since then, with values generally spanning between 2.0\,eV and 2.5\,eV (see Tab.~\ref{tab:GeSn_BowingParam}, discussed in Sec.~\ref{sec:GeSn_PhysicProp}).
The positive value of $b$ implies that the \gls{bg} crossover occurs at lower $x$ compared to what predicted by Vegard's law, explaining the early disagreement between theoretical and experimental data.
The discovery of a  positive \gls{bg} bowing behavior significantly boosted the prospects of the GeSn alloy, as it signified that the Ge \gls{bg} can be strongly red-shifted with  only a few \atp{} of Sn. This makes GeSn well-suited for optoelectronic applications in the \gls{swir} range and potentially longer wavelengths in spite of its metastability.

\subsubsection*{First CVD-grown GeSn films}
To overcome the limitations of \gls{mbe} associated with low-temperature growth and kinetic roughening~\cite{Desjardins1999}, in the 2000s the attention was shifted towards alternative growth methods.
\Acrfull{ms} was successfully employed to grow few-hundred-nm-thick GeSn films on Ge(001) up to 14\atp{} Sn~\cite{LadrondeGuevara2003,LadrondeGuevara2004}, while \gls{cvd} growth was demonstrated in 2001 by Taraci\etal{} thanks to the introduction of stable Sn precursors~\cite{Taraci2001,Bauer2002,Bauer2003}; films up to 200\,nm with 20\atp{} Sn were obtained on Si(100) by \gls{uhv} \gls{cvd}~\cite{DCosta2006}.
At the same time, the first SiGeSn films were demonstrated~\cite{Bauer2003a} with the potential of delivering a higher thermal stability and a decoupling of lattice parameter and \gls{bg} energy~\cite{Wirths2016}. 

\subsubsection*{First GeSn-based (opto)electronic devices}
These promising advances in epitaxy led to a growing interest in GeSn in the research community, evident in Fig.~\ref{fig:GeSnHistory} from the increase in number of published works on GeSn from 2009 onwards.
With this increase in the number of studies, researchers gained a deeper understanding of GeSn synthesis processes and properties. For example, it was found that the Ge surface reconstructs in presence of Sn~\cite{Yamazaki2008}, and that the misfit strain in GeSn films influences Sn precipitation during post-growth thermal processing~\cite{Takeuchi2008}. A method to calculate the alloy \gls{bg} dependence  on composition and strain was also proposed~\cite{Gupta2013}.
Furthermore, the introduction of doping methods of GeSn~\cite{Chizmeshya2006} led to the first demonstration of a GeSn-basedoptoelectronic device in 2009 by Mathews\etal~\cite{Mathews2009}, with the fabrication of a \gls{cvd}-grown \emph{n-i}-Ge$_{0.98}$Sn$_{0.02}$/\emph{p}-Si \gls{pd}, sensitive up to 1750\,nm. This detector possessed quantum efficiencies lower than 0.1\% but, only 2 years later, an improved design in \gls{mbe}-grown \emph{p-i}-Ge$_{0.97}$Sn$_{0.03}$/\emph{n}-Si demonstrated a responsivity in the \gls{swir} comparable to commercial Ge devices~\cite{Su2011}.
In 2011, GeSn p-type \glspl{mosfet} was demonstrated with a higher hole mobility than conventional Ge \gls{mosfet}s~\cite{Han2011,Gupta2011}.
Furthermore, following demonstration of GeSn \gls{pl}~\cite{Soref2006}, the first GeSn-based \gls{led} was fabricated, demonstrating room-temperature electroluminescence and a clear red-shift in light emission with respect to Ge~\gls{led}s~\cite{Oehme2011, Roucka2011}.
In the same year, Vincent\etal{} achieved GeSn epitaxial growth by \gls{apcvd} with commercially available precursors, demonstrating the viability of vacuum-free  alloy growth, thus significantly simplifying the synthesis process~\cite{Vincent2011}.
The growth of epitaxial GeSn on Si(100) with industry-compatible \gls{rpcvd} was published in 2013 by Wirths\etal~\cite{Wirths2013}.
These studies catalyzed additional momentum in the field, which gained even more interest from scientists and engineers. This heightened interest is evident from the corresponding growing number of publications dedicated to GeSn, appreciable in Fig.~\ref{fig:GeSnHistory}.

\subsubsection*{GeSn-based lasers}
One of the most remarkable steps in the development of GeSn-based technology was the demonstration of the first optically-pumped GeSn laser operating at $T<90$\,K with 12\atp{}   Sn at wavelengths of $\sim$2.3\,\textmu m~\cite{Wirths2015}.
This was the last long-sought building block required for \gls{swir} monolithic group-IV on-chip Si photonics~\cite{Soref2010}, setting off a race towards the achievement of an electrically pumped GeSn laser operating at room temperature, yet to be realized.
Breakthroughs were reached in 2020 with the first electrically-injected laser operating up to 100\,K~\cite{Zhou2020}, and in 2022 with room-temperature lasing obtained in optically-pumped Ge$_{0.86}$Sn$_{0.14}$ devices~\cite{Buca2022}. These works currently set the state-of-the-art of the operating temperatures in the respective injection modes.
Factors limiting laser performance are the low carrier lifetimes in GeSn due to the presence of defects, and the limited direct nature of the GeSn \gls{bg}. The latter, known as ``directness'' of the bandgap, is quantified via the difference in energy between the indirect L- and direct \textGamma-valleys~\cite{Wirths2015}.
These factors have the effect to increase the lasing threshold and to decrease the maximum operating temperature~\cite{Marzban2022}.
Common strategies to increase the \gls{bg} directness, and thus the lasing temperature, are to induce tensile strain in the GeSn gain material or increase the Sn fraction in the alloy~\cite{Buca2022,Elbaz2020,Thai2018}. The former is preferable to avoid the increase in defect concentration associated with large Sn contents~\cite{Buca2022}.
Material defects affecting the lasing thresholds may originate at the bulk~\cite{Marzban2021}, surface~\cite{Marzban2023} and interface~\cite{Elbaz2020} of the active material. Some approaches have been proposed to limit the detrimental effects of these defects.
Surface traps can be prevented via proper passivation of the material, though no absolute passivation scheme has been established for GeSn, resulting in each group using different methods~\cite{Dong2015,Gupta2016,Morea2017,Tran2019}.
Interface defects arise from the arrays of misfit dislocations forming during GeSn epitaxy. They are physically unavoidable when growing on mismatched substrates. In microdisk laser architectures, misfit dislocations have been etched away to improve the lasing threshold~\cite{Elbaz2020}. Alternatively, type-I GeSn/SiGeSn and GeSn/Ge \gls{mqw} structures have been employed to confine carriers away from the epitaxial interface~\cite{Stange2018,Marzban2021,Chen2023}.
On the other hand, there is no clear strategy to decrease the material bulk defect concentration, mainly because GeSn bulk defects and electrical properties are poorly understood, as elaborated later in Sec.~\ref{sec:GeSn_ElecProp}.

\subsubsection*{Other GeSn-based devices}
The highlighted challenges prevent room-temperature operation of electrically-injected GeSn lasers, hindering the commercialization of these devices.
The very same factors limit the performance of GeSn as active material in photodetectors, though promising improvements have been achieved in the latest years.
GeSn \gls{pd}s have been fabricated using various methods, including \gls{cvd}~\cite{Xu2019,Atalla2022}, \gls{mbe}~\cite{Dong2016,Zhang2018,Wanitzek2022}, and \gls{ms}~\cite{Zheng2014}. An extensive review of GeSn \glspl{pd} is present in Ref.~\cite{Chen2022}.
At wavelengths of 2\,\textmu m, GeSn \glspl{pd} have demonstrated a responsivity of several tens of mA/W~\cite{Zhou2020k,Wang2021} and operation frequencies up to 10~GHz~\cite{Xu2019,Wang2021}. Dark currents  as low as a few tens of mA/cm$^2$ were obtained~\cite{Xu2019,Tsai2020}, also thanks to the use of pseudomorphic Ge/GeSn \gls{mqw} absorber stacks to suppress strain relaxation of the GeSn alloy~\cite{Xu2019,Huang2017i}.
Trying to extend the wavelength sensitivity range with increasing Sn contents results in the degradation of GeSn quality due to increased defect concentration~\cite{Moutanabbir2021}.
Atalla \emph{et al.}~\cite{Atalla2022} achieved sensitivity up to 2.6\um{} with a Ge$_{0.885}$Sn$_{0.115}$ \emph{p-i-n} \gls{pd}.
At a reverse bias of 0.5~V, they obtained a peak responsivity of 0.3~A/W at 2\um, with a noticeable drop in above 2.25\um that reached $\sim$60~mA/W at 2.6\um.
They achieved high cut-off frequency of 7.5~GHz at -5~V, which reduced to $\sim$1~GHz at -1~V. However, as expected from the higher Sn content, the average dark current in these \gls{pd}s was 6.5~A/cm$^2$, demonstrated by the authors to come from bulk material defects.
Currently, the dark currents due to \gls{srh} generation mechanisms in GeSn alloys are too high to be competitive with III-V technology~\cite{Atalla2023}.
Despite the potential of GeSn suggested by theoretical studies to reach the \gls{mwir} (i.e., 3\um-8\um) wavelengths~\cite{Chang2022,Gupta2013}, the inherent epitaxial compressive strain on Ge and Si substrates increases the \gls{bg} energy, significantly complicating the experimental realization of \gls{mwir} devices. As a consequence, there are only a handful of studies focusing on operation in this wavelength range~\cite{Atalla2022, Luo2024}.
Strain engineering is  thus an essential point to consider to extend the wavelength operation range into the \gls{mwir}~\cite{Moutanabbir2021}.
Waveguides~\cite{Ghosh2022, Liu2022b,Tsai2021d} and photonic crystals~\cite{Li2023} have been proposed to enhance absorption and thus improve responsivity.
To the same goal, GeSn \glspl{apd} have been fabricated~\cite{Dong2016,Zhang2018,Wanitzek2022}, demonstrating responsivities close to 5\,A/W at 1550\,nm. On the other hand, GeSn \glspl{spad} have been proposed theoretically~\cite{Soref2019,Chen2021}, but are yet to be demonstrated experimentally.

While GeSn laser devices fabricated to date employ exclusively \gls{cvd}-grown materials, \gls{led}s have been realized with both \gls{cvd} and \gls{mbe} methods~\cite{Reboud2021}.
Emission at 3.3\um~has been achieved with Ge$_{0.85}$Sn$_{0.15}$/Ge heterostructures~\cite{Cardoux2022}, and a simple gas detector has been demonstrated employing a GeSn \gls{led}~\cite{Casiez2020}. Though comparable with commercial \gls{mwir} \gls{led}s~\cite{Moutanabbir2021}, the GeSn \gls{led} quantum efficiency remains suboptimal due to trap recombination mechanisms caused by material defects~\cite{Casiez2020,Stange2017}.
GeSn-based \glspl{led}, including alternative architectures such as waveguides~\cite{Bansal2023} and SiGeSn/GeSn \gls{mqw}~\cite{Stange2017,Peng2020}, have been reviewed in Refs.~\cite{Moutanabbir2021,Atalla2024}.
For a comprehensive review of GeSn-based optoelectronic devices, the reader is referred to the work of Reboud\etal~\cite{Reboud2024}.

In the realm of GeSn electronic devices, a significant milestone was recently achieved by Liu\etal~\cite{Liu2023}, who successfully fabricated GeSn \gls{fet}s for \gls{cmos} beyond-Si electronics with improved performance with respect to the pure Ge counterpart.
In the past decade, significant advancements have also been made in the field of GeSn nanostructures, which include nanowires, quantum wells and quantum dots. These works have been thoroughly reviewed by Doherty\etal~\cite{Doherty2020}. Finally, GeSn has also been proposed for thermoelectric applications~\cite{Spirito2021}, as a platform to host quantum spin Hall phases~\cite{Ferrari2023a}, and as a THz emitter~\cite{Chen2022a}.

\section{Physical Properties of GeSn}
\label{sec:GeSn_PhysicProp}
In the following section, we introduce the basic physical properties of GeSn. We review the GeSn phase diagram and the metastability of the alloy, and then introduce recent discoveries on the arrangement of Sn solute atoms in the Ge matrix. Finally, we describe the band structure of GeSn, and its dependence on the alloy composition and strain state of the material.

\subsection{GeSn: a Metastable Alloy}
\begin{figure*}[htb]
	\centering
	\includegraphics[width=\textwidth]{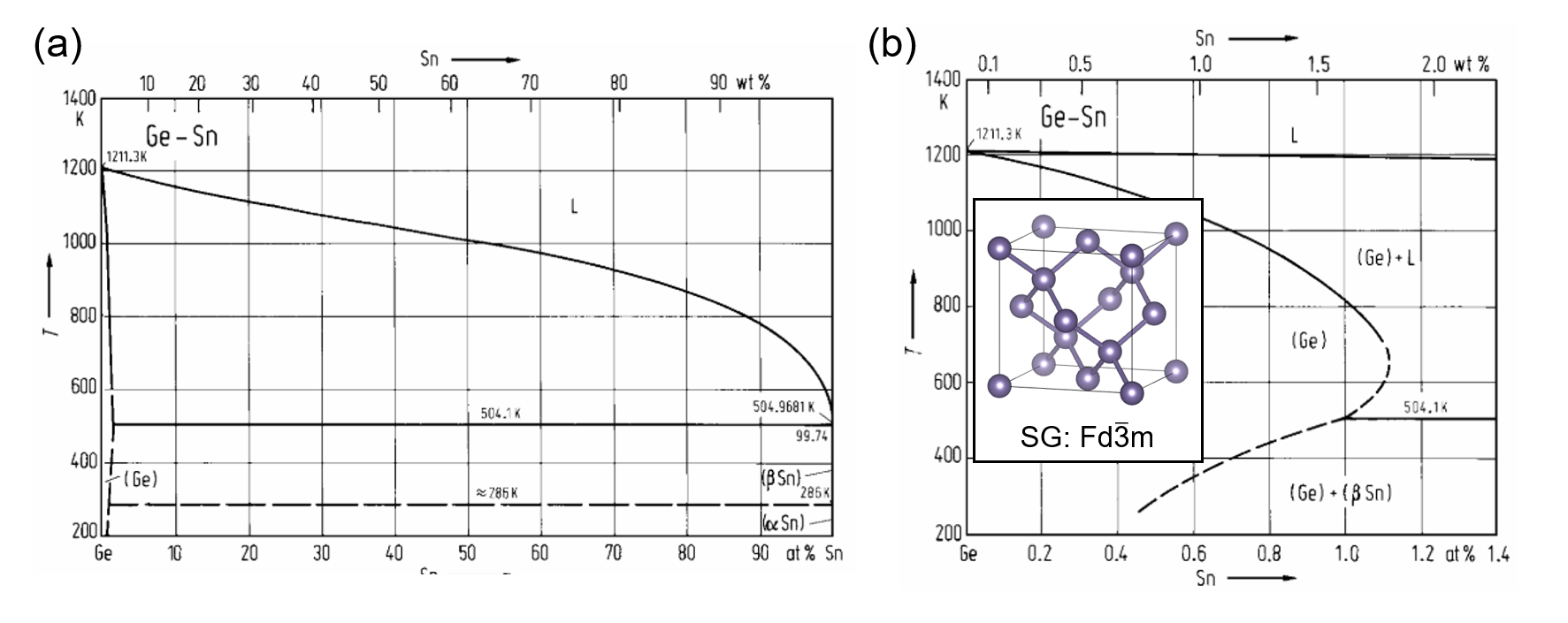}
	\caption{(a) Phase diagram of the Ge-Sn alloy, and (b) zoom on the Ge-rich compositions. In the inset of (b) is the diamond \gls{fcc} crystal unit cell with \gls{sg} $Fd\overline{3}m$, produced with the software \textit{VESTA}~\cite{Momma2011}. Figures (a,b) reproduced with permission from Predel~\cite{Predel1996}, \copyright\,1996 \emph{Springer Nature}.}
	\label{fig:GeSnPhaseDiagram}
\end{figure*}

GeSn is an alloy obtained from a solid solution of two group-IV elements, namely Ge and Sn. The Ge-Sn phase diagram is shown in Fig.~\ref{fig:GeSnPhaseDiagram}(a).
Ge is an indirect-\gls{bg} semiconductor with diamond \gls{fcc} crystal structure (\gls{sg}: $Fd\overline{3}m$), shown in the inset of Fig.~\ref{fig:GeSnPhaseDiagram}(b), with lattice constant of 5.658\,\AA~\cite{Goodman1982}.
Sn instead can be found in two phases, the high-temperature metallic \textbeta Sn phase, and the low-temperature semi-metallic \textalpha Sn phase, with a thermodynamic phase transition of 13.2\degree C.
In the case of GeSn alloys, \textalpha Sn is the phase of reference, since it also has a diamond \gls{fcc} crystal structure, with lattice constant of 6.489\,\AA~\cite{Goodman1982}.
Ge and Sn can thus be mixed in a solid solution to obtain a non-polar semiconductor (or semi-metal, depending on the composition), with the same diamond \gls{fcc} crystal structure of the two elemental materials. 
A consequence of the large lattice mismatch between Ge and \textalpha Sn of 14.7~\% is that the solubility of one element in the other is low.
Evident from the Ge-rich zoomed region of the phase diagram in Fig.~\ref{fig:GeSnPhaseDiagram}(b), the solubility of Sn in Ge is limited to a bare 1.1\atp~at 400\degree C, and drops well below 1\atp~ at room temperature.
Ge-Sn are mostly immiscible, with an eutectic temperature of 231\degree C. The single-phase GeSn is therefore metastable across most of its composition range.
Hence, phase-separation during growth or post-growth thermal processing is of concern.
If not kinetically hindered by out-of-equilibrium processes, Sn will tend to segregate out of the Ge crystal, forming a metallic \textbeta Sn phase that is detrimental for optoelectronic devices.
In this review, we discuss the Ge-rich phase of the GeSn alloy, as it is technologically more relevant for the targeted optoelectronic applications in the \gls{swir}, \gls{mwir}, and \gls{lwir} wavelength ranges.

\subsection{Crystal Lattice}
The diamond \gls{fcc} lattice parameter of GeSn  follows Vegard's law ~\cite{Polak2017,Xu2019a,Xu2017,Magalhaes2022}, i.e., it varies linearly with composition between the lattice constants of Ge and \textalpha Sn.
In agreement with \emph{ab-initio} calculations~\cite{Shen1997}, \gls{exafs} studies found that the Sn-induced strain in the GeSn alloy is accommodated by both bond stretching and bond bending, with a slightly larger contribution from the latter~\cite{Gencarelli2015,Beeler2011}.

For years, GeSn has been considered to be a homogeneous random solid solution following both computational predictions of the alloy properties~\cite{Gupta2013,Polak2017} and experimental studies~\cite{Assali2018}. A fully random solution implies the absence of any short- or long-range atomic order of the solute species.
However, recent works showed that the GeSn alloy may possess a \gls{sro}~\cite{Cao2020,Lentz2023}.
Combining statistical sampling and \emph{ab initio} calculations, Cao\etal~\cite{Cao2020} suggested the presence of a \gls{sro} at Sn atoms, with the first coordination shell nearly devoid of other Sn atoms. With this assumption, they provided better predictions of the GeSn \gls{bg} energy at high Sn contents, comparing it with experimental results previously fit with the random alloy assumption. With the latter model, the reduction in bandgap associated with increasing Sn content tends to be overestimated.
The \gls{sro} was later demonstrated experimentally by \gls{exafs} characterization of strain-free Ge$_{0.906}$Sn$_{0.094}$ nanowire shells grown around a compliant Ge core by \gls{cvd}~\cite{Lentz2023}.
The observed \gls{sro} may be rationalized in terms of the larger size of Sn atoms compared to Ge.  One can expect that self-repulsion of Sn atoms is favored to drive local strain accommodation with reduced bond distortions. This intuition is backed by a study from Fuhr\etal~\cite{Fuhr2013}, which through \gls{dft} computations observed a significant Sn-Sn repulsion in the first coordination cell of Sn~\cite{Fuhr2013}. The repulsion energy quickly drops moving away from the first coordination shell, becoming negligible at the fourth coordination cell~\cite{Fuhr2013}.

A previous experimental study by \gls{apt} reported the absence of \gls{sro} in \gls{cvd}-grown Ge$_{0.82}$Sn$_{0.18}$~\cite{Assali2018}.
The lack of observation of \gls{sro} has been suggested to be owed to the high temperatures employed during growth of the material~\cite{Windl2022}. In Ref.~\cite{Windl2022}, Windl\etal{} employed Monte Carlo methods to simulate the effect of growth temperature on the alloy by relaxing an 8000-atom cell with average composition corresponding to Ge$_{0.86}$Sn$_{0.14}$; they observed a significant reduction in \gls{sro} at the growth temperatures of 300--400\degree C, typically used in \gls{cvd} epitaxy of GeSn.
Furthermore, Lentz\etal~\cite{Lentz2023} suggested the employed growth rate may also play a role in the final alloy \gls{sro}, as their GeSn layer was grown at 275\degree C at a slow rate of 1\,nm/min in contrast with the rate of 1.5--2.8\,nm/min employed in the GeSn film stack studied by \gls{apt} in Ref.~\cite{Assali2018}.
Additional experimental studies on GeSn films are required to confirm the presence of \gls{sro} and to assess its dependence on growth conditions.

Lastly, \gls{apt} studies suggested that compressive strain in GeSn may favor the presence of Sn-Sn bonds~\cite{Liu2022a}, in contrast with the \gls{sro} observed in strain-free GeSn~\cite{Lentz2023}.
If confirmed, this would suggest the importance of strain engineering to prevent phase separation, especially in the case of GeSn with Sn content at the high end. It should be kept in mind, however, that compositional measurement artifacts may arise during \gls{apt} characterization of compound semiconductors and alloys~\cite{Muller2012}. For instance, the thermal energy introduced into the system during \gls{apt} sample preparation and measurements could promote the (surface) diffusion of Sn, leading to atomic rearrangement with respect to the original sample, and possibly local Sn segregation.

\subsection{Band Structure}
\begin{figure*}[htb]
	\centering
	\includegraphics[width=0.8\textwidth]{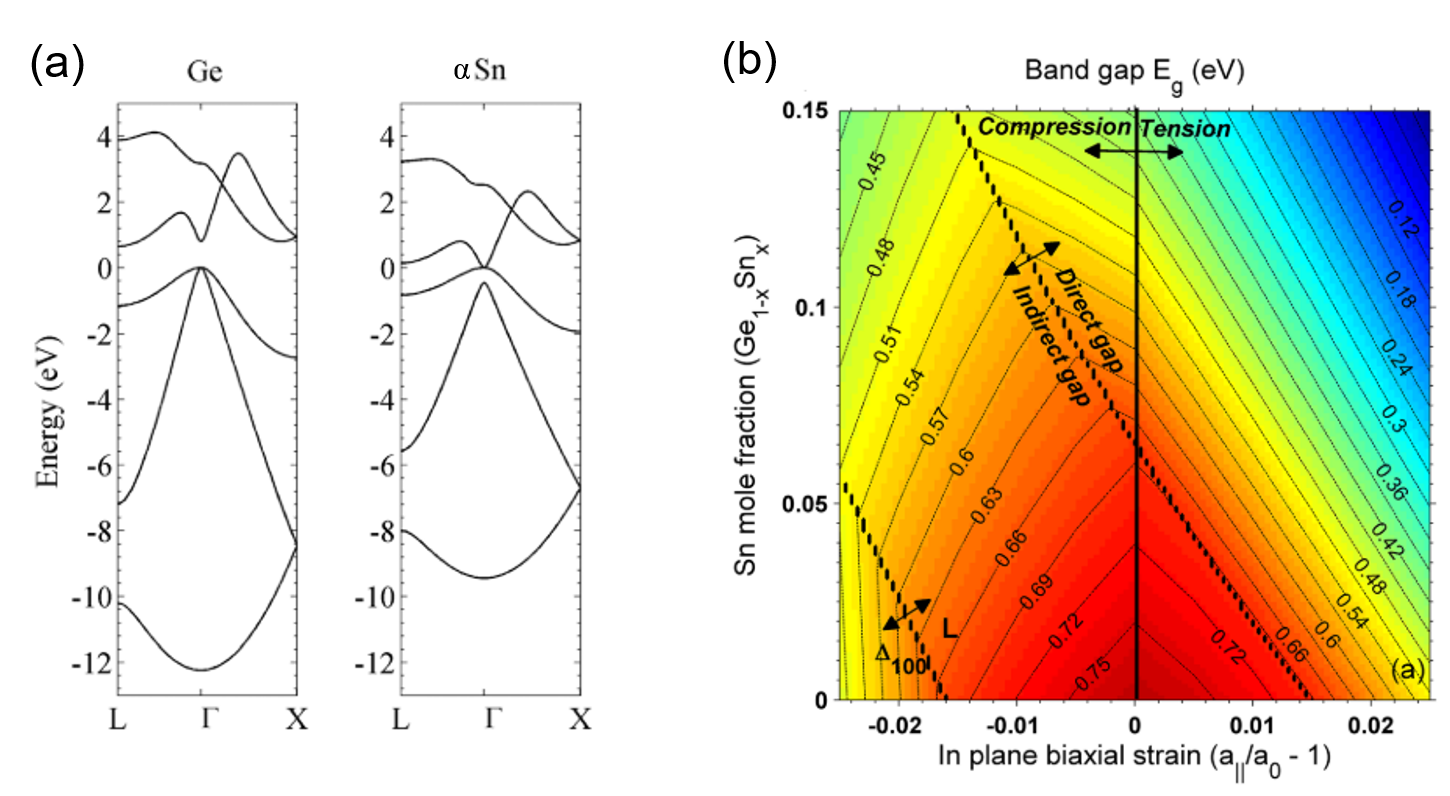}
	\caption{(a) Band structures of Ge and \textalpha Sn. (b) Calculations of \bg~energy dependence on strain, using $b_\Gamma=2.1$\,eV. Figure (a) adapted with permission from Moontragoon\etal~\cite{Moontragoon2012}, \copyright\,2012 \emph{AIP Publishing}. Figure (b) reprinted with permission from Gupta\etal\cite{Gupta2013}, \copyright\,2013 \emph{AIP Publishing}.
	}
	\label{fig:GeSnBandStructure}
\end{figure*}

The evolution of the band structure of GeSn  with the composition can be qualitatively understood first by extrapolating from the band structure of the constituent elements -- Ge and \textalpha Sn -- shown in Fig.~\ref{fig:GeSnBandStructure}(a).
Ge is an indirect-\gls{bg} semiconductor with \gls{bg} energy of 0.66\,eV at the L-valley, and direct \textGamma-valley energy of 0.80\,eV~\cite{Song2020}. \textalpha Sn, on the other hand, is a semi-metal with direct \gls{bg} energy of -0.41\,eV, and \gls{bg} energy at the L-valley of 0.09\,eV~\cite{Song2020}.
The Ge-Sn energy difference at \textGamma- and L-valleys are respectively of 1.21\,eV and 0.55\,eV.
Assuming the bandgap energies vary linearly with compositions, one can therefore expect that by adding Sn to Ge the \textGamma-point energy will red-shift faster than the L-point, inducing a cross-over from indirect to direct \bg~behavior in the material.
Experimentally, this is indeed the case, though  we have seen in Sec.~\ref{sec:GeSn_HistPersp} that the cross-over occurs at lower Sn fractions compared to the linear prediction, with the bandgap energies following a bowing behavior described by eq.~\eqref{eq:EnergyBGbowing}.
The bowing behavior of the GeSn alloy is the result of coupling of states through a non-diamond-like potential~\cite{Yin2008}. This asymmetric potential originates from the difference in electronegativity of the constituent elements, and the lattice distortions (i.e., bond stretching and bending~\cite{Gallagher2015}) due to their different atomic sizes~\cite{Yin2008}.
This explains why early works using highly symmetric potential-averaged virtual crystal approximations, such as that from Jenkins\etal~\cite{Jenkins1987}, failed to capture the large bowing behavior of the material.
On a side note, \gls{dft} studies by Yin\etal~\cite{Yin2008} found that the total bowing behavior originates equally from Ge-Sn charge and structural differences, while Chibane\etal~\cite{Chibane2010} determined a dominant contribution from the latter. Discrepancies in the result may have arisen from the different GeSn compositions and supercell sizes considered in the studies (see Tab.~\ref{tab:GeSn_BowingParam}).

In Tab.~\ref{tab:GeSn_BowingParam}, we report the measured and computed values of bowing parameters of the GeSn \textGamma- and L-energy gaps ($b_\Gamma$, $b_L$). From these results, it is clear that there is no close agreement on these values in the literature. 
In general, 2\,eV\,$<b_\Gamma<$\,3\,eV, while $b_L\sim1$\,eV.
The large variability of the reported experimental values of $b_\Gamma$ comes from the complexity of the material system: Besides possible experimental systematic errors coming from the different employed techniques, the atomistic structure of the alloy may vary depending on the employed growth technique and parameters, resulting in different physical properties.
For example, Sn-clustering has been observed in GeSn~\cite{Kumar2013,Liu2022a}, and it has been predicted to lower the material bandgap~\cite{Chibane2010}, consequently inducing an apparent larger bowing parameter.
Additionally, the strain state of the material affects \bg~energies shifts, as shown in Fig.~\ref{fig:GeSnBandStructure}(b). In particular, compressive strain has the opposite effect of Sn alloying on the \gls{cb} valleys~\cite{Gupta2013,Song2020a}, increasing the \bg~energy, and thus decreasing the apparent bowing behavior.
Strain also splits the \gls{hh}-\gls{lh} degeneracy, with compressive strain lifting the \gls{hh}, and tensile strain producing the opposite effect~\cite{Chang2010,Michel2010}.
Failure of appropriately taking into account the strain state of the material will result in experimental errors.
Concerning theoretical computations of the alloy bowing parameter, \gls{dft} model on strain-free material considered the GeSn alloy to be fully random, while recent works demonstrated the presence of a \gls{sro} in GeSn~\cite{Cao2020,Lentz2023}. The absence of Sn atoms in the first coordination shells of Sn yields a larger \bg{} energy~\cite{Chibane2010,Cao2020}, and thus $b$ is overestimated when fitted over the entire alloy spectrum; this is the case for Refs.~\cite{Polak2017,Eckhardt2014}, which report $b_\Gamma>3$\,eV.
Lastly, Gallagher\etal~\cite{Gallagher2014} found their data was better fitted with composition-dependent bowing parameter, also proposed with \gls{dft}-based calculations by Chibane\etal~\cite{Chibane2010}.
On the other hand, Yin\etal~\cite{Yin2008} computed by \gls{dft} a negligible dependence on composition of the bowing parameter. Furthermore, D'Costa\etal~\cite{DCosta2016} argued that with the composition-dependent $b_\Gamma$ from Gallagher\etal{} pseudomorphic GeSn would not show a direct-bandgap behavior for any composition. Fernando\etal~\cite{Fernando2018} predict the indirect-to-direct crossover behavior of pseudomorphic GeSn on Ge(001) to be at 26\atp, though extrapolated from samples with $x_{Sn}<0.11$, and calculated without considering any \gls{sro}. The latter is expected to further push the crossover composition to larger Sn contents~\cite{Cao2020}. More theoretical work is necessary to determine the exact behavior of the bowing parameter.
On the other hand, one needs to consider that experimentally derived GeSn bowing parameters may  exhibit variations because of inherent differences in the atomic configurations of the alloy, which may be influenced by the specific growth technique and growth parameters utilized~\cite{Lentz2023}.

\begin{table*}
	\renewcommand{\arraystretch}{1.5}
	\centering
	\caption{\label{tab:GeSn_BowingParam}Summary of bowing parameters determined by experiments and computations. In the header row, $x_{Sn, max}$ refers to the maximal Sn fraction considered in the study. Acronyms: \glsfirst{cer}, \glsfirst{vase}, \glsfirst{pr}, \glsfirst{ot}, \glsfirst{dsr}.}
	\begin{tabularx}{0.9\textwidth}{c m{0.16\textwidth}P{0.07\textwidth}P{0.08\textwidth}P{0.1\textwidth}cm{0.35\textwidth}}
		\toprule
		& \centering \arraybackslash Ref. & $b_\Gamma$~(eV) & $b_L$~(eV) & Method & $x_{Sn, max}$ & \centering \arraybackslash Comments \\
		\midrule
		\rowcolor{lightgrey}
		\cellcolor{white} & Zelazna, 2015 \cite{Zelazna2015} & 1.8 & -  & \gls{cer} & 0.11 & \makecell[{{p{\linewidth}}}]{\gls{mbe}, pseudomorphic on Ge. Compressive strain compensated in fits.} \\
		
		\parbox[b][5mm][t]{3mm}{\multirow{7}{*}{\rotatebox[origin=c]{90}{Experiments}}} & Gallagher, 2014 \cite{Gallagher2014} & \makecell[c]{$2.66-$\\$5.4x_{Sn}$} & \makecell[c]{$1.11-$\\$0.78x_{Sn}$}  & \gls{pl} & 0.11 & \makecell[{{p{\linewidth}}}]{\gls{uhv}-\gls{cvd}, on Ge/Si. Compressive strain compensated in fits.} \\
		
		  &\cellcolor{lightgrey} Jiang, 2014 \cite{Jiang2014} &\cellcolor{lightgrey} 2.46 & \cellcolor{lightgrey}1.03  &\cellcolor{lightgrey} \gls{pl} &\cellcolor{lightgrey} 0.06 & \makecell[{{p{\linewidth}}}]{ \cellcolor{lightgrey} \gls{cvd}, on Si. Compressive strain compensated in fits.} \\
		
		 & Lin, 2012 \cite{Lin2012} & 2.43 & -  & \gls{pr} & 0.064  & \makecell[{{p{\linewidth}}}]{\gls{mbe}, on lattice-matched In$_y$Ga$_{1-y}$As/GaAs.} \\
		
		
		& \makecell[{{L{\linewidth}}}]{\cellcolor{lightgrey} P\'erez Ladr\'on, 2007 \cite{LadrondeGuevara2007}} & \cellcolor{lightgrey} 2.30 & \cellcolor{lightgrey} -  & \cellcolor{lightgrey} \gls{ot} & \cellcolor{lightgrey} 
		0.14 &  \makecell[{{p{\linewidth}}}]{\cellcolor{lightgrey} \gls{ms}, on Ge(100). Strain unspecified, thickness of few hundred nm.}  \\
		
		& D'Costa, 2006 \cite{DCosta2006} & 2.3 & -  & \makecell[c]{\gls{vase}\\ \& \gls{pr}} & 
		0.20 & \makecell[{{p{\linewidth}}}]{\gls{uhv}-\gls{cvd}, on Si(100). Slight compressive (\gls{dsr}$>90$\%).} \\
		
		\rowcolor{lightgrey}
		\cellcolor{white} & He, 1997 \cite{He1997} & 2.8 & -  &  \gls{ot} &  0.15 & \makecell[{{p{\linewidth}}}]{ \gls{mbe}, on Ge/Si(100). Fully relaxed within experimental error.} \\
		
		\addlinespace
		\midrule
		\addlinespace
		
		& Song, 2020 \cite{Song2020} & 2.28 & 2.86  & \makecell[c]{30-band\\ k$\cdot$p} & 0.3 & Strain-free. Less precise than \gls{dft}. \\
		
		\parbox[b][5mm][t]{3mm}{\multirow{6}{*}{\rotatebox[origin=c]{90}{Computations}}}
		& \cellcolor{lightgrey} Polak, 2017 \cite{Polak2017} &\cellcolor{lightgrey}  3.02 &\cellcolor{lightgrey}  1.23  & \cellcolor{lightgrey} \gls{dft} &\cellcolor{lightgrey}  1 &\cellcolor{lightgrey}  Strain-free. 54-atom supercell, fully random alloy \\

		& Zelazna, 2015 \cite{Zelazna2015} & 1.97 & 0.26  & \gls{dft} & 0.08 & Strain-free. 54-atom supercell, non-random \\
		
		& \cellcolor{lightgrey} Eckhardt, 2014 \cite{Eckhardt2014} & \cellcolor{lightgrey} 3.1 & \cellcolor{lightgrey} -  &  \cellcolor{lightgrey} \gls{dft} & \cellcolor{lightgrey} 1 & \cellcolor{lightgrey} Strain-free. 216-atom supercell, fully random alloy \\
		
		& Chibane, 2010 \cite{Chibane2010} & 1.94 & 0.90  & \gls{dft} & 0.25 & \makecell[{{p{\linewidth}}}]{Strain-free. 16-atom supercell, Sn-clustered \& non-clustered cells.} \\
		
		& \cellcolor{lightgrey} Yin,2008 \cite{Yin2008} &\cellcolor{lightgrey} 2.55 &\cellcolor{lightgrey} 0.89  &\cellcolor{lightgrey} \gls{dft} &\cellcolor{lightgrey} 
		0.75 &\cellcolor{lightgrey} Strain-free. 64-atom supercell, fully random alloy. \\
		\bottomrule
	\end{tabularx}
\end{table*}

Until recently, GeSn was expected to show a hard crossover from an indirect to direct~\bg{} behavior. Several groups have tried to measure and/or compute the crossover alloy composition.
Inevitably, the large scatter in the reported values of bowing parameter (see Tab.~\ref{tab:GeSn_BowingParam}) resulted in an equally large scatter in the reported alloy crossover compositions, spanning between 6\atp{} and 10\atp{} Sn~\cite{Liu2019}.
In fact, recent works demonstrated that GeSn has a soft transition from an indirect to a direct~\bg{} behavior due to band mixing~\cite{Eales2019,OHalloran2019,Moutanabbir2021}.
In GeSn, the \gls{cb} edge states were found to consist in a linear combination of the $\mathrm{\Gamma}$ ($\mathrm{\Gamma_{7c}}$) and L states ($\mathrm{L_{6c}}$). This behavior originates from the large differences in covalent radius and electronegativity of the alloy constituent elements and was demonstrated by measuring the GeSn bandgap hydrostatic pressure coefficient ($dE_g/dp$) of GeSn~\cite{Eales2019}:
In Ge, the pressure coefficient of the direct \bg{} ($\mathrm{\Gamma_{7c}-\Gamma_{8v}}$) is 3 times that of the indirect \bg{} ($\mathrm{L_{6c}-\Gamma_{8v}}$). By measuring $dE_g/dp$, it is thus simple to discern the \bg{} behavior of the material.
In GeSn, experimental values of $dE_g/dp$ progressively increase from $dE_L/dp$ to $dE_\Gamma/dp$ for increasing Sn content ($x_{Sn}<0.15$~\cite{Moutanabbir2021}), indicating that the band mixing evolves progressively with $x_{Sn}$~\cite{Eales2019}.
\gls{dft} computations confirmed the progressive increase of $dE_g/dp$ in GeSn, corroborating the experiments~\cite{Eales2019,OHalloran2019}.
Furthermore, O'Hallaran\etal~\cite{OHalloran2019} found evidence of band mixing also in the previously reported \gls{dft} calculations, e.g., from Polak\etal~\cite{Polak2017}, emphasizing the importance of atomistic computations in capturing the electronic behavior of the GeSn alloy.
The band mixing behavior may explicate to some extent the variability in the reported experimental crossover alloy compositions.

In general, we can conclude that the lattice ordering and band structure of the GeSn alloy are fairly well understood qualitatively, but still lack accurate quantitative interpretation. Nevertheless, the data acquired until now allowed significant progress in the performance of GeSn-devices, reviewed in Sec.~\ref{sec:GeSn_HistPersp}.

\section{Challenges of the GeSn Alloy}
\label{sec:GeSn_Challenges}
The main challenges associated with GeSn processing are discussed in the following two sections. First, we provide a brief introduction to the challenges related to epitaxial strain relaxation via dislocation formation. This subject is further elaborated in Secs.~\ref{sec:GeSn-Ge_StrainRelax} and \ref{sec:GeSn_ElecProp}. Secondly, we delve into the processes of Ge-Sn phase separation and Sn segregation, which may occur during and after growth as a result of the alloy metastability. We review the efforts towards understanding the driving forces at play in this process. 

\subsection{Epitaxial Relaxation Defects}
\label{sec:Challenges_EpitaxialRelaxation}
In this section, we introduce the mechanism for epitaxial strain relaxation via the formation of misfit and threading dislocations. We furthermore review simple models for critical thickness prediction for the GeSn-on-Ge material system. Additionally, we address the challenges associated with the presence of dislocations in GeSn, and a few methods proposed to overcome them.

To maximize the material performance, GeSn should be integrated in devices as a monocrystal, since group-IV grain boundaries are known to be sources of trap states~\cite{Landwehr1985,Vladimirov2018,Tsurekawa2005,Imajo2022} and they act as scattering centers during charge transport~\cite{Grundmann2010}.
Monocrystalline GeSn can be obtained by epitaxial growth on substrates with a suitable crystal structure and similar lattice parameter.
Undoubtedly, the ideal substrate from the technological point of view is Si, which is widely commercially available and has a diamond cubic \gls{fcc} structure with  a   lattice constant of 5.431\,\AA{}, which corresponds to a lattice mismatch of 4.18\% with pure Ge.
GeSn, possessing a lattice parameter larger than Ge, will thus grow in a compressive strain on Si substrates.
With such large lattice mismatch, as the epitaxial film grows coherently on the substrate, it accumulates enough elastic energy to overcome the nucleation energy of dislocations, and the strain is thus (partially) relaxed via the formation of \acrfullpl{md} and \acrfullpl{td}, as shown in Fig.~\ref{fig:GeSnCriticalThickness}(a).
The \gls{tcr} on Si is only a few\,nm~\cite{Jesser1993}, and thus relaxation defects in epitaxial GeSn are unavoidable for any film thickness that would be technologically relevant for optoelectronic applications.

\subsubsection*{Critical thickness for strain relaxation ($t_{cr}$)}
\begin{figure*}[htb]
	\centering
	\includegraphics[width=0.9\textwidth]{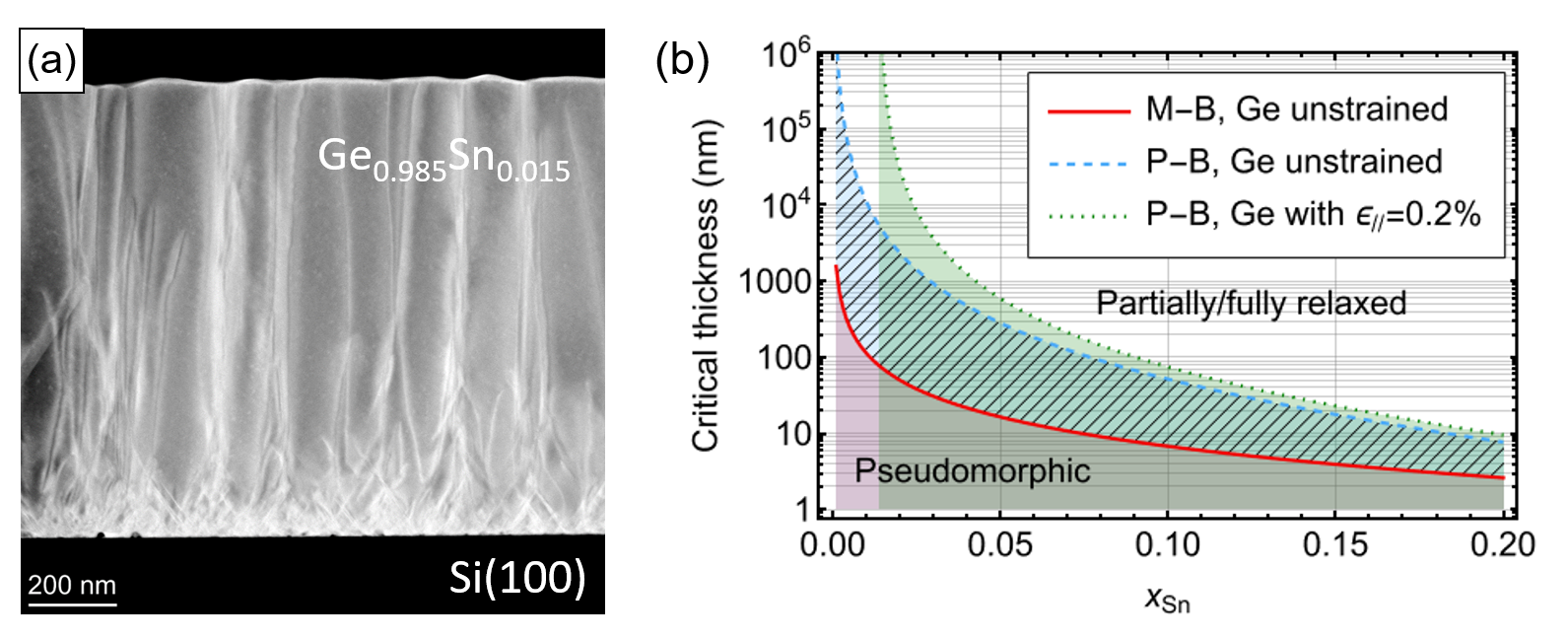}
	\caption{(a)\,\Glsfirst{xtem} of Ge$_{0.985}$Sn$_{0.015}$ epitaxially grown on Si(100) by \gls{ms} by the authors. The film is rich in dislocations due to epitaxial strain relaxation phenomena, elucidated in detail in Sec.~\ref{sec:GeSn-Ge_StrainRelax}. (b) In red, \acrfull{pb} thermodynamic model for GeSn critical thickness (\gls{tcr}) on Ge(100) substrates. The blue, dashed line represents the \gls{pb} model calculated for GeSn growth on a tensile-strained Ge layer, with in-plain strain ($\epsilon_{//}$) of 0.2\%. On tensile-strained Ge, the GeSn \gls{tcr} is increased as a consequence of the reduced lattice mismatch.}
	\label{fig:GeSnCriticalThickness}
\end{figure*}

To partially accommodate compressive strain, GeSn is grown on Ge-buffered Si substrates (also called \gls{vge}), or directly on Ge wafers. This yields considerably larger \gls{tcr}, which however drops as the lattice mismatch increases with the Sn fraction in the alloy. The dependence of  \gls{tcr} on GeSn composition is plotted in Fig.~\ref{fig:GeSnCriticalThickness}(b) using the \gls{mb} and \gls{pb} models, and assuming the strain is released by 60\degree \glspl{md}. The solid coloring indicates regions where the film grows pseudomorphically -- i.e., no strain relaxation occurs in the film -- according to the corresponding model.
The \gls{mb} model is based on a mechanical equilibrium approach and assumes that threading dislocations are already present in the film, inherited from the substrate~\cite{Matthews1974}. This is the case of GeSn grown on \gls{vge}.
In this model, \gls{tcr} is found by equating the force felt by a \gls{td} segment under misfit stress to the tension in the dislocation core.
The \gls{mb} model is an \emph{equilibrium} model. It turns out that it does not describe accurately the relaxation process and the critical thickness of group-IV semiconductors~\cite{Nix1989} because their \gls{tcr} is in fact dominated by the kinetics of dislocation nucleation and motion. These  kinetic factors, which are dependent on growth temperature and rate, can be taken into account with more complex  models capable of describing metastable strain states of the epitaxial films, reviewed in Ref.~\cite{Houghton1991}.
An alternative, simpler model has been proposed by People and Bean to describe the metastable critical thickness of epitaxial group-IV films~\cite{People1985,People1986}. The \gls{pb} model is based on a thermodynamic energy-balance  approach, where the \gls{tcr} is obtained by equating the misfit strain energy density to the self-energy of a dislocation. This model assumes that this condition is sufficient for the nucleation of a dislocation, ignoring the effective kinetics of dislocation nucleation.
Despite of this, the \gls{pb} model is widely used. It predicts relatively accurately the experimental critical thickness of GeSn grown on Ge~\cite{Wang2015,LadrondeGuevara2003}.
The higher accuracy of the \gls{pb} model compared to the \gls{mb} can be explained by the fact that GeSn growth occurs far from equilibrium, where kinetics of dislocations play a major role in determining strain relaxation. This leads to GeSn films retaining a metastable strain state, implying that strain relaxation may occur when the material is heated to temperatures exceeding those utilized during its growth.

\subsubsection*{The need for dislocation engineering}
Dislocations not only limit the thermal stability of GeSn~\cite{Nicolas2020,Mukherjee2021}, but they are also a source of electronic deep trap states~\cite{Giovane2001,Kondratenko2020,Tetzner2021}, which are detrimental for the performance of GeSn-based optoelectronic devices.
For example, in GeSn \glspl{pd}, trap states increase the dark currents via \gls{srh} and \gls{tat} carrier generation mechanisms in the junction depletion regions~\cite{Sze2006}.
Hence, epitaxial relaxation defects require appropriate management for the optimization of optoelectronic device performance.
Extensive research on dislocation engineering showed that the \gls{tdd} in Ge and GeSn films can be reduced via nano- and micro-patterning of the substrate~\cite{Park2007,Luan1999}, thick (graded) buffers~\cite{Aubin2017}, thermal processing~\cite{Loo2009,Wang2009}, or any combination of these strategies~\cite{Luan1999}.
A more thorough discussion of strain relaxation mechanisms and defects will follow in Secs.~\ref{sec:GeSn-Ge_StrainRelax} and~\ref{sec:GeSn_ElecProp}. Because of the technological relevance of Ge buffers in the realization of GeSn devices, the growth of Ge, its relaxation phenomena, and its electrical properties are presented as well throughout this review.

\subsubsection*{The Nanowire case}
Finally, we should add some words on a particular case that enables the growth of GeSn without dislocations and or stacking faults. This is the case of GeSn obtained around ultra-thin Ge nanowires (NWs). NWs are filamentary crystals with a tailored diameter ranging from the few to few tens of nm. Their aspect ratio renders them new properties and enables a large variety of functional applications that include electronics, optoectronics and sensing~\cite{Yuan2021,Guniat2019,McIntyre2020,Thelander2006}. Their reduced diameter also enables the combination of materials along or across the axis that would not otherwise be possible in a planar substrate. This is due to the more facile strain relaxation both axially and laterally~\cite{Glas2006}. 

The most common method to obtain NWs is the so-called Vapor-Liquid(Solid)-Solid growth (VLS/VSS). Here, a liquid (or solid) metal droplet is used as a catalyst to preferentially gather and decompose growth precursors. Upon supersaturation of the droplet, there is precipitation underneath. Continuous precipitation leads to the formation of the filamentary crystal~\cite{Wagner1964}. Free-standing GeSn NWs with Sn content up to 4\% have been obtained by VLS using a mixture of AuAg or Au that facilitates the dissolution of both Ge and Sn~\cite{Doherty2019,Haffner2017}. 
More commonly, Ge NWs are used as a template to grow GeSn shells around them~\cite{Meng2016,Assali2017}. Using ultra-thin Ge NWs strongly avoids strain relaxation via structural defects~\cite{Albani2018,Meng2019}. As a consequence, Ge/GeSn core/shell NW systems constitute a benchmark for the functional properties of GeSn free of extended defects. Several studies reveal that indeed the properties of such GeSn are better than in thin films, including room temperature luminescence, photodetection and high carrier mobility~\cite{Meng2020,Lentz2023,Kang2021,Luo2024}. The NW geometry brings advantages both in the understanding of GeSn as a semiconductor and in the foresight and potential of GeSn technology.

\subsection{Sn Segregation and Thermal (In)stability}
\label{sec:GeSn_Challenges_Segregation}
In this section, we discuss the driving forces at play in Sn segregation in metastable GeSn, with the implications on the material thermal stability.
We review the current understanding of Sn diffusion phenomena, the mediating role of defects in GeSn, and the efforts to elucidate the interplay between the alloy composition, strain relaxation and Sn out-diffusion.

\begin{figure*}[htb]
	\centering
	\includegraphics[width=\textwidth]{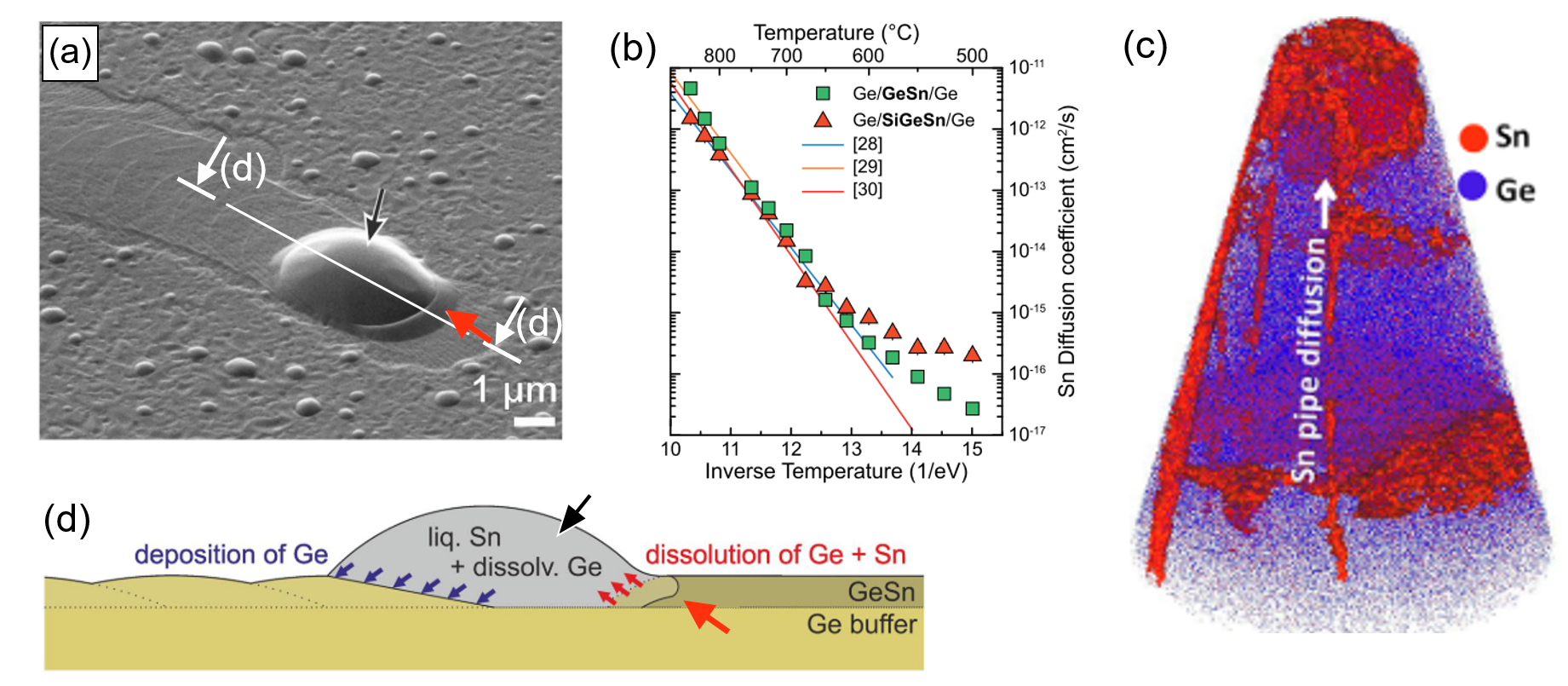}
	\caption{(a)\,\gls{sem} image of a \textbeta Sn segregation droplet  on the surface of a GeSn film. (b)\,Activation energy of Sn diffusion in GeSn alloys in function of the annealing temperature, evidencing two regimes of diffusion: (1) vacancy-mediated Sn diffusion at high temperatures, and (2) metastability-enhanced Sn diffusion at low temperatures. References [28,29,30] in this figure correspond to Refs.~\cite{Riihimaki2007,Kringhoj1994,Friesel1995} here. (c) Pipe diffusion of Sn along a dislocation core, as measured by \gls{apt}. (d) Cross-section schematic view of a Sn droplet on surface. Figures (a,d) adapted from Ref.~\cite{Groiss2017}, under terms of the CC-BY license. Figure (b) reprinted from Ref.~\cite{VondenDriesch2020}, under terms of the CC-BY license.  Figure (c) reprinted with permission from Nicolas\etal~\cite{Nicolas2020}, \copyright\,2020 \emph{American Chemical Society}.}
	\label{fig:GeSnSegregation}
\end{figure*}

GeSn is metastable for all compositions with $x_{Sn}>1$\atp.
According to the Ge-Sn phase diagram in Fig.~\ref{fig:GeSnPhaseDiagram}, at room temperature thermodynamics predict a phase separation of GeSn into a Ge-rich phase with less  1\atp~Sn, and a \textbeta Sn phase.
This process is kinetically hindered by the low atomic diffusivity of Sn and Ge at room temperature.
On the other hand, upon thermal processing of the material, may it be for annealing, doping, or any \gls{cmos} post-growth process that requires heating, Sn atoms may acquire sufficient thermal energy to diffuse to the surface of the material~\cite{VondenDriesch2020,Li2013}, and/or cluster in the bulk into a \textbeta Sn phase~\cite{Li2013}.
Sn segregation may also occur during growth if the substrate temperatures are too elevated~\cite{Groiss2017,Kuchuk2022,Wu2020}. The resulting film will show several Sn droplets on surface, similar to the one reported in the \gls{sem} image of Fig.~\ref{fig:GeSnSegregation}(a).
The Sn out-diffusion from the Ge matrix increases the effective bandgap of the material, defeating the purpose of alloying with Sn. Additionally, the metallic behavior of the segregated \textbeta Sn phase is detrimental for optoelectronic devices.
Hence, in general, it is necessary to prevent the Sn segregation by maintaining a low thermal budget both during growth and post-growth processes. Exceptions  where the Ge-Sn phase separation is sought exist~\cite{Schulte-Braucks2017}.

\subsubsection*{Driving forces for Sn segregation}
Several experimental investigations have elucidated the driving forces for Sn segregation and the influence of material defects on the latter.
It is widely accepted that Sn diffusion in Ge is mediated by vacancies~\cite{Friesel1995,Riihimaki2007,Panayiotatos2017}, whose formation energy in group-IV semiconductors is known to depend on the strain state of the material~\cite{Antonelli1989,Aziz1997,Choi2010}.
In particular, the vacancy formation energy decreases under compressive strain. Therefore, in compressively strained GeSn films, the associated increase in vacancy concentration is  expected to increase the Sn diffusion coefficient, facilitating Sn segregation.
However, a study from von den Driesch\etal~\cite{VondenDriesch2020} in nearly dislocation-free GeSn films showed that this cannot be the only contributor to Sn out-diffusion.
In fact, they observed that given an equal Sn composition and film compressive strain, Sn out-diffusion in Si$_{0.040}$Ge$_{0.895}$Sn$_{0.065}$ is accelerated compared to Ge$_{0.94}$Sn$_{0.06}$ due to the larger metastability of the ternary alloy, concluding that the main driving force for Sn out-diffusion is given by the thermodynamic instability of the material.
In support of this conclusion, they further observed two regimes of Sn diffusion depending on the annealing temperature, shown in Fig.~\ref{fig:GeSnSegregation}(b).
By gradually increasing the annealing \gls{t} from 500\degree C, they first found a low activation energy\,\footnote{\,The atomic diffusion coefficient follows the Arrhenius-type law $D=D_0\exp(-E_a/kT)$, where $E_a$ is the activation energy for diffusion.} regime with enhanced diffusion due to the large metastability of the material. Around 650\degree C, as the Sn progressively diffused out of the Ge matrix, its diffusion activation energy increased and  the diffusion coefficient matched that of Sn in non-metastable, Sn-doped Ge.
The latter is plotted in Fig.~\ref{fig:GeSnSegregation}(b) using activation energies from three different references~\cite{Riihimaki2007,Kringhoj1994,Friesel1995}, and it is shown to match the Sn diffusion coefficients measured during annealing of Si$_{0.040}$Ge$_{0.895}$Sn$_{0.065}$ and Ge$_{0.94}$Sn$_{0.06}$ above 650\degree C in Ref.~\cite{VondenDriesch2020}.
The authors attributed the increase in activation energy at high temperatures to the loss of metastability-enhanced diffusion as the Sn concentration decreased to values within the solubility limit in Ge~\cite{VondenDriesch2020}. They ruled out this effect was purely from a composition-dependent diffusion coefficient as they did not observe the typical Gaussian shapes of composition-dependent diffusion profiles~\cite{VondenDriesch2020}.

As a consequence of the strong influence of the alloy metastability on Sn segregation, the thermal stability of GeSn decreases with increasing Sn content, making it progressively more difficult to achieve large Sn fractions in GeSn.
Zaumseil\etal~\cite{Zaumseil2018} showed that the decrease is almost linear with Sn content, with phase-separation temperatures reported to be between 600\degree C and 350\degree C for Sn fractions respectively  between 4.8\atp{} and 11.7\atp. This is in close agreement with other reports~\cite{Xu2022,Conley2014a}.

\subsubsection*{Influence of extended defects on Sn segregation}
The presence of extended defects in GeSn has been found to strongly affect the Sn segregation behavior during thermal treatments.
In pseudomorphic GeSn, Sn segregation occurs gradually via vacancy-mediated diffusion~\cite{Zaumseil2018,VondenDriesch2020}, while in strain-relaxed GeSn the segregation process is considerably different due to the presence of \glspl{md} and \glspl{td} in the film.
By investigating the atomic structure of segregated GeSn by \gls{apt}, Mukherjee\etal~\cite{Mukherjee2021} observed that during annealing of GeSn -- with step-graded increase in Sn content from 8\atp{} to 18\atp{} -- Sn atoms tend to diffuse and accumulate at the cores of both at \glspl{md}~\cite{Zaumseil2018} and \glspl{td}~\cite{Mukherjee2021,Nicolas2020}.
This is rationalized in terms of larger space available for Sn atoms at the dislocation cores, with consequent lower lattice local strains with respect to substitutional Sn atoms in the bulk.
Fig.~\ref{fig:GeSnSegregation}(c) shows a Sn-decorated \gls{td}, as reconstructed from \gls{apt} in the work of Nicolas\etal~\cite{Nicolas2020}.
The upward increase in Sn concentration in the \gls{td} core along the epitaxial growth direction allowed to confirm that \glspl{td} act as preferential pathways for Sn diffusion to the film surface, a mechanism referred to as \emph{pipe diffusion}~\cite{Mukherjee2021}.
The Sn diffusion coefficient is estimated to increase up to 4 orders of magnitude in presence of linear defects~\cite{Zaumseil2018,Nicolas2020}.
Moreover, Mukherjee\etal~\cite{Mukherjee2021} observed that, for a short annealing time, while Sn was accumulating at dislocation cores, the GeSn film was pristine far from the core, with no sign of Sn clustering. This observation allowed to conclude that dislocations do not merely facilitate Sn segregation, but they also act as initiators of phase separation as a consequence of the local strain fields generated around their core.
Similarly, it has been suggested that during strain relaxation Sn is harvested by \glspl{td} as they diffuse in the film, further accelerating Ge-Sn phase separation~\cite{Stanchu2023}.
Hence, Sn segregation is enhanced in presence of relaxation defects, also confirmed by other systematic studies: Stanchu\etal\cite{Stanchu2021} showed that Sn segregation increased with higher density of dislocations, while Bonino\etal~\cite{Bonino2022} demonstrated an improved thermal stability by etching relaxation defects in microstructured GeSn, achieving stability of Ge$_{0.831}$Sn$_{0.169}$ for temperatures as high as 400\degree C. This thermal stability range is considerably higher than what observed by Zaumseil\etal{} for relaxed films of similar alloy compositions~\cite{Zaumseil2018}.
Finally, the presence of a GeSn surface oxide -- formed by annealing GeSn in rough vacuum -- has also been found to affect the Sn segregation behavior, inhibiting Sn out-diffusion and improving the material thermal stability~\cite{Braun2022}.

\subsubsection*{Model of Sn segregation}
The Sn segregation behavior during annealing of GeSn has been modeled by Groiss\etal~\cite{Groiss2017}. The authors showed that, once a consistent amount of Sn atoms out-diffuse and accumulate on surface, they nucleate liquid Sn droplets. These droplets move on surface through a self-maintained segregation process, schematized in Fig.~\ref{fig:GeSnSegregation}(d); liquid droplets of Sn have a tendency to dissolve the metastable GeSn film at their leading edge, which leads to the deposition of a Ge layer at the trailing edge as a result of Ge supersaturation in the Sn droplet. This Ge layer contains a Sn concentration near its solubility limit.
Due to a better wetting of Sn on the GeSn film compared to Ge, the droplet tends to advance  in the opposite direction from the Ge deposited layer, often along the \hkl<110> and \hkl<100> crystal orientations. The wetting behavior explains the motion of droplets on the surface of the GeSn film, which leave a trail of precipitated Ge at the back, similarly to the Vapor-Liquid-Solid process~\cite{Azrak2018}. This results in characteristic trails associated with \textbeta Sn segregation droplets on the surface of phase-separated GeSn films, visible in Fig.~\ref{fig:GeSnSegregation}(a).
Besides the surface phenomena modeled by Groiss\etal~\cite{Groiss2017}, upon annealing of (partially) relaxed films, Sn clustering has been observed in the bulk~\cite{Li2013} and at the film/substrate interface~\cite{Zaumseil2018}.
Bulk Sn clustering has been linked with a large concentration of vacancies induced by the low-temperature growth of GeSn~\cite{Zaumseil2018}, while clustering at the interface occurs due to accumulation of Sn at the core of \glspl{md}~\cite{Zaumseil2018}.

\subsubsection*{Interplay between alloy composition, strain relaxation and Sn segregation}
In light of the above results, several studies tried to elucidate the interplay between alloy composition, strain relaxation and Sn segregation during annealing of GeSn.
With in-situ annealing \gls{xrd} studies, Zaumseil\etal~\cite{Zaumseil2018} observed a gradual Sn out-diffusion with increasing temperature in \gls{rpcvd}-grown Ge$_{0.95}$Sn$_{0.05}$  pseudomorphic films on fully relaxed \gls{vge}. In these pseudomorphic samples, the epitaxial strain was progressively released as a consequence of the reduction in lattice parameter of the film as Sn diffused out of the Ge matrix.
Comrie\etal~\cite{Comrie2016a} demonstrated, however, that strain relaxation in GeSn can occur without any Sn out-diffusion, and thus that the material metastability is not a driving force for plastic strain relaxation. The latter is governed purely by dislocation nucleation and diffusion dynamics. By annealing Ge$_{0.935}$Sn$_{0.065}$ on a \gls{vge} substrate, Comrie\etal{} observed by \gls{xrd} a change in lattice parameter for \gls{t}\,$>400$\degree C, with no change in Sn substitutional fraction, as evidenced by \gls{rbs} channeling contrast measurements. They could thus conclude that plastic strain relaxation in GeSn occurred without any Sn out-diffusion, which instead took place for \gls{t}\,$>600$\degree C.
In agreement with these results, Zaumseil\etal~\cite{Zaumseil2018} showed that partially relaxed films with Sn content from 5\atp{} to 12\atp{} underwent an initial strain relaxation stage starting at approximately 300\degree C, with no Sn segregation. This occurred via the elongation of \glspl{md} at the GeSn/Ge interface, and was followed by a  ``sudden'' complete Sn segregation  out of the Ge matrix when a higher critical temperature was reached~\cite{Zaumseil2018}. The latter critical temperature decreased with increasing Sn content, shifting from 640\degree C for 5\atp{} to 350\degree C for 12\atp{} of Sn.
The sudden Sn segregation was attributed to the activation of fast Sn pipe diffusion through \glspl{td}.
In the study from Zaumseil\etal~\cite{Zaumseil2018}, the onset of vacancy-mediated Sn out-diffusion occurred at the same temperature in pseudomorphic and metamorphic Ge$_{0.95}$Sn$_{0.05}$ ($T\sim500$\degree C), but the segregation initially took place at a higher rate in the former, suggesting the existence of a driving force from the compressive strain present in the pseudomorphic film. Despite of these differences, it is interesting to notice that both samples reached the equilibrium Sn concentration of 1\atp{} at approximately 650\degree C.
On the other hand, in metamorphic films with larger composition -- i.e., $x_{Sn}=0.09,0.12$ -- the critical temperature for pipe diffusion was lower than the onset of vacancy-mediated diffusion, and therefore the latter was not observed prior to full Sn segregation.
In agreement with Ref.~\cite{Zaumseil2018}, strain relaxation through solely Sn out-diffusion was also observed in Ref.~\cite{VondenDriesch2020} in pseudomorphic \gls{rpcvd}-grown GeSn with 6\atp{} and 9\atp~Sn on fully relaxed \gls{vge}. Here, no dislocation nucleation took place during the process.
On the other hand, in GeSn grown by \gls{mbe} on \gls{vge} substrates, relaxation of Ge$_{0.92}$Sn$_{0.08}$ pseudomorphic films was observed to occur first via dislocation formation up to temperatures of $\sim550$\degree C, and then also through Sn out-diffusion~\cite{Jia2020,Li2013}. Contrary to Ref.~\cite{Zaumseil2018}, no ``sudden'' segregation of Sn through pipe diffusion was observed, possibly due to the different experimental time scales in play (40-sec-\gls{rta} in Ref.~\cite{Jia2020} and 5-min-\gls{rta} in Ref.~\cite{Li2013}, as opposed to 15--20\,min for each 12.5\degree C step in \emph{in situ} \gls{xrd} in Ref.~\cite{Zaumseil2018}).
The different GeSn strain relaxation behavior in these works may arise from the different  \glspl{tdd} present in the employed Ge buffers. In particular, in Ref.~\cite{Zaumseil2018} they specify the Ge buffer \gls{tdd} to be $<1\cdot10^7$\,cm$^{-2}$, while in Refs.~\cite{Li2013} and~\cite{Jia2020} the Ge buffers have a thickness of respectively 210\,nm and 250\,nm, in which we can expect in the best case a \gls{tdd} of $\sim4\cdot10^8$\,cm$^{-2}$~\cite{Wang2009,Loo2009}.
This \gls{tdd} is transferred to the epitaxial GeSn, facilitating its strain relaxation via \gls{td} bending, as elucidated briefly in Sec.~\ref{sec:Challenges_EpitaxialRelaxation}.
In addition, to explain the discrepancies in the reported studies, the presence of vacancies generated by the low-temperature \gls{mbe} growth may also have played a role in the interplay between relaxation and segregation, favoring Sn diffusion or conversely decreasing the metastable driving force for segregation due to local stabilization through the formation of Sn-vacancy complexes. 
Altogether, the above-cited studies converge in that the critical temperatures for Sn out-diffusion and plastic strain relaxation in GeSn are independent of each other. Nevertheless, it is important to note that these two processes do exert mutual influences. On the one hand, strain can be relaxed purely via Sn out-diffusion in pseudomorphic films; on the other hand, the presence of dislocations can facilitate Sn out-diffusion via pipe diffusion phenomena. Lastly, systematic studies with comparable time scales are required to shed light on the role of point defects arising from the different growth conditions of GeSn.

\subsubsection*{Sn segregation during epitaxial growth}
Until now we have discussed phase separation of the metastable GeSn alloy during \gls{pda} of the material. It is however important to make a distinction with the Sn segregation behavior during the growth process itself.
While phase separation during \gls{pda} requires bulk Sn diffusion, during epitaxy Sn segregation can occur via adatom surface diffusion and  nucleation of \textbeta Sn droplets.
Surface adatom diffusion rates are considerably more elevated than Sn bulk diffusion, as the latter requires breaking of covalent bonds.
As a result, the processing temperatures withstood by GeSn during growth are lower than the thermal stability after growth~\cite{Tsukamoto2015a,Mukherjee2021}.
Evidently, Sn segregation is also facilitated by larger Sn nominal fractions  during growth as a result of the increased concentration of Sn adatoms and consequent larger \textbeta Sn nucleation probability~\cite{Tsukamoto2015a}.
On the contrary, increasing the growth rate of the film reduces the adatom diffusion length, hindering Sn segregation~\cite{Taoka2016}.
Sn surface segregation can also be enhanced during \emph{in situ} doping of GeSn due to co-segregation with dopant species~\cite{Giunto2023}.

In conclusion, the above-cited works illustrate the physics of Ge-Sn phase separation and its driving forces. The interplay between strain state of the material, defects, and alloy composition influences its thermal stability. During growth, the thermal stability of GeSn is decreased further.

\section{Epitaxial Growth of GeSn, Ge}
\label{sec:GeSn_Epitaxy}
In this section, we outline the primary advancements in GeSn epitaxy, and the efforts towards achieving a higher incorporation of Sn in the alloy for \gls{mwir} and \gls{lwir} devices. While epitaxial growth by \gls{cvd} and \gls{mbe} methods has been extensively reviewed~\cite{Wirths2016,Zaima2015,Zheng2018,Miao2021,Moutanabbir2021}, we aim to complement these works by reviewing the state of art of sputtering epitaxy of Ge and GeSn. Additionally, we outline the general strategy employed to grow Ge buffer layers on Si substrates, necessary to prevent 3D growth and islanding.

\subsection{Epitaxy of GeSn towards Higher Sn Content}
\subsubsection*{The need of out-of-equilibrium growth}
GeSn must be grown epitaxially in monocrystalline form to obtain the desired optoelectronic properties in the material. For technologically relevant studies, GeSn growth is investigated on Si(100) substrates, with or without Ge buffer. To study pseudomorphic GeSn free of linear defects, GeSn is sometimes grown directly on Ge(001).
Since the early studies on GeSn thin-film growth, researchers emphasized the need of out-of-equilibrium synthesis methods to prevent Sn segregation.
Several epitaxial techniques have been thereby successfully reported with compositions up to 10\atp{} Sn,  mostly focusing on \gls{cvd} and \gls{mbe} growth as widely recognized standard epitaxial methods, reviewed in Refs.~\cite{Wirths2016,Zaima2015,Miao2021,Zheng2018,Moutanabbir2021}.
A handful of groups employed different techniques, including \gls{ms} epitaxy~\cite{Zheng2014,Tsukamoto2015,Miao2018,Dascalescu2020,Huang2022}, \gls{spe}~\cite{Lieten2013,Sadoh2016,Moto2017,Dev2020} and a few more exotic methods such as \gls{lpe}~\cite{Gao2014,Inenaga2020} and \gls{fla} of Sn-implanted Ge~\cite{Prucnal2019}.
The main focus of the more recent studies on GeSn epitaxy has been to improve the material crystal quality and push the Sn content to access the \gls{mwir} and \gls{lwir} wavelengths. 
Assuming a strain-free material, Sn fractions  of $\sim$16\atp~and $\sim$26\atp~are respectively required to access these wavelength ranges~\cite{Song2020}. However, GeSn crystal quality and thermal stability quickly degrade with increasing Sn fractions, posing major challenges to the synthesis and use of these materials in actual devices~\cite{Moutanabbir2021}.

\subsubsection*{Strategies to increase Sn content}
In Tab.~\ref{tab:GeSn_HighCompositions}, the highest Sn compositions achieved in epitaxial, monocrystalline GeSn films are reported, together with a few representative films grown with more than 15\atp{} Sn.
A common strategy to maximize Sn incorporation avoiding segregation is the use of low substrate temperatures ($T<150$\degree C) in \gls{pvd} growth processes~\cite{He1996,Zheng2018a,Imbrenda2018,Oehme2014,Hickey2017}.
While low substrate temperatures prevent Sn segregation, the associated reduced adatom mobility is detrimental for the material crystal quality~\cite{Kasper2013,Mircovich2021} and limits the maximal epitaxial thicknesses due to kinetic roughening phenomena~\cite{Bratland2005,Desjardins1999}. 
The latter take place at low temperatures due to the presence of a potential barrier at atomic terrace steps, termed \emph{Erlich} barrier~\cite{Ehrlich1995}. With low thermal energy available from the system, adatoms cannot overcome this potential barrier, and are forced to remain on their atomic plane. This leads to an unbalance of flux between atomic planes, and to increased nucleation of 2D islands on terraces, with consequent progressive 3D roughening and faceting of the film~\cite{Desjardins1999}. After a certain \emph{critical thickness}, the film roughness prevents filling of trenches and the epitaxial growth breaks down, with the film growing first in highly defective~\cite{Bratland2005} or polycrystalline fashion, and then purely amorphous~\cite{Nerding2003}.
In presence of Sn, the Ge adatom mobility and interlayer mass-transport (i.e., adatom up- or down-stepping at terraces) is increased~\cite{Bratland2005}, and therefore kinetic roughening is partially suppressed~\cite{Bratland2005}. Nonetheless, with increasing Sn content, compressive strain increases in the film -- on Ge and Si substrates -- leading to a hybrid strain-roughening mode, where the reduced kinetics induce the surface oscillations necessary to allow strain relaxation through roughening~\cite{Desjardins1999}. Pure strain-induced roughening effects are normally observed at high temperature due to the large mass-transport involved~\cite{Cullis1996}.
Hybrid kinetic-strain roughening is exacerbated with larger Sn contents, as the compressive epitaxial strain increases  in the film~\cite{Bratland2005}.
This poses a fundamental limit to the maximal epitaxial thickness achievable with large Sn contents at low growth temperatures.
Hence, all experimental GeSn films with the highest Sn content ever achieved are only a few tens of nm thick, as reported in Tab.~\ref{tab:GeSn_HighCompositions}.
The epitaxial critical thickness increases with higher growth temperatures, yet these temperatures are constrained by the elevated tendency of Ge-Sn phase separation, in particular with higher Sn contents.

\begin{table*}
	\renewcommand{\arraystretch}{1.5}
	\caption{\label{tab:GeSn_HighCompositions}Summary of the epitaxial, monocrystalline GeSn films with the highest fractions of Sn achieved in the literature. Acronyms: graded GeSn buffer (grGeSn), \acrfull{ml}.}
	\begin{tabular}{cccccl}
		\toprule
		Sn\atp{} &    Growth method    &     Substrate     &                                                          Thickness (nm)                                                          & Growth \gls{t} (\degree C) &         \centering \arraybackslash Ref.         \\
		\midrule
		\rowcolor{lightgrey}
		35.4  & \gls{uhv}-\gls{cvd} &      Si(100)      &                                                                49                                                                &          220           & Mircovich, 2021 \cite{Mircovich2021} \\
		34   &  Ar$^+$ \gls{mbe}   &    Ge/Si(100)     &                                                                20                                                                &          150           &  He, 1996   \cite{He1996}     \\
		\rowcolor{lightgrey}
		30   & \gls{uhv}-\gls{cvd} &      Si(100)      &                                                                40                                                                &          245           & Xu, 2019    \cite{Xu2019b}    \\
		28   &     \gls{ms}      &  grGeSn/Ge(100)   &                                                                20                                                                &          100           & Zheng, 2018 \cite{Zheng2018a}   \\
		\rowcolor{lightgrey}
		27   &      \gls{mbe}      &      Ge(100)      &                                                               120                                                                &          100           & Imbrenda, 2018 \cite{Imbrenda2018}  \\
		25   &      \gls{mbe}      &      Si(100)      &                                                                48                                                                &          120           & Oehme, 2014  \cite{Oehme2014}   \\
		\rowcolor{lightgrey}
		22.3  &     \gls{rpcvd}     & grGeSn/Ge/Si(100) & Few \acrshortpl{ml}\footnote{\,The composition of this film is not uniform over its thickness. Only the top few \acrshortpl{ml} of the film showed the max Sn content of 22.3\atp{} reported in the table.} &          200           &  Dou, 2018  \cite{Dou2018}    \\
		18.3  &      \gls{mbe}      &      Ge(100)      &                                                               100                                                                &           90           & Hickey, 2017 \cite{Hickey2017}   \\
		\rowcolor{lightgrey}
		18   &  \acrshort{lpcvd}   & grGeSn/Ge/Si(100) &                                                                40                                                                &          280           & Assali, 2019 \cite{Assali2019}   \\
		
		16  &      \gls{mbe}      &    Ge/Si(100)     &                                                               250                                                                &          150           & Rathore, 2021 \cite{Rathore2021}  \\
		\rowcolor{lightgrey}
		15   & \gls{uhv}-\gls{cvd} &      Si(100)      &                                                               245                                                                &          285           &  Xu, 2019  \cite{Xu2019b} \\
		\bottomrule
	\end{tabular}
\end{table*}

On the other hand, \gls{cvd} growth can take place at higher deposition temperatures thanks to the different growth dynamics involving gaseous species. This has been the motivation since the early studies on \gls{cvd} growth of GeSn~\cite{Taraci2001}.
Common precursors are Ge and Sn hydrides/chlorides~\cite{Wirths2016}. In \gls{cvd} growth, Sn incorporation depends on the strain state of the film and increases for decreasing compressive strain, a phenomenon known as strain-relaxation enhancement of Sn incorporation~\cite{Assali2019,Aubin2017a,Dou2018,Stanchu2020a}. This does not occur in \gls{mbe} films~\cite{Rathore2021}.
Strain engineering is therefore key in maximizing Sn incorporation in \gls{cvd}-grown GeSn~\cite{Moutanabbir2021}.
Hence, graded GeSn buffers with progressively increasing Sn content have been proposed to gradually relax the film, preventing Sn segregation, to maximize the Sn content at the top of the GeSn stack~\cite{Senaratne2016,Dou2018a,Gallagher2015,Aubin2017,Assali2019}.
On the other hand, other groups argued that growing directly on Si(100) allows for increased strain relaxation, which leads to a higher degree of Sn incorporation during growth by \gls{cvd}~\cite{Mircovich2021}, and additionally reduces strain-induced roughening, leading to a larger critical epitaxial thickness~\cite{Oehme2014}.
Recently, by growing directly on Si(100) and switching the Sn precursor from SnD$_4$ to SnH$_4$, Mircovich\etal~\cite{Mircovich2021} demonstrated a record 35.4\atp{} Sn content in \gls{uhv}-\gls{cvd} 49-nm-thick GeSn.

\subsubsection*{Accessing MWIR and LWIR ranges}
Achieving the large compositions from Tab.~\ref{tab:GeSn_HighCompositions} is not necessarily sufficient to access the \gls{mwir} (3\,\um\,--\,8\,\um) and \gls{lwir} wavelengths (8\,\um\,--\,15\,\um).
In fact, the inherent compressive strain induced by epitaxial mismatch during growth on Ge or Si substrates causes a blue-shift in the GeSn film \gls{bg} energy~\cite{Song2020a,Gupta2013,Attiaoui2014}. This implies that a compressive strained film requires higher Sn content to red-shift its \gls{bg} to the desired energies.
Strain engineering thus becomes essential in achieving full relaxation of the compressive strain to operate at long wavelengths~\cite{Mircovich2021}.
In addition, the crystal quality of GeSn is known to degrade with increasing Sn contents~\cite{Su2011}, complicating the realization of optoelectronic devices.
For these reasons, while numerous GeSn film-based optoelectronic devices have been demonstrated to operate in the \gls{swir} wavelengths  (1.4\,\um\,--\,3.0\,\um), only a handful of them could  reach the \gls{mwir}~\cite{Atalla2021,Assali2018} despite the large Sn contents achieved in GeSn epitaxial films, as reported in Tab.~\ref{tab:GeSn_HighCompositions}.
The alloy compositions in the works reported in Tab.~\ref{tab:GeSn_HighCompositions}  are also in principle sufficient to access the \gls{lwir}; nevertheless, up to date, there exists only a few studies of bandgap energy measurements that barely reach the edge of the \gls{lwir} wavelength range~\cite{Xu2019b,Imbrenda2018}, and the fabrication of devices in the \gls{lwir} is lacking.
Employing GeSn of such large Sn contents in devices remains an open challenge.

The factor currently hindering the commercialization of GeSn-based devices is the material crystal quality.
The main challenge remains the  thorough understanding and management of linear and point defects in GeSn alloys. While the former can be avoided to some extent by engineering the growth processes, e.g., with (graded) buffer layers, the latter are poorly understood.
In Sec.~\ref{sec:GeSn-Ge_StrainRelax} and \ref{sec:GeSn_ElecProp}, we review the research done towards the understanding of defect nucleation, and their influence on carrier trap states in the material.

\subsection{State-of-Art of GeSn Sputtering Epitaxy}
\subsubsection*{MS as alternative method for GeSn growth}
Despite its poor reputation for thin film epitaxy, \acrfull{ms} may represent a plausible  solution for industrial scale-up of GeSn. Compared to mainstream epitaxial methods such as \gls{mbe} and \gls{cvd}, \gls{ms} tools  are relatively simple, and they allow large growth rates, film homogeneity over large substrates and the use of non-toxic Ge and Sn targets.
An additional advantage related to \gls{ms} growth of GeSn is that ion impingement enhances the incorporation of Sn adatoms via collisional mixing inhibiting segregation, as demonstrated in early works with ion-assisted \gls{mbe}~\cite{He1995a} and sputtering growth~\cite{Shah1987}.
This allows to employ higher substrate temperatures compared to other \gls{pvd} methods such as \gls{mbe}, leading in principle to a lower concentration of point defects~\cite{Ueno2000,Knights2001}.
Experimental verification is required to confirm this hypothesis.
However, while \gls{cvd} and \gls{mbe} have been extensively studied for epitaxy of GeSn,  \gls{ms} was investigated only by a handful of groups~\cite{LadrondeGuevara2003,Zheng2014,Tsukamoto2015,Miao2018,Dascalescu2020,Qian2020,Huang2022,Lin2023,Shah1987}.
A summary of the works on sputtering epitaxy of GeSn is reported in Tab.~\ref{tab:GeSn_Sputtering}.
Monocrystalline GeSn with Sn content up to 28\atp{} has been demonstrated~\cite{Zheng2018a}, among the highest Sn fractions observed in the literature with any growth technique.
GeSn films are generally sputtered using Ar gas and substrate temperatures between 100\degree C and 300\degree C, which are higher than \gls{mbe}, but lower than \gls{cvd}.
Macroscopic structural properties of the layers are well documented, with reported values of alloy composition, film roughness, and \gls{dsr}.
However, while structural characterization may give good indications of the promise of a material, it does not offer a full picture of its potential for use in optoelectronic devices. The latter is given exclusively by accurate optoelectronic characterization of the layer, currently lacking in the literature for sputtered epitaxial GeSn.
Up to date, there has been only one report demonstrating direct-\gls{bg} in Ge$_{0.87}$Sn$_{0.13}$ grown on Si, Ge and GaAs substrates~\cite{Qian2020}.
No study focused on the \gls{bg} crossover composition.
Room-temperature \gls{pl} was shown in \gls{ms} Ge$_{0.97}$Sn$_{0.03}$ by Miao\etal~\cite{Miao2018} and in Ge$_{0.938}$Sn$_{0.062}$ by Lin\etal~\cite{Lin2023a}.
Zheng\etal~\cite{Zheng2016} demonstrated the use of \gls{ms} GeSn in an optoelectronic device. They fabricated a  \emph{p-i-n} photodetector using Ge$_{0.94}$Sn$_{0.06}$ as active material and showed performances comparable with devices realized with \gls{mbe} or \gls{cvd} at the time. Despite the promising results, only one additional GeSn-based device made from \gls{ms} has been documented in literature since then~\cite{Zhu2024}.

\begin{adjustwidth}{-0.3cm}{-0.3cm}
\begin{turnpage}
	\begin{table*}
		\caption{\label{tab:GeSn_Sputtering}Summary of monocrystalline GeSn films grown by the \acrfull{ms} method, except for Ref.~\cite{Shah1987}, where \gls{dc} diode sputtering was employed. Acronyms: \acrlong{tf} (\gls{tf}), graded GeSn buffer (grGeSn), low-temperature GeSn buffer (LT-GeSn), \acrlong{tf} (\gls{gr}), \acrlong{tf} (\gls{rrms}), \acrlong{tf} (\gls{rf}),   \acrlong{tf} (\gls{hipi}), \acrlong{tf} (\gls{sims}).}
		\begin{tabular}{m{0.15\linewidth}P{0.053\linewidth}P{0.072\linewidth}ccP{0.035\linewidth}P{0.035\linewidth}P{0.054\linewidth}cccm{0.3\linewidth}}
		\toprule
	\centering\arraybackslash Ref. 		& \makecell{Comp. \\ Sn\atp} 		& Substrate 		& \makecell{Growth \gls{t}\\ (\degree C)}& \makecell{\gls{tf}\\ (nm)}& \multicolumn{2}{c}{\makecell{Power (W)\\Ge~~~~~Sn}} & \multicolumn{2}{c}{\makecell{~~Gas~~~~~~~p~\\ ~~~~~~~~~(mTorr)}}		&  \makecell{\gls{gr} \\ (nm/min)} & \makecell{$R_{rms}$\\ (nm)} & \centering\arraybackslash Claims \\
		\midrule
		
		
		\addlinespace
		\rowcolor{lightgrey}
		Lin, 2023 \cite{Lin2023a}&
		6.2 	&  \gls{vge} 		& 405 			& 266 		& 	\makecell[c]{\gls{dc}\\86.6} & \makecell[c]{\gls{dc}\\7.3}	&	Ar & 3.75		& 26.5 & 0.93 & \makecell[{{p{\linewidth}}}]{Obtained room-temperature \gls{pl}, enhanced by high \gls{t} growth.} \\

		\addlinespace
		 Lin, 2023 \cite{Lin2023}&
		16.2 	&  \makecell[c]{(grGeSn/)\\/Ge(001)} 		& 259 			& 	60,~120	& 	\makecell[c]{\gls{dc}\\51} & \makecell[c]{\gls{dc}\\6.5}	&	Ar & 	3.7	&12 & $\sim1$ &\makecell[{{p{\linewidth}}}]{High \gls{dsr}. Relaxation mechanisms w/o grGeSn buffer.}\\
		
		\addlinespace
		\rowcolor{lightgrey}
		Tsukamoto, 2023 \cite{Tsukamoto2023}&
		9 	&  Ge(001) 		& 215 -- 250 			& 70,~390 		& 	\makecell[c]{\gls{dc}\\70} & \makecell[c]{\gls{dc}\\5}	&	\makecell[c]{Ar +\\ 5\% H$_2$} & 3		&24 & $\sim0.4$ &\makecell[{{p{\linewidth}}}]{Uniform Sn profile, and limited diffusion in substr.}\\

\addlinespace
		Huang, 2022 \cite{Huang2022}&
		13 	&  Ge/Si(001) 		& 180 -- 300 			& 350 		& 	\makecell[c]{\gls{dc}\\140} & \makecell[c]{\gls{rf}\\40}	&	Ar & 3		& 8.4 & $>4$ & \makecell[{{p{\linewidth}}}]{Highest $x_{Sn}$ on Si(100) by sputtering.}\\
		
		\addlinespace
		\rowcolor{lightgrey}
		Khelidj, 2021 \cite{Khelidj2021}&
		$\sim10$	&  Ge/Si(001)		& $<360$	&  	100	& 	\makecell[c]{\gls{dc}\\150} & \makecell[c]{\gls{dc}\\15}	& Ar & 2.25		&62.5 & 1.24 & \makecell[{{p{\linewidth}}}]{96-sec growth during cooling from 360\degree C. O, C contamination $<2\cdot10^{19}$\,cm$^{-3}$.}\\

\addlinespace
		Qian, 2020 \cite{Qian2020}&
		$\sim13$	&  \makecell[c]{Ge(001)\\Si(001)} 		& 	-	&  	330	& 	\makecell[c]{\gls{rf}\\100} & \makecell[c]{\gls{rf}\\20}	& Ar & 3		& 5.5 & $\sim1$ & \makecell[{{p{\linewidth}}}]{Direct-\gls{bg} demonstrated via \gls{ot} and \gls{pl}.}\\
		
		\addlinespace
		\rowcolor{lightgrey}
		Dascalescu, 2020 \cite{Dascalescu2020}&
		$\sim12.5$ 	&  Ge/Si(001) 		& 200 -- 250 			&  	$\sim120$	& 	\makecell[c]{\gls{hipi}} & \makecell[c]{\gls{dc}}	& \makecell[c]{Ar +\\ 5\% H$_2$} & 3		& 1 & - & \makecell[{{p{\linewidth}}}]{First growth by \gls{hipi} sputtering. Polycrystalline defects.}\\

\addlinespace
		Tsukamoto, 2019 \cite{Tsukamoto2019}&
		9 	&  \makecell[c]{(LT-GeSn)\\/Si(001)}  		& 300 			& 100 		& 	-& -	&	\makecell[c]{Ar +\\ 5\% H$_2$} & -		&- & $\sim1.7$ &\makecell[{{p{\linewidth}}}]{Low-\gls{t} GeSn buffer allows to prevent Sn segregation.}\\
		
		\addlinespace
		\rowcolor{lightgrey}
		Yang, 2019 \cite{Yang2019}		&
		7		&  Ge(100) 	& 160 	& 400		&	\makecell[c]{\gls{rf}\\120} & \gls{rf} &	 Ar & 3		& 12 -- 30 & 0.8 & \makecell[{{p{\linewidth}}}]{\gls{tdd} $\sim10^9$\,cm$^{-2}$.}\\
		
		\addlinespace
		Miao, 2018 \cite{Miao2018}		&
		$<3$		&  Ge/Si(100) 	& 160 	& 400		&	\makecell[c]{\gls{rf}\\120} & \gls{rf} &	 Ar &3		& 12 -- 30 & 0.74 & \makecell[{{p{\linewidth}}}]{\gls{sims} shows non-uniform Sn distribution along thickness. Showed GeSn \gls{pl}.}\\
		
		\addlinespace
		\rowcolor{lightgrey}
		Zheng, 2018 \cite{Zheng2018a}		&
		\makecell[c]{20\\ 28} 			&  \makecell[c]{grGeSn/ \\ /Ge(100)} 	& 100 -- 160 	& 20 	&	\makecell[c]{\gls{rf}\\ 70}& \makecell[c]{\gls{rf}\\9--30}	& Ar& 3	& 10 -- 14.5	&$<0.5$ & \makecell[{{p{\linewidth}}}]{Thermal stability of 350\degree C for Ge$_{0.80}$Sn$_{0.20}$. Highest $x_{Sn}$ achieved by sputtering.}\\

\addlinespace
		Zheng, 2017 \cite{Zheng2017}		&
		12		&  \makecell[c]{grGeSn/ \\ /Ge(001)} 	& 140 	&140 	&	-& -	& \makecell[c]{Ar + H$_2$\\ 0 -- 40\%} & 3	& -	& - & \makecell[{{p{\linewidth}}}]{Sn uniform along $z$. H$_2$ decreases Sn diff. length and helps strain relaxation.}\\
		
		\addlinespace
		\rowcolor{lightgrey}
		Tsukamoto, 2015 \cite{Tsukamoto2015a}&
		\makecell[c]{8.5\\11.5} 	&  Si(001) 		& 250 -- 300 			& 100 		& 	\gls{dc} &\gls{dc}	&	\makecell[c]{Ar +\\ 5\% H$_2$} &-		&- & - &\makecell[{{p{\linewidth}}}]{Sn segregation occurs more easily during  growth. }\\
		
		\addlinespace
		Tsukamoto, 2015 \cite{Tsukamoto2015}&
		$\leq11.5$	 &  Si(001) 	& 200 -- 300 			& 100 		& 	\gls{dc}	&\gls{dc}	& \makecell[c]{Ar +\\ 5\% H$_2$}&- & 25.2 & $<1$ &\makecell[{{p{\linewidth}}}]{ High \gls{gr} allows high gr. \gls{t} => better cryst. quality. }\\
		
		\addlinespace
		\rowcolor{lightgrey}
		Zheng, 2014 \cite{Zheng2014}&
		$\leq6$	 &  Ge/Si(001) 	& 150 			& 450 		& 	\makecell[c]{RF\\ 70}&\makecell[c]{RF\\5--15}	& Ar &3 & $\sim10.2$ & - & \makecell[{{p{\linewidth}}}]{Thermal stability of 700\degree C for Ge$_{0.975}$Sn$_{0.025}$, 500\degree C for Ge$_{0.94}$Sn$_{0.06}$.} \\
		
		\addlinespace
		P\'erez Ladr., 2003 \cite{LadrondeGuevara2003}		&
		$\leq14$		&  Ge(001) 	& 150 -- 170 	& $\leq 800$		&	\gls{rf} & - &	 Ar & -		& - & - & \makecell[{{p{\linewidth}}}]{\gls{pb} model \cite{People1985} describes well $t_{cr,GeSn}$ on Ge.}\\
		
		\addlinespace
		\rowcolor{lightgrey}
		Shah, 1987 \cite{Shah1987}		&
		$\leq8$		&  \makecell[c]{Ge(001)\\GaAs(001} 	& 90 -- 150 	& 	1000	&	\gls{dc} & \gls{dc} &	 Ar & 25	& 12 &  & \makecell[{{p{\linewidth}}}]{First epi-GeSn. Absence of magnetron required substrate bias for low-energy ion bombardment.}\\
		\bottomrule
		\end{tabular}
	\end{table*}
\end{turnpage}
\end{adjustwidth}

Historically, sputtering machines frequently lacked the capability for substrate heating, hindering their use for epitaxial growth.
In addition, sputtering tends to have a poor reputation as technique for epitaxy.
This is likely owed to the association of the ion impingement occurring during sputtering to the detrimental effects taking place during ion implantantion processing, e.g. for material doping.
In the latter, implanted species accelerated at energies of few keV are known to generate defects and even amorphization in the implanted layer~\cite{Napolitani2015}.
However, energies of sputtered atoms impinging on surface during \gls{ms} are considerably lower, of a few tens of eV~\cite{Drusedau2000,Gudmundsson2020}, and thus the interaction with the film is fundamentally different.
With these kinetic energies, impinging atoms will only influence the surface layers of the film, inducing some atomic displacement~\cite{Takagi1984}. In addition, ion impingement can both enhance~\cite{Gudmundsson2020} or suppress~\cite{He1995a} surface diffusion, depending on the ion impingement rate and energies.
It is true that ions from the working gas (i.e., Ar) neutralized and reflected at the target can possess higher energies up to a few hundred of eV~\cite{Drusedau2000,Gudmundsson2020}, but with appropriate cathode design the acceleration voltages at the target can be kept low enough to limit the Ar energy to values similar to the sputtered atoms~\cite{Drusedau2000}, limiting bombardment and implantation effects.
Furthermore, being the growing film and substrate heated during growth, the thermal energy will allow for crystal reorganization and annihilation of the eventual defects caused by ion surface impingement.
Auret\etal~\cite{Auret2007,Auret2010} showed that while Ar plasma exposure of Ge at room temperature generated some trap levels in Ge, no electrically active Ar-related defects could be observed in the material. Additionally, all traps generated by the process were removed by annealing at 250\degree C.
Hence, with growth temperatures well above 250\degree C and appropriate cathode design, \gls{ms} should yield films free of ion-impingement-induced defects, though this conclusion should be experimentally verified for GeSn alloys.

\subsubsection*{Takeaways from literature on GeSn growth by MS}
From the works reported in Tab.~\ref{tab:GeSn_Sputtering}, the investigations of GeSn \gls{ms} epitaxy have allowed to establish a few guidelines for optimization of the material crystal quality:
\begin{itemize}
	\item \textbf{Substrate $\mathbf{T}$:} As explained in the above paragraph, it should be kept as high as possible while avoiding Ge-Sn phase separation~\cite{Tsukamoto2015}.
	\item \textbf{Sputtering gas:} H$_2$ is often added to the Ar sputtering gas, as reactive H species from the plasma are expected to passivate dangling bonds and prevent Sn segregation~\cite{Zheng2017}.
	In \gls{mbe}, the presence of H$_2$ during growth has been found to be beneficial in reducing adatom diffusion length~\cite{Sakai1994,Asano2015}. Via \gls{dft}-based calculations, Johll\etal~\cite{Johll2015} showed that hydrogenated epi-surface promotes Sn incorporation in the film, preventing Sn segregation.
	H$_2$ has also been found to induce strain relaxation in sputtered films~\cite{Zheng2017}.
	\item \textbf{Growth rate:} Large growth rates reduce adatom diffusion length, promoting Sn incorporation, hindering phase separation~\cite{Tsukamoto2015}.
\end{itemize}
To the best of our knowledge, systematic studies on the effect of gas pressure have not been reported, despite its role in determining the ion impingement energy through scattering mechanisms in the gas phase. 
Several other questions regarding GeSn \gls{ms} epitaxy remain open.
For example, contrasting observations have been reported regarding Sn distribution uniformity along the epitaxial thickness, with studies showing both uniform~\cite{Tsukamoto2023} and non-uniform~\cite{Miao2018} distributions.
Both theoretical and experimental investigations in defect levels and electronic properties in sputtered GeSn are missing.

\subsection{Ge Buffer Epitaxial Growth on Si Substrates}
Epitaxial growth of Ge has been investigated for years, with well-established recipes for \gls{mbe} and \gls{cvd} methods~\cite{Ye2014}.
As for the GeSn alloy, \gls{ms} epitaxy of Ge has been investigated to a lesser extent.

\subsubsection*{General strategies for Ge growth on Si}
Due to the large epitaxial mismatch of 4.2\%, strain-induced roughening and islanding tend to occur during Ge heteroepitaxy on Si~\cite{Desjardins1999,Cunningham1991}.
Ge is therefore generally grown in a 2-step method, introduced by Colace\etal~\cite{Colace1998}. A first flat Ge layer is grown at \gls{lt}  as the film relaxes the epitaxial strain. The \gls{lt} serves to reduce the adatom mobility to hinder 3D growth. This layer is typically thinner than 100\,nm, since $t_{cr,Ge}$ on Si is of the order of few\,nm~\cite{Jesser1993}. A second Ge layer is then grown to the desired final thickness at a \gls{ht} to maximize the crystal quality.
Growth of the Ge buffer is followed by thermal annealing at \gls{t} typically higher than 800\degree C to fully relax the film and reduce the \gls{tdd}~\cite{Ye2014}, as elucidated in the next section.
Besides the reduced \gls{tdd}, the annealing of the Ge buffer has an additional benefit for GeSn epitaxy. Thanks to the high temperatures used during annealing, after cooling, the Ge buffer will be left with a $\sim0.2$\% in-plane tensile strain due to the differential \glspl{tec} of the Ge film and the Si substrate~\cite{Cannon2004,Hartmann2004,Luong2013}.
The tensile strain built in the Ge layer will provide a decrease in lattice mismatch for GeSn epitaxy, increasing the \gls{tcr} for strain relaxation and thus extending the thickness limits for GeSn pseudomorphic growth.
This is important to avoid additional nucleation of relaxation defects to optimize the material performance in optoelectronic devices.
For a qualitative understanding of the effect of tensile-strained \gls{vge} substrate, the improvement in \gls{tcr} in epi-GeSn is shown in Fig.~\ref{fig:GeSnCriticalThickness}(b), using the \gls{pb} model for GeSn  grown on Ge with 0.2\% in-plane tensile strain.
One should however note that relaxation of GeSn films on Ge-buffered Si will be eased by the presence of  \glspl{td} inherited from the Ge buffer layer~\cite{Houghton1991}. Hence, one can expect a lower \gls{tcr} of GeSn on \gls{vge} substrates compared to Ge substrates.

\subsubsection*{Takeaways from literature on Ge growth by MS}
We now briefly review the literature on Ge  \gls{ms} epitaxy summarized in Tab.~\ref{tab:Ge_Sputtering}.
Surprisingly, we did not find any study reporting the growth via a 2-step deposition, despite being well known to yield higher crystal quality.
Monocrystalline Ge films on Si(100) have been achieved using both \gls{dc} and \gls{rf} power sources.
With \gls{dc} sputtering, \acrfull{rrms} $\sim0.3$\,nm were obtained, optimal for the use of Ge as an epitaxial buffer layer. \gls{rrms} for sputtered Ge were noticeably better than for GeSn (see Tab.~\ref{tab:GeSn_Sputtering}) likely due to the higher Ge growth \gls{t} that limits kinetic roughening.
In a few studies, the resistivity of the Si substrate has been found to play a major role in achieving flat Ge layers~\cite{Hanafusa2012,Tsukamoto2013}. In particular, highly doped Si substrates, independently on their doping type, reduced Ge surface adatom mobility, preventing Si-Ge intermixing and strain-induced roughening at 350\degree C. This phenomenon is not yet well understood and would need further investigation.
A reduction in adatom mobility was also observed at high sputtering powers as a result of the increased growth rate~\cite{Tsukamoto2015b}, analogously to GeSn~\cite{Tsukamoto2015}.
Zeng\etal~\cite{Zeng2021} showed an improvement in crystal quality with the application of a positive substrate bias to reduce ion bombardment by decelerating Ar$^+$ ions from the plasma. This work highlighted the importance of tuning the ion impingement energy, an aspect often overlooked.
The annealing of sputtered Ge was demonstrated to reduce \gls{tdd}~\cite{Liu2018,Yeh2014,Pietralunga2009} and annihilate \acrfull{sf}s~\cite{Otsuka2017}.
A few studies reported also the electrical properties of the films, showing \emph{p-type} \acrfull{undop} of $10^{16}-10^{17}$\,cm$^{-3}$~\cite{Yeh2014,Steglich2013}, and room-temperature \gls{mup} values of $1000$\mobun~\cite{Yeh2014}, which are not too far from intrinsic bulk Ge \gls{mup} of $\sim1900$\mobun. These properties are in line with other epitaxial techniques, as discussed in Sec.~\ref{sec:GeSn_ElecProp}.

\begin{turnpage}
	\centering
	\begin{table*}
		\renewcommand{\arraystretch}{2}
		\caption{\label{tab:Ge_Sputtering}Summary of monocrystalline Ge films grown by the \acrfull{ms} method, except for Ref.~\cite{Bajor1982}, where \gls{rf} diode sputtering was employed. Acronyms: \acrlong{gr} (\gls{gr}), \acrlong{pdc} (\gls{pdc}), \acrlong{cw} (\gls{cw}), \acrlong{sf} (\gls{sf}), \acrlong{undop} (\gls{undop}), \acrlong{gsmbe} (\gls{gsmbe}).}
		\begin{tabular}{m{0.15\linewidth}P{0.055\linewidth}cP{0.06\linewidth}cP{0.025\linewidth}ccm{0.45\linewidth}}
		\toprule
		\centering\arraybackslash Ref. 	& \makecell[c]{Growth\\\gls{t} (\degree C)} & \makecell{\gls{tf}\\ (nm)} & \makecell{Power\\ (W)} & \multicolumn{2}{c}{\makecell{~~Gas~~~~~p\\ ~~~~~~(mTorr)}}		&  \makecell{\gls{gr} \\ (nm/min)} & \makecell{\gls{rrms}\\ (nm)} & \centering\arraybackslash Claims \\
		\midrule
		\rowcolor{lightgrey} Zeng, 2021 \cite{Zeng2021} & 600 & 450 & \makecell[c]{\gls{rf}\\75--125} & \makecell[c]{Ar + H$_2$\\ 0--15\%} & 1.2 & - & 0.58 & \makecell[{{p{\linewidth}}}]{Improved crystal quality and $R_{rms}$ with positive substrate bias to reduce Ar$^+$ ion impingement energy. H$_2$ enhanced relaxation.}\\
		
		Liu, 2018 \cite{Liu2018} & 400 & 100--200 & \gls{rf} & - & 1.1 & 4 & - & \makecell[{{p{\linewidth}}}]{Reduction of 3 orders of magnitude in \gls{tdd} upon \gls{cw} laser annealing above $T_{m,Ge}$.}\\
		
		\addlinespace
		\rowcolor{lightgrey}Otsuka, 2017 \cite{Otsuka2017} & 350 & 6 & \gls{dc} & Ar & 2 & 17.4 & 0.39 & \makecell[{{p{\linewidth}}}]{\gls{rta} at 720\degree C annihilates \gls{sf}s in Ge film, though slightly degrading $R_{rms}$ to 0.54\,nm.}\\
		
		\addlinespace
		Tsukamoto, 2015 \cite{Tsukamoto2015b} & 450 & 200 & \makecell[c]{\gls{dc}\\100} & \makecell[c]{Ar +\\ 5\%H$_2$} & 3 & 20 & 0.23 & \makecell[{{p{\linewidth}}}]{TDD$\sim2\cdot10^9$\,cm$^{-2}$. Higher sputtering power decreases adatom diffusion length, limiting islanding. Si-Ge interface intermixing observed at low power.}\\
		
		\addlinespace
		\rowcolor{lightgrey}Yeh, 2014 \cite{Yeh2014} & $<360$ & 1000 & \makecell[c]{\gls{dc}\\100} & Ar & 2.5 & 126 & - & \makecell[{{p{\linewidth}}}]{Demonstrate \gls{tdd} reduction with annealing at 800\degree C. \gls{mup} improves to $>1000$\mobun; \gls{undop} decreases at \gls{lt}, but increases at $T>\sim150$\,K.}\\
		
		\addlinespace
		Tsukamoto, 2013 \cite{Tsukamoto2013} & 350 & $\leq100$ & \makecell[c]{\gls{dc}\\10} & \makecell[c]{Ar +\\ 5\%H$_2$} & 3 & 2 & 0.3 & \makecell[{{p{\linewidth}}}]{B-doping of Si sub. has same effect as P in Ref.~\cite{Hanafusa2012}. Doping type of sub. not important.}\\
		
		\addlinespace
		\rowcolor{lightgrey}Steglich, 2013 \cite{Steglich2013} & 380 & 120 & \makecell[c]{\gls{dc}\\10} & Ar & 7.5 & 4.5 & - & \makecell[{{p{\linewidth}}}]{Oxide contamination on surface did not prevent epitaxy at 380\degree, but did at 320\degree C. Measured \gls{undop}(300K)$\sim10^{17}$\,cm$^{-3}$.}\\
		
		\addlinespace
		Hanafusa, 2012 \cite{Hanafusa2012} & 350 & $\leq60$ & \gls{dc} & \makecell[c]{Ar +\\ 5\%H$_2$} & 3 & 2.2 & 0.31 & \makecell[{{p{\linewidth}}}]{Si sub. P-doping reduces adatom diff. length, influencing Ge islanding and strain relxn. Si-Ge  intermixing observed at low Si doping. Sub. doping has no effect in \gls{gsmbe}.}\\
		
		\addlinespace
		\rowcolor{lightgrey}Pietralunga, 2009 \cite{Pietralunga2009} & \makecell[c]{170 \\ 370} & 100--200 & \makecell[c]{\gls{pdc}\\550,800} & Ar & \makecell[c]{13\\17} & 4.5 & $\sim1$ & \makecell[{{p{\linewidth}}}]{Unintentional p-type doping observed in Ge. \gls{rta} at 400\degree C improved crystal quality.}\\
		
		\addlinespace
		Bajor, 1982 \cite{Bajor1982} & 470 & 1500 & \gls{rf} & Ar & 15 & 14.5 & - & \makecell[{{p{\linewidth}}}]{Strong Si-Ge interface intermixing, likely due to initial ion etching. \gls{undop}(300K)$\sim10^{17}$\,cm$^{-3}$, \gls{mup}$=1280$\,cm$^2$/Vs}\\
		\bottomrule
		\end{tabular}
	\end{table*}
\end{turnpage}

\section{Strain Relaxation Mechanisms and Defects in GeSn, Ge}
\label{sec:GeSn-Ge_StrainRelax}
In this section, we explore the defects typically found in Ge and GeSn films. A  thorough comprehension of electrically-active defects is crucial for assessing the electrical properties of these materials.
We begin by outlining the extended defects that emerge during epitaxial growth of Ge on Si substrates. Subsequently, we analyse the annealing processes proposed to reduce the density of these defects. Annealing is commonly performed on Ge buffer films prior to GeSn growth. Next, we discuss defect formation during epitaxial growth of GeSn on Ge. Finally, we review the point defects that may form in both Ge and GeSn.

For an extensive summary of the different types of dislocations in the group-IV \gls{fcc} lattice, the reader is directed to the work from Arroyo Rojas Dasilva\etal~\cite{ArroyoRojasDasilva2017}. It contains detailed explanations on the different types of extended defects in group-IV materials, and their visualization at \gls{tem}.

\subsection{Strain Relaxation during Epitaxial Growth of Ge on Si(001)}
\label{sec:GeSn-Ge_StrainRelax_GeRelax}
This section provides a brief review of the phenomena occurring during strain relaxation in epitaxial Ge growth on Si(001) substrates.

\subsubsection*{60\degree{} and 90\degree{} misfit dislocations}
\begin{figure*}[htb]
	\centering
	\includegraphics[width=\textwidth]{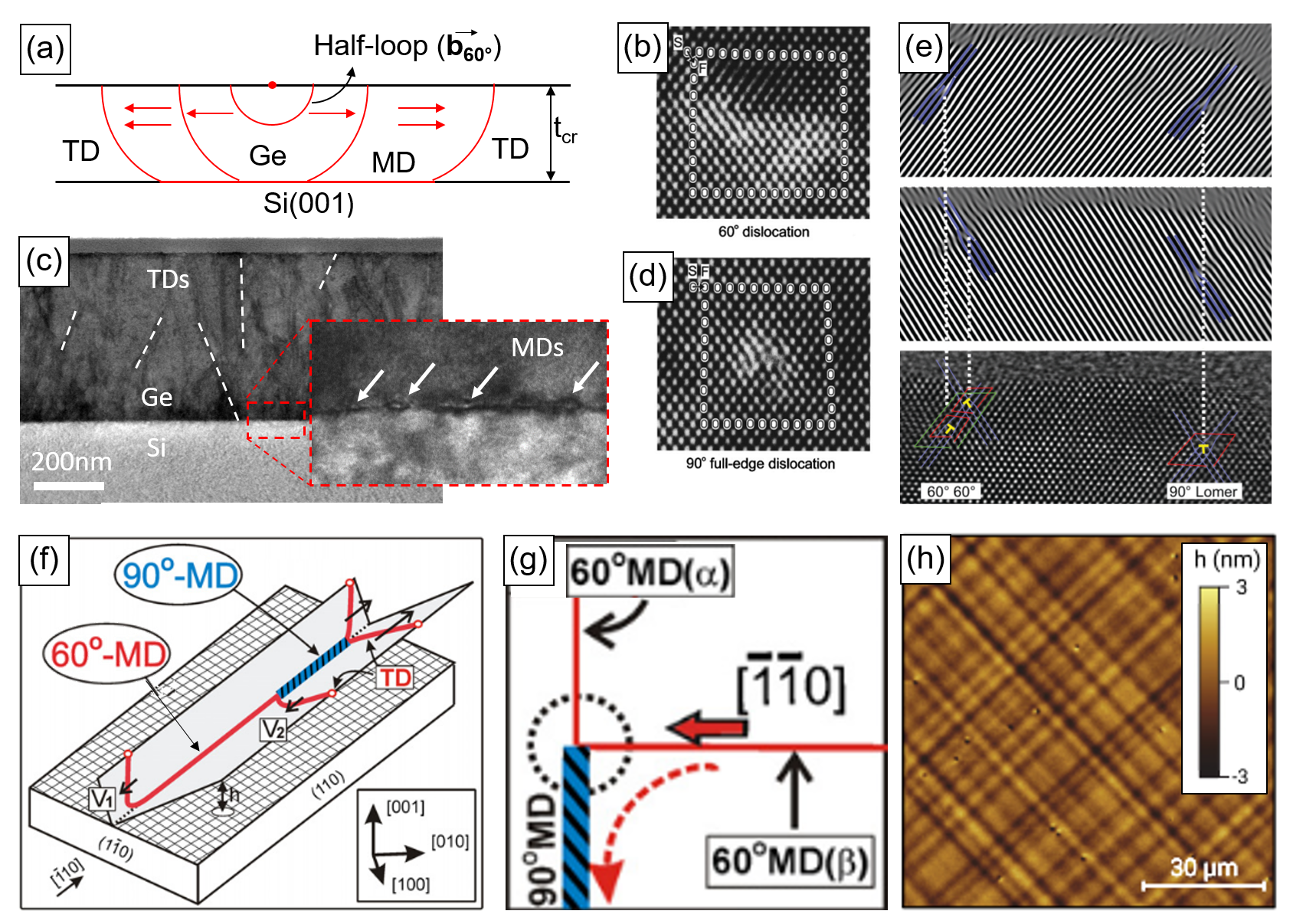}
	\caption{(a)\,Schematics of 60\degree{} dislocation half-loops nucleating on surface at \gls{tcr}, and gliding to the Ge/Si interface to form a 60\degree{} \gls{md} + two screw \glspl{td}. (b, d)\,\gls{xtem} image parallel to the dislocation core (zone axis \hkl<110>) of respectively a 60\degree{}  and 90\degree{} \gls{md} showing their Burger's circuit and \gls{bvec}. (c)\,\gls{xtem} of epi-Ge on Si(001), grown by \gls{ms} by the authors. (e)\,\gls{xtem} showing the subtle difference between a 90\degree{} \gls{md} and a pair of 60\degree{} \glspl{md} having the same \gls{bvec}. (f,g)\,Schematics of two 60\degree \glspl{md}, with (f) parallel and (g) perpendicular dislocation cores, recombining in one 90\degree \gls{md}. (h)\,Cross-hatch patterns on the surface of an annealed Ge film grown on Si(001), visualized by means of \gls{afm}.
	Figures (b,d) are adapted from Ref.~\cite{Sakai1997} with permission from the authors, \copyright\,1997 \emph{AIP Publishing}. (e) is reprinted with permission from Marzegalli\etal~\cite{Marzegalli2013}, \copyright\,2013 \emph{American Physical Society}. (f,g) are adapted with permission from Bolkhovityanov\etal~\cite{Bolkhovityanov2012}, \copyright\,2012 \emph{IOP Publishing}. (h) is reprinted with permission from Rovaris\etal~\cite{Rovaris2019}, \copyright\,2019 \emph{American Physical Society}.}
	\label{fig:GeStrainRelaxation}
\end{figure*}

Models of critical thickness  of strain relaxation predict a \gls{tcr} of few nm for epitaxial growth of Ge on Si~\cite{People1985,Matthews1974,Houghton1991}. These models have been briefly introduced in Sec.~\ref{sec:Challenges_EpitaxialRelaxation}.
During growth, as the epitaxial film reaches its \gls{tcr}, 60\degree{} \glspl{md} nucleate on surface to plastically release the misfit strain.  Driven by the misfit stress field, the  60\degree{} \glspl{md} form dislocation half-loops and in this configuration they glide down along the \hkl{111} planes ~\cite{Bolkhovityanov2012}. This picture is represented schematically in Fig.~\ref{fig:GeStrainRelaxation}(a).
When a half-loop reaches the Ge/Si interface, it will form a \acrfull{md} with \gls{bvec} along one of the \hkl<110> directions. The dislocation will thread to the surface via two segments of mostly screw character~\cite{Bolkhovityanov2012}, the latter being known as \acrfullpl{td}.
The Burger's circuit around a 60\degree{} \gls{md} dislocation core with  \gls{bvec}$=a/2 \hkl<011>$ is shown in Fig.~\ref{fig:GeStrainRelaxation}(b). Since \gls{bvec} has an in-plane edge component, the formation of a such \glspl{md} will contribute to the release of the epitaxial strain. On the other hand, due to their antiparallel \gls{bvec}, the \gls{td} pair does not release misfit strain~\cite{Speck1996}. However, \glspl{td} are electrically active, i.e., source of trap states, and are therefore undesired.

A representative \gls{xtem} image of a Ge epitaxial film on Si(001) along the \hkl<110> zone axis is shown in Fig.~\ref{fig:GeStrainRelaxation}(c), with \glspl{md} and \glspl{td}  indicated by arrows and lines, respectively.
Besides the aforementioned 60\degree{} \glspl{md} and \glspl{td}, at the Si-Ge interface we also find several 90\degree{} full edge dislocations with \gls{bvec}=$\pm a/2\hkl[1-10]$ or \gls{bvec}=$\pm a/2\hkl[110]$, known as Lomer dislocations~\cite{Bolkhovityanov2012}.
A 90\degree{} \gls{md} is imaged by \gls{xtem} in Fig.~\ref{fig:GeStrainRelaxation}(d), with its Burger's circuit showing that \gls{bvec} lies in the \hkl(001) plane. Thanks to their full edge character, Lomer dislocations release twice the misfit strain of 60\degree{} \glspl{md}~\cite{Marzegalli2013}.
However, while 60\degree{} \glspl{md} are glissile as their slip plane\,\footnote{\,The \emph{slip plane} is the plane containing both the dislocation line and \gls{bvec}, within which a dislocation can \emph{glide}~\cite{Hull2011}. Gliding involves the switching of inter-atomic bonds, with no diffusion~\cite{Bolkhovityanov2012}.} is the \hkl{111} family, Lomer \glspl{md} are sessile as their slip plane is the \hkl(001) plane, which is not a close-packed slip plane of the \gls{fcc} lattice.
90\degree{} \glspl{md} therefore cannot nucleate directly as half-loops on surface and glide to the Si/Ge interface~\cite{Marzegalli2013}.
At high processing temperatures, Lomer dislocations can move perpendicularly to their slip plane by \emph{climb}, involving vacancy-assisted atomic diffusion, but this does not explain their presence at the Ge/Si interface at the growth \gls{t} used for Ge epitaxy~\cite{Bolkhovityanov2012}, since their \emph{climbing} would require the presence of extremely large concentrations of vacancies.
In fact, Lomer \glspl{md} have been found to be the result of the reaction
\begin{equation}
	a/2 \hkl[10-1] + a/2 \hkl[011] = a/2 \hkl[110]
\end{equation}
which describes the recombination of two 60\degree{} \glspl{md} with parallel dislocation cores~\cite{Bolkhovityanov2012}, shown in Fig.~\ref{fig:GeStrainRelaxation}(f) after Ref.~\cite{Bolkhovityanov2012}. Perpendicular 60\degree{} \glspl{md} may also recombine into a Lomer \gls{md}, as schematized in Fig.~\ref{fig:GeStrainRelaxation}(g) for specific combinations of \gls{bvec}~\cite{Bolkhovityanov2012}.

The large occurrence of 90\degree{} \glspl{md} in epitaxial Ge on Si(001) has been explained by the fact that in high-misfit epitaxial systems the nucleation of a first 60\degree{} \gls{md} induces the nucleation of a second 60\degree{} \gls{md} in the near vicinity. 
In a 3-nm Ge thin film grown on Si(001) by \gls{cvd}, Marzegalli\etal~\cite{Marzegalli2013} showed the formation of 60\degree{} \gls{md} pairs at the film interface in Fig.~\ref{fig:GeStrainRelaxation}(e). They reported an occurrence of 88\% of these pairs, with the remaining 12\% being isolated 60\degree{} or 90\degree{} \glspl{md}, and they elucidated this phenomenon with dislocation dynamics simulations.
They found that after a first \gls{md} has nucleated and glided to the interface, its strain field is sensed at the film surface, and the surface nucleation of a \emph{complementary} 60\degree{} \gls{md} becomes energetically favored. Its most likely position is off the gliding plane of the first 60\degree{} \gls{md}, explaining the large occurrence of the \gls{md} pair, with dislocation cores distanced by $\sim1$\,nm. On the other hand, nucleation of the complementary dislocation can statistically still occur in the mirror-like gliding plane of the \emph{parent} \gls{md}, with consequent gliding and recombination into a Lomer \gls{md}.
At high processing temperatures, e.g. during growth of the \gls{ht} Ge layer in a 2-step process, or during annealing, the 60\degree{} \gls{md} pair can recombine via short-range climbing, explaining the predominance of Lomer dislocations observed at high processing temperatures~\cite{Marzegalli2013}.
The authors, however, point out that the experimental reports of statistics of Lomer \glspl{md} could be affected by the difficulty of discerning a 60\degree{} pair from a Lomer dislocation. In Fig.~\ref{fig:GeStrainRelaxation}(e), they show that the \acrlongpl{bvec} of a 60\degree{} \gls{md} pair and a Lomer \gls{md} are the same, and that the two can be easily confused given the small distance between the dislocation cores in the 60\degree{} \gls{md} pair.
Lastly, it should be mentioned that the elucidated mechanism of 60\degree{}-\gls{md}-induced nucleation is typical of only largely mismatched heteroepitaxial systems, such as the 4.2\% of Ge on Si substrates. At low misfit ($<1.5$\%~\cite{Kvam1990}), such as that of Si-rich SiGe grown on Si, the \gls{tcr} is too large for this phenomenon to occur; as the first 60\degree{} \gls{md} nucleates and glides to film/substrate interface, its strain field will not be felt at the surface. Hence, there will be no preferential nucleation of a complementary \gls{md}. In this system, Lomer dislocations are thus not observed after growth~\cite{Bolkhovityanov2012}.

\subsubsection*{Stacking faults}
\Acrfullpl{sf} has also been observed in epitaxial Ge on Si(001) in few-nm-thick films~\cite{Sakai1994,Otsuka2017}, Ge islands~\cite{Zou2002}, as well as on growth on patterned substrates~\cite{Huangfu2013,Leonhardt2011}.
The presence of \glspl{sf} is due to the splitting of a 60\degree \gls{md} into two Shockley partial dislocations (90\degree $+$ 30\degree)~\cite{ArroyoRojasDasilva2017}. This process is driven by the reduction in energy associated with the \gls{md} self-energy, proportional to \gls{bvec}$^2$~\cite{Wegscheider1992}.
After dissociation of the 60\degree \gls{md}, the two partials split apart, connected by a \gls{sf}, reaching an equilibrium distance determined by the interplay between the mutually exerted repulsive force from two partials, and the attractive force due to energetically unfavorable lengthening of the \gls{sf}~\cite{Wegscheider1992}.
Of course, the misfit strain also plays a role in the stability of this dislocation complex. In particular, in compressively strained Ge films on Si(001), after a critical thickness of a few~nm the two partials are expected to glide under the driving force of the misfit stress and recombine~\cite{Gutkin2001}. This explains the presence of \glspl{sf} in thin Ge films on Si(001), and their absence in thick films. Alternative mechanisms of \gls{sf} formation in few-monolayer-thick Ge films have been proposed~\cite{Maras2016}.
It has been found that \glspl{sf} in thin films disappear after annealing due to recombination of the partial dislocations into a \gls{md}~\cite{Sakai1994,Otsuka2017}.

\subsubsection*{Threading dislocations}
As a result of the elucidated strain relaxation phenomena, the as-grown Ge epitaxial films on Si(001) present numerous \glspl{td} and short segments of \glspl{md} at the Ge/Si interface, as shown in Fig.~\ref{fig:GeStrainRelaxation}(c).
In the ideal case, an array of regularly spaced Lomer dislocations with a pitch of $\sim9$\,nm with no \gls{td} is sufficient to fully relax the misfit strain of 4.2\%~\cite{Hartmann2004,Bolkhovityanov2012}. This configuration, however, cannot be achieved through spontaneous strain relaxation during the epitaxial process.
At the onset of plastic relaxation, i.e., \gls{tf}$=$\gls{tcr}, 60\degree{} \glspl{md} nucleate at stress concentrators, which consist in impurities or surface steps~\cite{Speck1996}. Their density  thus determine the initial \gls{tdd}, typically of the order of $10^{11}$--$10^{12}$\,cm$^{-2}$~\cite{Bolkhovityanov2012}.
As the film grows, the \glspl{td} propagate in the new monolayers and, developing mostly in inclined directions with respect to the (001) growth plane, there is an increasingly high probability that they meet a neighboring \gls{td} as the film grows~\cite{Speck1996}. When two \glspl{td} with opposite components of \gls{bvec} come within each other's strain field, they can glide/climb under the effect of mutual elastic interactions. They can repel each other, or fuse/annihilate, decreasing the average \gls{tdd} in the film~\cite{Romanov1997}. As a consequence, it is observed that the thicker the Ge film is grown, the lower its final \gls{tdd} is~\cite{Loo2009}. For very large thicknesses of few \textmu m, this geometric effect saturates due to the low \gls{tdd} and consequent decrease in probability of interaction~\cite{Skibitzki2020}.
To reduce the final \gls{tdd} without thermal processing, SiGe buffer layers or Si$_{1-x}$Ge$_x$/Si$_{1-x}$Ge$_x$ superlattice structures may be introduced to act as dislocation filters, causing bending of existing \glspl{td} into \glspl{md} at the hetero-interfaces~\cite{Skibitzki2020, Hull1989}. Buffer layers also favor nucleation of Lomer \glspl{md} over 60\degree{} \glspl{md}, facilitating a more efficient relaxation of misfit strain with lower \gls{md} densities~\cite{Bolkhovityanov2012}.

During growth of a Ge film, dislocations will tend to diffuse (i.e., glide or climb) under the effect of the resolved shear strain in their slip plane, in an Arrhenius-like thermally activated process~\cite{Freund2004}.
In the case of epitaxial Ge on Si, strain acting on dislocations is induced by the lattice mismatch, and by the interaction with other dislocations, which may be of attractive or repulsive character.
\glspl{td} glide by thermally activated motion of kinks along the threading segment~\cite{Freund2004}.
\gls{td} pairs tend to move apart under the effect of misfit strain, elongating the associated \gls{md} core at the Ge/Si interface. They can remain blocked by the repulsive field of a perpendicular \gls{md}~\cite{Gillard1994, Pichaud1999}, or by a parallel \gls{td} with opposite \gls{bvec} that causes cross-slip~\cite{Bolkhovityanov2012,Stach2000}.
Dislocations may thus remain pinned by the interactions with other dislocations, or by point defects~\cite{Loo2009}.
Depending on the growth temperature, kinetic barriers of dislocation diffusion may not be overcome, and the Ge film may be left with residual global -- i.e., lattice mismatch -- and local strains -- i.e., dislocation-dislocation interactions -- acting on dislocations.
Annealing the film at high temperatures, typically $\geq800$\degree C, activates dislocation diffusion driven by the felt shear stresses. The resulting recombination of dislocations allows to improve the structural and electrical properties of the material, and is thus essential in the absence of different strain engineering strategies.
Effective strain relaxation in the film, enabled by high growth temperature or annealing, is recognized by cross-hatch surface roughness patterns, shown in Fig.~\ref{fig:GeStrainRelaxation}(h). These patterns are associated with surface Ge adatom diffusion driven the strain induced by buried networks of elongated \glspl{md} at the Ge/Si interface~\cite{Rovaris2019}.

The next section is dedicated to explaining the dynamics of dislocations in Ge films during annealing.

\subsection{Annealing of Ge buffer}
\label{sec:GeSn-Ge_StrainRelax_GeAnneal}
\subsubsection*{Reactions between threading dislocations}
\begin{figure*}[htb]
	\centering
	\includegraphics[width=\textwidth]{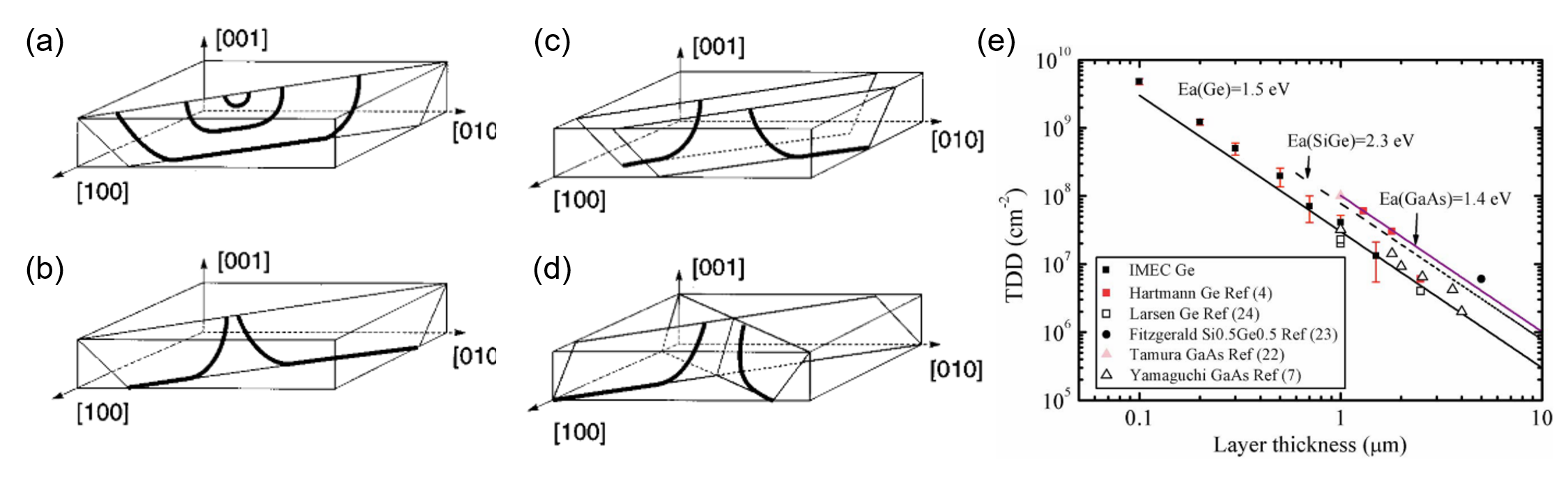}
	\caption{Modes of \glspl{td} recombination: (a)\,Self-annihilation; (b)\,On same slip plane by glide; (c)\,On parallel slip planes by glide+cross-slip or climb; (d)\,Non-parallel slip planes by glide and/or climb. (e)\,Prediction of \emph{quasi-equilibrium} \gls{tdd} according to the model from Ref.~\cite{Wang2009}. Figures (a-d)\, reprinted with permission from Speck\etal~\cite{Speck1996}, \copyright\,1996 \emph{AIP Publishing}. Figure (e) reproduced with permission from Wang\etal~\cite{Wang2009}, \copyright\,2009 \emph{AIP Publishing}.
	}
	\label{fig:GeAnnealing}
\end{figure*}

As elucidated in the previous section, as-grown epitaxial Ge films on Si(001) contain several segments of \glspl{md} with their associated \glspl{td}, and possibly residual misfit strain.
\glspl{td} additionally sense mutual stress fields from neighboring \glspl{td}, which can extend with a radius of 50\,nm from their dislocation core~\cite{Speck1996}. Despite the sensed stress, dislocation motion is impeded by the lack of sufficient thermal energy to overcome kinetic diffusion barriers.
Upon annealing, \glspl{td} glide and climb under the effect of global -- i.e. misfit -- and local -- i.e. neighboring \glspl{td} -- stress fields, pulling their associated \gls{md} segment behind.
\gls{td} pairs will be pushed apart by the misfit stress, and the associated \glspl{md} will be elongated, resulting in the relaxation of misfit strain.
As \glspl{td} sense the stress field from neighboring \glspl{td}, they may repel or attract each other, depending on their relative signs of \gls{bvec}. \glspl{td} can fuse into one \gls{td}, or fully annihilate if they have antiparallel \gls{bvec}.
The film \gls{tdd} will therefore decrease as a result of \gls{td} recombination, which can take place in 4 different modalities~\cite{Speck1996}, illustrated in Fig.~\ref{fig:GeAnnealing}(a-d):
\begin{enumerate}[(a)]
	\item Dislocation loop self-annihilation, which can only occur upon reversal of film strain.
	\item \glspl{td} on the same slip plane recombine by glide.
	\item \glspl{td} on parallel slip planes recombine by glide+cross-slipping, or by climb.
	\item \glspl{td} on non-parallel slip planes recombine by glide, climb, or a combination of the two.
\end{enumerate}
Due to the inclined angle of \glspl{td}, their interaction probability increases with the film thickness.
Wang\etal~\cite{Wang2009} constructed a simple thermodynamic model considering a balance of the energies involved in reaction\,(b) to calculate the extent of \gls{tdd} reduction in function of the Ge layer thickness.
Their argument is that in absence of external stress, which is the case after the misfit stress is fully relaxed, the interaction energy has to be larger than the Peierls barrier  -- i.e. the energy barrier for dislocation glide -- for \glspl{td} to glide under the effect of mutual stress fields, independently on their attractive or repulsive interaction.
Therefore, in their model, they calculate the \emph{quasi-equilibrium} distance between two interacting \glspl{td} dislocations by equating their interaction energy, dependent on the distance between 2 threading segments, to the \glspl{td} Peierls barrier.
Their model, presented in Fig.~\ref{fig:GeAnnealing}(e), was found to  describe well the experimental \gls{tdd} measured in annealed Ge films on Si(001), with the average \gls{tdd} scaling with the inverse square of the film thickness.
They referred to the calculated average \gls{tdd} as the \emph{quasi-equilibrium \gls{tdd}} -- and not \emph{equilibrium \gls{tdd}} -- as \glspl{td} are not equilibrium defects that minimize the free energy of the system. Furthermore, the thermal/processing history of the film can influence the final \gls{tdd}.
For example, micro-patterning of films is  beneficial for achieving an improved \gls{tdd}, as \glspl{td} can glide and annihilate at the edges of microstructures: In  10-\textmu m mesas of Ge, Luan\etal~\cite{Luan1999} have demonstrated an order of magnitude improvement in \gls{tdd} with respect to equally-thick Ge films.
Additionally, also cyclic annealed Ge films are found to achieve lower \gls{tdd} values than the quasi-equilibrium \gls{tdd}~\cite{Wang2009}, as discussed in the following paragraph.

\subsubsection*{Experimental recipes for Ge annealing}
In Tab.~\ref{tab:Ge_Annealing}, we summarize the results of a  few experimental studies of annealing of Ge epitaxial films on Si(001).
Ge is always annealed post-growth at $T\ge800$\degree C for a few minutes. Terzieva\etal{} have shown that a higher annealing temperature of 850\degree C results in the quasi-equilibrium \gls{tdd} with shorter annealing times~\cite{Terzieva2008}.
On the other hand, temperature-swing annealing has been found to be a more effective process with respect to single-temperature annealing~\cite{Luan1999,Yamamoto2018}, with the final \gls{tdd} being potentially lower than the quasi-equilibrium \gls{tdd}. This is understood as an enhancement of \gls{td} diffusion -- and probability of interaction -- due to cycled compressive-tensile strains developing during cooling and warming owed to the differential \gls{tec} of the Ge film and Si substrate\,\footnote{\,At 300\,K, the \glspl{tec} of Ge and Si are respectively 5.9$\cdot10^{-6}$\,\degree C$^{-1}$ and 2.6$\cdot10^{-6}$\,\degree C$^{-1}$}~\cite{Bolkhovityanov2012}.
However, as clearly visible from Tab.~\ref{tab:Ge_Annealing}, the high and low temperature limits in temperature-swing annealing are chosen arbitrarily, despite respectively having a major influence on the Si-Ge inter-diffusion and dislocation glide velocity. Knowing that the activation energy for \gls{td} glide increases linearly with the Si fraction in SiGe~\cite{Hull1993}, the Si-Ge intermixing occurring at excessively high annealing \gls{t} may hinder the efficacy of the annealing process in reducing the \gls{tdd}~\cite{Wang2009}.
A systematic study on the effect of the high and low-temperature limits in cyclic annealing should yield to finding the optimal temperatures to employ in the process.

\begin{turnpage}
	\centering
	\begin{table*}
		\renewcommand{\arraystretch}{1.5}
		\caption{\label{tab:Ge_Annealing}Summary of a few works reporting Ge films annealing processes. Films are annealed after growth unless differently specified. Disclaimer: this table does \textbf{not report} an exhaustive list of all works on Ge annealed films.}
		
		\begin{tabular}{p{0.12\linewidth}P{0.06\linewidth}cP{0.065\linewidth}P{0.065\linewidth}P{0.28\linewidth}p{0.3\linewidth}}
			
			\toprule
			
			\centering \arraybackslash Ref. 	& \makecell[c]{Growth\\method} & \makecell{Thick.\\ (\textmu m)}  & \multicolumn{2}{c}{\makecell{\gls{tdd} (cm$^{-2}$)\\ As-grown~~After anneal.}}		& \makecell[c]{Annealing\\process}  &  \centering\arraybackslash Comments \\
			
			\midrule
			
			\rowcolor{lightgrey} 
			Wei, 2020 \cite{Wei2020} & \gls{mbe} & 0.5 & $\sim10^9$ & $\sim8\cdot10^7$ & 800\degree C & \makecell[{{p{\linewidth}}}]{Same \gls{tdd} obtained at 900\degree C, but with more Si-Ge intermixing.}\\

			\addlinespace
			
			 &  & 1 & -& $1.1\cdot10^7$ &  & \\
			& & 5 &  -& $2.0\cdot10^6$ & \multirow{-2}{*}{5 cycles: 750\degree C, 850\degree C} & \\
			\multirow{-3}{*}{\makecell[c]{Yamamoto,\\ 2018 \cite{Yamamoto2018}}} & \multirow{-3}{*}{\gls{rpcvd}} & 5 & -& $1.3\cdot10^6$ & 850\degree C cycles during growth & \multirow{-3}{*}{Annealing during growth yields lower \gls{tdd}.} \\
			
			\addlinespace
			
			\rowcolor{lightgrey} 
			Yeh, 2014 \cite{Yeh2014} & \gls{ms} & 1 & \sout{$2.6\cdot10^6$} & \sout{$\sim10^4$} & 800\degree C (30\,min) & Likely error in \gls{tdd} measurements.\\
			
			\addlinespace
			
			Tan, 2012 \cite{Tan2012} & \gls{rpcvd} & 0.98 &  - & $6\cdot10^6$ & 8 cycles: 825\degree C (10\,min), 680\degree C (10\,min)& \\

			\addlinespace
			
			\rowcolor{lightgrey} 
			 &  &  &  & $6\cdot10^7$ & 800\degree C cycles during growth & \\
			 \rowcolor{lightgrey} 
			&  &  \multirow{-2}{*}{1}  & \multirow{-2}{*}{$1.6\cdot10^9$} & $6\cdot10^7$ & 800\degree C post-growth & \\
			\rowcolor{lightgrey} 
			&  &  &  & $7\cdot10^5$ & 800\degree C cycles during growth & \\
			\rowcolor{lightgrey} 
			\multirow{-4}{*}{\makecell[c]{Yamamoto,\\2011 \cite{Yamamoto2011}}} & \multirow{-4}{*}{\makecell[c]{\gls{rpcvd}}} & \multirow{-2}{*}{4.7}  & \multirow{-2}{*}{$4\cdot10^8$} & $8\cdot10^6$ & 800\degree C post-growth & \multirow{-4}{*}{\makecell[cl]{Cyclic annealing only effective in films with \\ $t>1$\textmu m.}} \\
			
			\addlinespace
			
			\multirow{2}{*}{\makecell[c]{Shah, 2011 \\ \cite{Shah2011}}} & \multirow{2}{*}{\makecell[c]{\gls{rpcvd}}} & 0.76  & $2.8\cdot10^8$ & $4.6\cdot10^7$ & \multirow{2}{*}{830\degree C (10\,min) in-situ} & \\
			&  & 1.2  & $1.7\cdot10^8$ & $1.0\cdot10^7$ & & \\

			\addlinespace
			
			\rowcolor{lightgrey} 
			Choi, 2008 \cite{Choi2008} & \gls{cvd} & 2.5 & - & $5\cdot10^6$ & Every 500\, of film growth: 800\degree C (30\,min)& \makecell[{{p{\linewidth}}}]{Annealing during growth is more effective than same process after growth.}\\
			
			\addlinespace
			
			\multirow{5}{*}{\makecell[c]{Terzieva,\\ 2008 \cite{Terzieva2008}}} & \multirow{5}{*}{\gls{cvd}} & \multirow{5}{*}{1.6}  & \multirow{5}{*}{$1\cdot10^8$} & $2.0\cdot10^7$ & 800\degree C (1\,min)& \multirow{5}{*}{\makecell[cl]{Higher annealing temperatures allow to reach \\quasi-equilibrium \gls{tdd} faster.}}\\
			& &  &  & $1.5\cdot10^7$ & 800\degree C (3\,min)& \\
			& &  &   & $1.2\cdot10^7$ & 800\degree C (30\,min)& \\
			& &  &   & $1.2\cdot10^7$ & 850\degree C (3\,min)& \\
			& &  &   & $1.2\cdot10^7$ & 850\degree C (20\,min)& \\

			\addlinespace
			
			\rowcolor{lightgrey} 
			 &  & 0.73  & $9\cdot10^8$ & $<2\cdot10^8$ &  &  \\
			\rowcolor{lightgrey} 
			\multirow{-2}{*}{\makecell[c]{Hartmann, \\ 2004 \cite{Hartmann2004}}}& \multirow{-2}{*}{\makecell[c]{\gls{rpcvd}}} & 1.66  & - & - & \multirow{-2}{*}{\makecell[c]{10 cycles: 750\degree C (10\,min),\\875\degree C (10\,min)}} & \multirow{-2}{*}{}\\ 
		
			\addlinespace
			
			Luan, 1999 \cite{Luan1999} & \gls{cvd} & 1  &$9.5\cdot10^8$ & $2.3\cdot10^7$ & 10 cycles: 900\degree C (10\,min), 780\degree C (10\,min) & In 10-\textmu m mesas, obtained \gls{tdd} is $2.3\cdot10^6$\,cm$^{-2}$.\\

			\bottomrule
		\end{tabular}
	\end{table*}
\end{turnpage}

Performing annealing steps during growth has been found to be more effective in reducing the final \gls{tdd} density with respect to post-growth annealing~\cite{Yamamoto2018,Choi2008}.
This is because the annihilated \glspl{td} do not thread in the following grown layers, decreasing the density of \gls{td} that have to recombine in the post-growth annealing step.
This process has been observed to be more effective in Ge films with thickness $>1$\textmu m~\cite{Yamamoto2011}, likely due to the saturation of the geometric effect at the low \gls{tdd} of thick films~\cite{Skibitzki2020}.
Lastly, Nayfeh\etal~\cite{Nayfeh2004} suggested that the presence of H$_2$ during annealing enhances Ge adatom mobility, enabling a decrease in surface roughness from 25\,nm to $\sim3$\,nm. Their as-grown 200-nm-thick Ge films, however, had suboptimal \gls{rrms} levels.
Similarly, annealing in \ce{H2} has been observed to reduce the roughness from 3.5\,nm to 0.7\,nm in Ge  layers epitaxially overgrown on patterned Si substrates~\cite{Yu2010}. Furthermore, \ce{H2} annealing was shown to induce monolayer terracing in Ge(001) wafers~\cite{Nishimura2014}, and to promote outdiffusion of oxygen impurities~\cite{Toriumi2018}.
On the other hand, with $\sim1$\,nm  \gls{rrms} in as-grown 150-nm-thick Ge, annealing in  \ce{H2} worsened the \gls{rrms} at $T>650$\degree C~\cite{Kobayashi2010}. Hartmann\etal{}~\cite{Hartmann2010} also observed the worsening of \gls{rrms} with increasing annealing time at 750\degree C in smooth 270-nm-thick Ge films on Si(001). In the same study, the authors showed that these annealing conditions had no effect on the roughness of 2.45-\textmu m-thick films, which however considerably roughened with prolonged cyclic annealing between 750\degree C and 890\degree C.
The data suggests that indeed \ce{H2} enhances surface adatom mobility. This may be detrimental however when the thin films are under strain, e.g. with residual compressive strain in thin Ge epitaxial films, or in strain developed during cyclic annealing due to differential \gls{tec}, as strain-induced roughening mechanisms may lead to worsening of \gls{rrms}.
Systematic studies of the performance of annealing with/without \ce{H2} in (un)strained films would help clarifying the role of \ce{H2}, often employed during Ge annealing~\cite{Choi2008,Tan2012,Yeh2014,Yamamoto2018}

\subsubsection*{Dislocation arrangement in annealed Ge }
After annealing and full strain relaxation of Ge on Si(001), arrays of regularly spaced \glspl{md} can be observed at the Ge/Si interface~\cite{Yeh2014,Choi2008,Sakai1997, Wei2020}, mainly of 90\degree character~\cite{Bolkhovityanov2012, Wei2020, Sakai1997}. 
This is a consequence of the strain-driven diffusion of \glspl{td}, which causes the elongation of \glspl{md} segments for efficient misfit strain relaxation. The arrangement in regular spacing of ~\glspl{md} is instead owed to gliding of 60\degree{} and vacancy-mediated climbing of 90\degree{} \glspl{md} due to self-interactions~\cite{Barbisan2022}.  A regular spacing between 9\,nm and 10\,nm is  expected to fully release the Ge-Si misfit strain of 4.2\%~\cite{Bolkhovityanov2012,Sakai1997,Hartmann2004,Wei2020}. 
The \gls{md} density will therefore increase as a result of annealing, which should be taken into consideration in case \glspl{md} were discovered to be electrically active. Currently, the electrical activity of \glspl{md} is not understood, as discussed later in Sec.~\ref{sec:ElecProp_TrapStates}.

\subsection{Strain Relaxation during Epitaxial Growth of GeSn on Ge buffer}
\label{sec:GeSn-Ge_StrainRelax_GeSnRelax}
The strain relaxation behavior of GeSn on Ge is similar to that of Ge on Si due to the materials having the same crystal structures. There are however some differences attributable to the lower lattice mismatch of GeSn on Ge, and the metastability of GeSn at Sn contents larger than 1\atp{}.

The impact of GeSn metastability and the associated Sn out-diffusion on strain relaxation in the material has been discussed in Sec.~\ref{sec:GeSn_Challenges_Segregation}. In summary, the alteration in alloy composition via Sn out-diffusion has been identified as a possible mechanism for strain relaxation in pseudomorphic GeSn thin films grown on Ge~\cite{VondenDriesch2020,Zaumseil2018}.
Conversely, bulk Sn clustering has been ruled out to contribute to strain relaxation by \gls{apt} studies~\cite{Kumar2015}. Sn clustering is only driven by the material metastability upon temperature increase~\cite{Kumar2015}, and it is enhanced in presence of linear defects~\cite{Stanchu2020,Mukherjee2021}.

\subsubsection*{Critical thickness for strain relaxation of GeSn on Ge}
Wang\etal~\cite{Wang2015} showed that GeSn relaxes on Ge(001) substrates following \acrfull{pb} model for critical thickness~\cite{People1985,People1986}, in agreement with the \acrfull{tcr} reported in other studies where GeSn was grown by \gls{cvd}~\cite{Assali2019}, \gls{mbe}~\cite{Bhargava2013}, and \gls{ms}~\cite{LadrondeGuevara2003}.
Slightly lower values of \gls{tcr} were found in GeSn grown on Ge buffer layers~\cite{Gencarelli2013}, possibly due to the presence of \glspl{td} propagating from the buffer layer that reduce the nucleation energy of \glspl{md} in GeSn~\cite{Bolkhovityanov2012}. One cannot exclude though that it may have been due to different growth parameters (i.e., growth rate and growth \gls{t} )~\cite{Wang2015}.
On the other hand, Cai\etal~\cite{Cai2022} reported the \gls{mbe} growth of GeSn on Ge(001) up to 9.7\atp\,Sn significantly exceeding the \gls{tcr} predicted by the \gls{pb} model. The authors claimed this originated due to the low growth temperature of 150\degree C, though in contrast with the systematic study of Wang\etal{}~\cite{Wang2015}, where GeSn films were deposited at the same growth rate and \gls{t}. The results from Ref.~\cite{Cai2022} may in fact be affected by Sn-Ge intermixing, clearly visible by the asymmetry of the Bragg-Brentano \gls{xrd} curves of GeSn films with thickness larger than the \gls{tcr} predicted by the \gls{pb} model. Intermixing may arise from strain relaxation, and is known to occur in the Si-Ge system~\cite{Hanafusa2012,Tsukamoto2015,Cheng2010,Huang2022a}. Hence, the mechanism of strain relaxation in Ref.~\cite{Cai2022} may have been Sn-Ge intermixing rather than nucleation of \glspl{md}, explaining the observed \gls{tcr} exceeding the values predicted by the \gls{pb} model.
In general, these results support the need to use a strain relaxation model that accurately considers all the kinetic physical processes at play~\cite{Houghton1991}.

\subsubsection*{Strain relaxation mechanisms}
Upon relaxation of GeSn on Ge at  \gls{tf}$=$\gls{tcr}, 60\degree{} half-loops will nucleate on surface~\cite{Assali2019,Stanchu2021}, gliding to the interface to form a network of \glspl{md}. The latter have been observed predominantly to be of 60\degree{} character~\cite{Dou2018,Lin2023}, in accordance with predictive models for low-misfit group-IV systems~\cite{Bolkhovityanov2012}.
In contrast with the Ge-on-Si(001) system, after relaxation, \glspl{sf} along the \hkl{111} planes are often observed at the GeSn/Ge interface, extending in short segments across the interface~\cite{Dou2018,Gupta2018,Gallagher2015a,Gallagher2015,Lin2023}.
Their origin is associated with two phenomena: (a) splitting of 60\degree{} \glspl{md} into the more energetically favorable Shockley partials, bound by a \gls{sf}~\cite{Gupta2018,Dou2018}, and (2) \gls{sf}-bound Frank partial dislocations associated with vacancy complexes~\cite{ArroyoRojasDasilva2017,Dou2018}. The latter may be characteristic of the GeSn material system, where Sn-vacancy complexes are expected to form~\cite{Zaima2015}.
In addition, Dou\etal~\cite{Dou2018} observed the formation of full-edge Lomer dislocations from the reactions between Shockley and Frank partials.
Relaxation by strain-induced roughening has also been observed~\cite{Assali2019}.

In low-temperature growth ($T\sim150$\degree C), relaxation of GeSn on Ge  occurs with slightly different mechanisms due to the limited thermal energy available in the system.
In particular, in both \gls{mbe}~\cite{Rathore2021,Wan2022,Wang2015} and \gls{ms}~\cite{Zheng2017}, it has been observed that only the upper part of GeSn films relaxes via the formation of dislocations, while the portion of the film close to the substrate remains fully strained.
A model for this  phenomenon was proposed by Wan\etal~\cite{Wan2022}. In their model, during growth of GeSn in kinetic roughening regime, surface roughness arises with the typical shape of mounds. Since surface roughness features can  give rise to stress concentrators, facilitating dislocation nucleation,  at $t=t_{cr}$, \gls{md} nucleation  is facilitated at the cusps formed between the mounds.
Dislocations propagate upwards as the film grows, while the downward propagation is believed to be hindered by the low temperatures~\cite{Wan2022}, explaining the fully strained bottom region of the film.

Furthermore, the relaxation behavior of GeSn is affected by the layers beneath. For example, nucleation of \glspl{md} is facilitated if the Ge buffer has a large \gls{tdd}~\cite{Bolkhovityanov2012,Ju2017}. In this situation, GeSn films are not expected to follow the \gls{pb} model for critical strain relaxation.
On the other hand, on graded GeSn buffers, dislocations tend to be confined in the first layers, with limited propagation of \glspl{td} to the surface~\cite{Dou2018,Assali2019,Castioni2022}.
This is understood as the result of dislocation bending at the buffers interfaces, with resulting enhanced probability of interaction with dislocations with opposite components of \gls{bvec}~\cite{Bolkhovityanov2012} and due to Hagen-Strunk multiplication mechanisms that induce the nucleation of complementary 60\degree{} \gls{md}, with resulting formation of Lomer \glspl{md}~\cite{Dou2018,Assali2019}.
Lastly, large amounts of \glspl{sf} have been found at the interface of GeSn with more than 20\atp{}\,Sn  grown directly on Si(001)~\cite{Xu2019a,Mircovich2021}. This is unexpected for such large compressive misfits~\cite{Gutkin2001}, and it may indicate different relaxation behaviors that remain to be understood to date.

Upon annealing GeSn films, strain relaxation may be activated in pseudomorphic or partially relaxed films.
Experimental data suggests that \glspl{md} elongate during strain relaxation, but that no new \glspl{md} are formed~\cite{Stanchu2021}. In pseudomorphic films, strain relaxation takes place rather by Sn out-diffusion~\cite{VondenDriesch2020}, although this behavior may be strongly composition-dependent~\cite{Zaumseil2018,Cai2022}.
\glspl{md} may also nucleate from \glspl{td} present in the underlying Ge buffer~\cite{VondenDriesch2020,Zaima2015}.

\subsection{Point Defects in Ge}
\subsubsection*{Intrinsic Ge point defects}
In pure Ge, theory predicts a formation energy for monovacancy (V) defects of about 2.9\,eV~\cite{Weber2015}, which yields a practically null vacancy concentration at temperatures typically employed for GeSn growth\,\footnote{\,The concentration of vacancies is given by $N=N_0*\exp(-E_f/kT)$, where $N_0$ is the density of sites the defect can occupy and $E_f$ is the formation energy. At 300\,K, with $N_0=4.41\cdot10^{22}$\,at/cm$^{-3}$, this expression yields $N<<1$\,at/cm$^{-3}$, while at 600\,K, typically used for GeSn growth, it yields $N\sim0.02$\,at/cm$^{-3}$.}.
Self-interstitials have even larger formation energies~\cite{Weber2015}.
At room temperature, these defects are therefore absent in intrinsic Ge~\cite{Sgourou2022}.

Vacancies and interstitial defects can arise from irradiation damage~\cite{Kuitunen2008,Markevich2009}, out-of-equilibrium growth processes~\cite{Knights2001,Gossmann1992,Jorke1989,Bai2012a,Shoukri2005,Ueno2000}, strain~\cite{Antonelli1989,Aziz1997,Choi2010}, and presence of impurities~\cite{Slotte2016,AvSkardi2002}.
Experimental studies of neutron-irradiated lightly n-doped Ge ($n_{Sb}\sim1.5\cdot10^{15}$\,cm$^{-3}$) by \gls{pas} showed that Ge monovacancies are unstable above 65\,K~\cite{Slotte2021}, as they tend to agglomerate into neutral divacancies~\cite{Slotte2012}. Divacancies are stable at room temperature, and tend to agglomerate in larger-sized vacancy clusters after annealing at 200\degree C~\cite{AvSkardi2002,ChristianPetersen2010}, while negatively-charge divacancies are stable up to 400\degree C~\cite{Kuitunen2008}. However, these defects disappear after annealing at 500\degree C~\cite{Kuitunen2008,Elsayed2015}, indicating that after annealing of a Ge buffer for \gls{tdd} reduction all intrinsic defects are expected to annihilate.
Ge self-interstitials caused by irradiation also annihilate at $T>150$\,K~\cite{Markevich2009}.
Point defects remaining in the Ge buffer after annealing are due to impurities present in the film, which may typically be H, N, O, C from the base vacuum pressure.

\subsubsection*{Ge-impurity complexes}
In germanium, H and O sit in interstitial positions~\cite{Weber2007,VandeWalle2008,Clauws1996}. N is often found in the Ge crystal in molecular form (i.e., \ce{N2}) occupying interstitial sites~\cite{Chambouleyron1998}, while C occupies mainly substitutional sites~\cite{Brotzmann2008}.
Furthermore, N, O, and C form stable complexes with vacancies in the Ge crystal~\cite{Chroneos2007,Chroneos2009a,Kuganathan2022,Markevich2003,Li2016,Vohra2019,Vohra2019a}.
These impurities and defect complexes evolve upon annealing. For example, \ce{VO} pairs annihilate at 160\degree C, forming \ce{VO2} complexes, which have different electrical activity~\cite{Carvalho2007}, while O interstitials (\ce{O_i}) can cluster into \ce{O4} during annealing~\cite{Clauws1996}.
The electrical activity of these defects will be reviewed in Sec.~\ref{sec:GeSn_ElecProp}.
Dopant elements, with the exception of B, also tend to form vacancy complexes that are more stable compared to Ge monovancies~\cite{Chroneos2008}.
On the other hand, the presence of Ar in the lattice is not expected to yield any electrically-active Ar-specific defect~\cite{Auret2007a}. In additon, surface defects due to Ar implantation are annealed at $T>250$\degree C~\cite{Auret2007}.

\subsection{Point Defects \& Sn Clustering in GeSn}
\label{sec:Defects_GeSn}

\subsubsection*{Sn occupies substitutional lattice sites}
When adding Sn to the Ge matrix, \emph{ab-initio} calculations predict that Sn occupies preferentially substitutional positions in the lattice~\cite{Hohler2005,Decoster2010}, in agreement with experimental reports~\cite{Kamiyama2014a}.
The fraction of Sn atoms occupying substitutional sites can be evaluated with \gls{rbs} in channeling geometry, comparing the $\chi_{min}$ -- i.e., ratio of \emph{aligned} to \emph{random} peak height -- values of Ge and Sn~\cite{Bhargava2013}.
It was found that Sn incorporates in substitutional sites for Sn contents up to 14\atp\,Sn in \gls{cvd}~\cite{Bauer2002,Rainko2016}, \gls{mbe}~\cite{Su2011a}\,\footnote{\,In Ref.~\cite{Su2011a}, they claim $\chi_{min}$ is equal for Sn and Ge, but precise values are not reported.},  and \acrfull{gsmbe}~\cite{Xu2017,Xu2022}.
Su\etal~\cite{Su2011a} measured a decrease in $\chi_{min}$ in \gls{mbe} GeSn with increasing Sn content or decreasing growth temperature, indicating an worsening of crystal quality.
A significant improvement in $\chi_{min}$ value in \gls{cvd}-grown Sn-doped Ge films was observed after \gls{pda} at 680\degree C by Roucka\etal~\cite{Roucka2011i}.
Bhargava\etal~\cite{Bhargava2013} reported substitutional Sn fractions of at least 90\% in \gls{mbe} GeSn with 2.3--14.5\atp\,Sn, with a decrease in substitutional fraction with increasing Sn content.
Nonetheless, one should note that small deviations from full substitutional Sn occupations measured by \gls{rbs} may in fact arise from measurement artifacts.
In fact, to accommodate the local strain induced by Sn atoms in the Ge matrix, Ge-Sn bonds are distorted ~\cite{Gencarelli2015,Beeler2011}. This yields to a reduced channeling in correspondence with Sn atoms even in a GeSn crystal with full Sn substitutional incorporation~\cite{Xu2017}.

\subsubsection*{Sn-vacancy complexes}
The presence of Sn in the Ge matrix correlates to an increase in point defect concentration, as Sn atoms act as vacancy sinks. Experimental studies of Sn-implanted Ge, backed by \emph{ab initio} computations, demonstrated that Sn-V defects are more stable than isolated vacancies in Ge~\cite{Decoster2010,Riihimaki2007,Chroneos2007,Fuhr2013}. Furthermore, due to opposite elastic fields, Sn-V pairs attract neighboring substitutional Sn atoms, favoring phase separation~\cite{Ventura2009}.
Intuitively, this can be understood as a vacancy accommodating local lattice strain induced by large Sn atoms in the Ge matrix~\cite{Shimura2010}.
The stability of Sn-V defects thus increases when Sn$_n$-V$_m$ complexes are formed~\cite{Chroneos2009}.
Sn-V pairs arrange in split-configuration -- i.e., Sn atom sitting at the bond-centered site -- and are stable up to at least 400\degree C~\cite{Decoster2010}. Considering the low thermal stability of GeSn alloys~\cite{Zaumseil2018,Xu2022,Conley2014a}, this finding suggests that it is not always possible to annihilate Sn-V pairs by annealing GeSn, as Ge-Sn phase separation may occur simultaneously.
As we will see in Sec.~\ref{sec:GeSn_ElecProp}, Sn-V vacancy complexes are electrically active. It is therefore of primary importance to limit the amount of vacancies induced during growth, e.g., by low-temperature growth~\cite{Gossmann1992}, excessive growth rate~\cite{Jorke1989} and compressive  strain~\cite{Antonelli1989,Aziz1997,Choi2010}, as they cannot be eliminated in post-growth processing.
On the other hand, as the processing temperature is increased, the concentration of Sn-V pairs predicted by thermodynamics also increases~\cite{Ventura2009}.
This would suggest a trade-off between high-temperature \gls{cvd} growth methods and low-temperature \gls{mbe} may be necessary for optimal optoelectronic properties of GeSn. The intermediate temperatures employed in \gls{ms} GeSn epitaxy may turn out to be favorable in this sense. In this context, an assessment of the electrical properties of \gls{ms} GeSn is required considering that its carrier concentration, mobility, and lifetime have not been documented before. 

Vacancies are difficult to characterize in a material and, for this reason, there exists only a limited amount of studies on vacancy-related defects in epitaxially grown GeSn, which do not allow to draw final conclusions on optimal growth conditions of the material.
\gls{pas} is the ideal technique to observe vacancy-related defects due to its versatility in measurement conditions and acceptable sensitivity range of 10$^{15}$--10$^{19}$\,cm$^{-3}$, depending on the charge state of the vacancy~\cite{Tuomisto2013}.
Assali\etal~\cite{Assali2019a} studied thoroughly the presence of vacancies in 500--700-nm-thick relaxed GeSn with Sn contents of 6.5--13\atp{}  grown by \gls{lpcvd}  at 330--300\degree C on \gls{vge}. With room-temperature \gls{pas} measurements, they observed the absence of monovacancies, and a predominance for divacancy complexes, with few higher-order vacancy clusters.
On the other hand, in the Ge buffer they measured a predominance of vacancy clusters, attributed to the diffusion of (di)vacancies and clustering during high-temperature annealing of the material prior to GeSn growth.
Interestingly, they also showed a decrease in vacancy clusters and concomitant increase in divacancies with increasing Sn fractions in the film, attributed to capturing of divacancies by the higher concentration of Sn atoms in the lattice.
Slotte\etal~\cite{Slotte2014} found somewhat contrasting results in \gls{pas} characterization  of 400--500\,nm GeSn with 6--12.6\atp\,Sn (\gls{dsr}$\sim70$\%). In a preliminary study, they observed a predominance of vacancy clusters over mono- or di-vacancies. The difference in claims between the two studies may arise from the growth conditions of the material, unspecified in Ref.~\cite{Slotte2014}.
Lastly, the \gls{pas} results from Kamiyama~\cite{Kamiyama2014} in 200-nm-thick pseudormorphic Ge(Sn) grown at 170\degree C by \gls{mbe} on Ge substrates also hint at a higher vacancy concentration in Ge$_{0.983}$Sn$_{0.017}$ with respect to pure homoepitaxial Ge. On the other hand, they observed  a lower positron lifetime -- hence, vacancy concentration -- with 0.1\atp\,Sn, corroborated by electrical measurements, but they did not provide explanations of this result.

\subsubsection*{Sn clusters}
Another type of defect commonly observed in GeSn are few-atom-sized Sn clusters, considered to be the onset of phase separation in metastable GeSn.
Atomistic calculations predict a repulsion between Sn substitutional defects in Ge~\cite{Fuhr2013,Ventura2009}, which may explain the \gls{sro} recently experimentally observed by Lentz\etal~\cite{Lentz2023}.
Sn clusters are therefore expected to be stable only in the \textbeta-Sn phase.
Sn clustering is favored by the presence of vacancies in the film, as Sn-V pairs attract neighboring substitutional Sn atoms~\cite{Ventura2009}, and is also favored by compressive strain~\cite{Liu2022a}.
Calculations from Chroneos\etal~\cite{Chroneos2009} predict Sn$_n$-V$_m$ complexes to be more stable with respect to simple Sn-V pairs.
Experimentally, Sn clusters have been indirectly detected by \gls{apt} in \gls{apcvd} GeSn grown at 320\degree C with nominal Sn content of 5\atp{}~\cite{Kumar2013}, where the authors deduced the presence of Sn$_2$V, Sn$_3$V, and Sn$_4$V$_2$ complexes.
They also observed a higher concentration of these defects in a relaxed film, likely associated to solute segregation at dislocations~\cite{Mukherjee2021,Nicolas2020}.
With successive investigations, the same research group concluded that Sn clustering is not involved in strain relaxation mechanisms, but it is rather driven by the material metastability~\cite{Kumar2015}.
Consistent with the tendency of Sn to segregate on surface, Liu\etal~\cite{Liu2022a} found by \gls{apt} a higher concentration of Sn clusters towards the surface of their \gls{cvd} Ge$_{0.86}$Sn$_{0.14}$ films.
On the other hand, Rathore\etal~\cite{Rathore2021} employing \gls{apt} found no evidence of Sn clustering in their \gls{mbe}-grown Ge$_{0.84}$Sn$_{0.16}$ films up to thicknesses of 250\,nm.
This suggests that appropriate growth parameters allow to prevent Sn clustering.
On the other hand, one must consider that compositional artifacts may originate from \gls{apt} characterization~\cite{Muller2012}. For example, due to the thermal energy deposited in the material, diffusion of Sn may occur during \gls{apt} sample preparation and measurements, resulting in a change of atomic structure with respect to the as-grown sample. This could potentially lead to the emergence of Sn clusters in the \gls{apt} sample, which were not present in the as-grown film. Further work is required to shed light on these results.

\subsubsection*{Sn-impurity complexes}
The presence of Sn in the GeSn film has also been found to stabilize impurities.
Sn-V pairs tend to attract impurities such as C~\cite{Chroneos2007}, O~\cite{Chroneos2011}, forming complexes that are more stable compared to impurity-V pairs.
H/H$_2$ species have also been reported to interact with Sn-V complexes~\cite{Nakatsuka2013} while, to the best of our knowledge, no study has been reported on the behavior of N impurities in GeSn. 
Interestingly, substitutional Sn atoms have been found to attract C~\cite{Kamiyama2015} and repel O impurities~\cite{Chroneos2011}.
Finally, due to the strong binding energy of Sn with vacancies, Sn-doping of Ge has been proposed as a method to prevent the formation dopant-V pairs that limit dopant activation in Ge~\cite{Vohra2019,Khanam2020}.
However, both As-V~\cite{Khanam2020} and P-V pairs~\cite{Markevich2011,Markevich2011a,Vohra2019,Vohra2020} seem to be more stable than Sn-V pairs.

\section{Optoelectronic Properties of Ge \& GeSn}
\label{sec:GeSn_ElecProp}
In this section, we provide an overview of the current understanding of the optoelectronic properties of Ge and GeSn.
First, we briefly discuss the optical absorption coefficient of GeSn, focusing on the wavelength of 1.55\,\um{}. We highlight a significant scattering in the data, especially evident in experimental results.
Subsequently, we explore the electronic trap states measured in Ge and GeSn, summarizing extensively the data reported in the literature and presenting it in a table format for enhanced clarity.
We then undertake a similar process concerning experimental data on unintentional doping levels, carrier lifetime, and mobility in both Ge and GeSn films. For each property, we summarise the main findings to establish a coherent understanding of the subject.
Control of these properties is key in achieving optimal performance in GeSn-based devices~\cite{Ghosh2023}.

We anticipate that the variations in the reported optoelectronic properties of GeSn thin films reflect a significant influence of the growth technique, parameters, and purity conditions. These aspects are absolutely crucial for controlling the performance of GeSn-based devices.


\subsection{Absorption Coefficient of GeSn}
\begin{figure*}[htb]
	\centering
	\includegraphics[width=\textwidth]{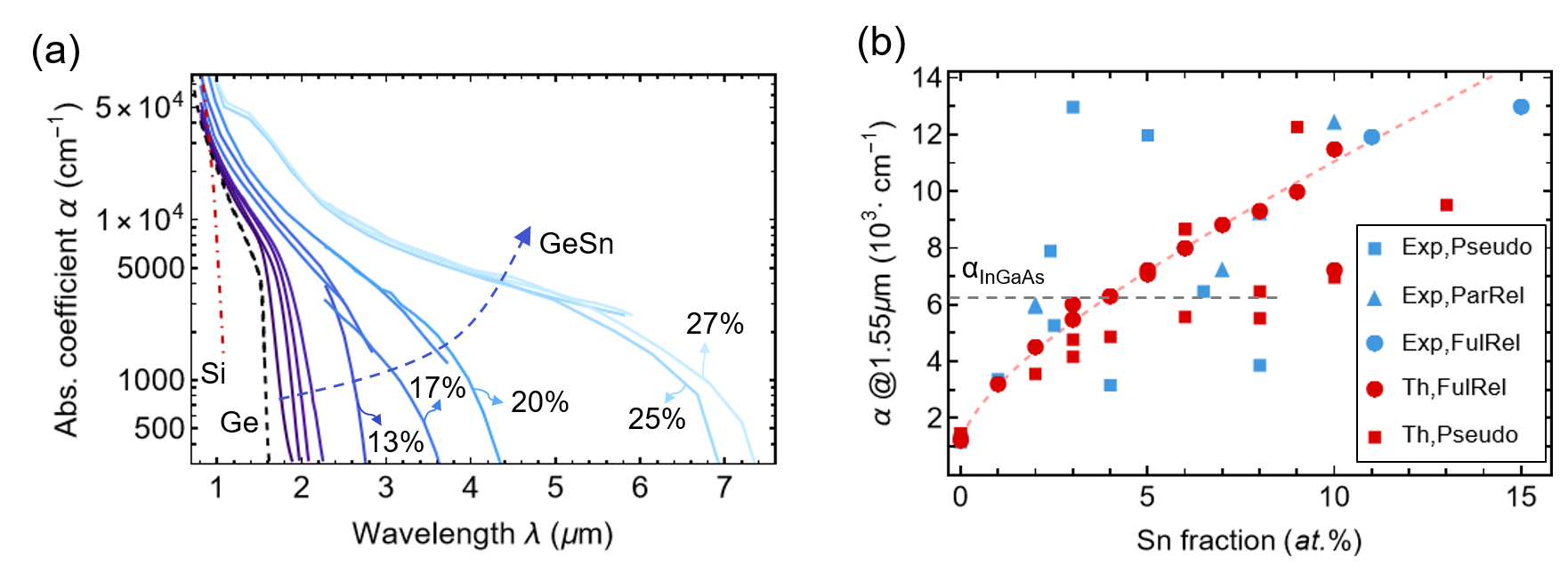}
	\caption{(a)\,Shift of absorption edge in GeSn with increasing Sn \atp. Data plotted from Refs.~\cite{Xu2019b,Imbrenda2018,Xu2019c}.
	(b)\,Absorption coefficient of GeSn at 1.55\,\textmu m plotted from both theoretical (i.e., \emph{Th.}) and experimental (i.e., \emph{Exp}.) studies~\cite{Zhang2015,Tran2016,Li2016a,Menendez2009,Kumar2018,Pandey2018,Pandey2018a,He1997,Roucka2011a,Chang2016a,Dong2016,Wang2016,Tao2020}. The abbreviations \emph{Pseudo}, \emph{ParRel}, and \emph{FulRel} stand respectively for \emph{pseudomorphic}, \emph{partially relaxed}, \emph{fully relaxed}. The grey, horizontal, dashed line indicates the absorption coefficient of In$_{0.53}$Ga$_{0.47}$As from Ref.~\cite{Adachi1989}, while the red,  dashed line is a fit for the computed absorption coefficient of Ge$_{1-x}$Sn$_x$ from Ref.~\cite{Hsieh2012}.}
	\label{fig:GeSnAbsorption}
\end{figure*}

The optical properties of GeSn have been extensively investigated with both theory and experiments.
Following the bandgap predictions reported in Sec.~\ref{sec:GeSn_PhysicProp}, the absorption edge of GeSn shifts towards the infrared for increasing Sn content, as shown in Fig.~\ref{fig:GeSnAbsorption}(a) with experimental data plotted from Refs.~\cite{Xu2019b,Imbrenda2018,Xu2019c}.
\gls{mwir} wavelengths ($\sim$3--8\um) can be accessed starting from 15/16\atp\,Sn, while by extrapolation of the curves in  Fig.~\ref{fig:GeSnAbsorption}(a) we can expect the absorption edge to reach the \gls{lwir} for $x_{Sn}>0.3$.
Fig.~\ref{fig:GeSnAbsorption}(b) shows experimental and theoretical values of GeSn absorption coefficient at 1.55\um{}, wavelength of interest e.g. for telecommunication and \gls{lidar} applications.
The red, dashed line is a fit of theoretical absorption coefficient of unstrained GeSn from Ref.~\cite{Hsieh2012}\,\footnote{\,From Ref.~\cite{Grundmann2010}, the dependence of the absorption coefficient can be approximated to $\alpha \propto (E-E_g)^{-\sfrac{1}{2}}$. Given the bowing behavior of $E_{g,GeSn}$, the data is fitted using the square root of a second-degree polynomial $f(x_{Sn})$.}, and serves as a guide to the eye.
In spite of the considerable scattering in the data, the absorption coefficient tends to increase with the Sn fraction in the alloy.
In Fig.~\ref{fig:GeSnAbsorption}(b), we report with a grey, horizontal, dashed line, the absorption coefficient of  In$_{0.53}$Ga$_{0.47}$As, the absorber material employed in commercial III-V technologies. Multiple experimental data suggests that a few \atp{}\.Sn in relaxed GeSn are sufficient to surpass the absorption coefficient of In$_{0.53}$Ga$_{0.47}$As, demonstrating the potential of GeSn for detector applications.
On the other hand, in GeSn films the epitaxial compressive strain induces a blue-shift in bandgap energy, and is thus expected to limit the absorption coefficient. This effect however remains hidden by the large scattering in both experimental and computed data in Fig.~\ref{fig:GeSnAbsorption}(b).
We conclude that a systematic analysis of the material absorption coefficient as a function of strain and material synthesis method is required to resolve the inconsistencies arising from the different experimental and computational methods employed.

\subsection{Trap states in Ge \& GeSn}
\label{sec:ElecProp_TrapStates}
Trap energy levels in the bandgap of an intrinsic semiconductor determine its electrical properties. Shallow trap levels act as acceptors/donors, releasing free carriers in the material -- i.e., unintentional doping concentration -- while deep trap levels induce \gls{srh} and \gls{tat} generation/recombination mechanisms, detrimental for device performance.
The \gls{srh} recombination rate can be expressed as
\begin{widetext}
\begin{equation}
	\label{eq:srhGen}
	R_{srh}(T)=\frac{pn-n_i^2}{\tau_{r,p}\left(n+N_C\cdot exp\left(\frac{E_t-E_C}{kT}\right)\right)+\tau_{r,n}\left(p+N_V\cdot exp\left(-\frac{E_t-E_V}{kT}\right)\right)}
\end{equation}
\end{widetext}
where $n_i$ the intrinsic material concentration, $p$ ($n$) is the hole (electron) concentration, $\tau_{r,p}$ ($\tau_{r,n}$) the hole (electron) lifetime, $E_V$ ($E_C$) is the valence-band (conduction-band) energy. $E_f$ is the Fermi energy, $E_t$ is the trap energy level, $k$ is the Boltzmann constant and \gls{t} is the lattice temperature~\cite{Grundmann2010}.
From eq.~\eqref{eq:srhGen}, one can deduce that \gls{srh} carrier generation has a strong dependence on temperature and on the trap energy level. Traps closer to the middle of the \gls{bg} will facilitate \gls{srh} generation of carriers, as they will ensure the lowest possible rate-determining energetic barrier.
\gls{tat} mechanisms have an analogous dependence on trap levels~\cite{Simoen2009}.
It is thus fundamental to review trap states that may arise from impurities and defects in intrinsic Ge and GeSn. There exist multiple studies reporting electronic levels in both materials, typically measured by \gls{dlts}.

\subsubsection*{Trap states in pure Ge}
In Ge, the energy position of electronic levels originating from defects and impurities have been partially reviewed in Refs.~\cite{Zaima2015,Grundmann2010,Claeys2016,Madelung2004}.
Trap states observed in intrinsic Ge and GeSn are reported in Tab.~\ref{tab:TrapLevels}.
Intrinsic point defects in Ge are charged and in principle affect the material electrical properties.
However, as discussed in Sec.~\ref{sec:GeSn-Ge_StrainRelax}, these defects annihilate at temperatures lower than those employed for Ge annealing~\cite{Markevich2009,Kuitunen2008,Elsayed2015}. Consequently, intrinsic Ge point defects are not expected to persist in the annealed Ge buffer layers.
Threading dislocations generating from epitaxial strain relaxation are however electrically active.
Due to the complexity of the system involved, to date electronic trap states in Ge have not been unambiguously assigned to the different linear defect types.
In fact, while mid-gap trap states in Ge are beyond doubt associated with the presence of \glspl{td}~\cite{Simoen2009a,Simoen2019,Tetzner2021,Claeys2016,Giovane2001}, a reduction in \gls{tdd} does not always correspond to a proportional reduction in trap concentrations~\cite{Gonzalez2011,Eneman2009,DiLello2012}, clearly indicating the presence of other sources of deep traps~\cite{Simoen2009a}.
Deep trap states can be generated by point defects, dislocations, and any combination thereof, substantially complicating the assignment of the measured trap levels to specific defects.
A number of different mid-gap levels have been experimentally measured by \gls{dlts}, and reported in Tab.~\ref{tab:TrapLevels}.
It has been proposed  in Ref.~\cite{Simoen2019} that mid-gap traps arise from the interaction of dislocation cores with point-defect clouds~\cite{Simoen2019,Tetzner2021,Grillot1996}, while clean \glspl{td} are expected to generate shallower levels, which may contribute to unintentional doping of the material. This remains however to be verified, since the referred works do not verify if dislocations are clean, or interacting with point defects~\cite{Simoen1985,Sande1986}. 
Studies in which the \gls{tdd} is reduced by thermal annealing, such as those in Refs.~\cite{Gonzalez2011,Eneman2009}, may thus be influenced by the diffusion and clustering of point defects, which may explain the sub-proportional reduction in trap states with \glspl{tdd}.
The elongation of \glspl{md} upon annealing may additionally play a role in changing the electrical properties of annealed Ge. However, to the best of our knowledge, the investigation of the electrical activity of \glspl{md} in epitaxial Ge on Si has never been reported.
Lastly, electronic states in the bandgap arise also from the interaction between dislocations~\cite{Shevchenko1999} and from dangling bonds at grain boundaries~\cite{Imajo2022,Simoen2019}.
It is clear that further systematic studies backed by computational works are required to discern among trap states induced by \glspl{td}, 60\degree{} \glspl{md}, 90\degree{} \glspl{md}, \glspl{sf} and partial dislocations.


\setlength{\LTcapwidth}{\textwidth}
\begin{longtable*}[htb]{p{0.08\textwidth}lp{0.01\textwidth}m{0.17\textwidth}m{0.18\textwidth}p{0.01\textwidth}m{0.36\textwidth}}
	\caption{\label{tab:TrapLevels}Trap levels generated by defects in Ge and GeSn measured by \gls{dlts}, unless differently specified. This table was inspired from that of Ref.~\cite{Zaima2015} and significantly expanded to include data for impurities and extended defects in Ge and GeSn. Additional trap levels for Ge may be found in Ref.~\cite{Madelung2004}. Used abbreviations: \acrlong{spc} (\gls{spc}), \acrlong{vge} (\gls{vge}), PD:~point defect, D:~dislocation, ED:~extended defect (i.e. \gls{sf} + partial Ds), GB:~grain boundary, DB:~dangling bonds, BL:~band-like states, int.:~interface, rlx.:~(partially) relaxed, str.:~fully strained, Cz:~Czochralski, e:~electron, p:~proton, n:neutron, \textgamma:~gamma-ray.}\\
	\toprule
	\multicolumn{2}{l}{Defects} & & \multicolumn{2}{l}{Activation energy (eV)} && Sample condition\\
	\cmidrule{1-2}
	\cmidrule{4-5}
	\cmidrule{7-7}
	&             							 &&  Electron  	&  Hole  	&&       				\\
	\midrule
	
	\multirow{8}{*}{\makecell[c]{Point \\defects \\ in Ge}} 
	&  \ce{Ge_i}	 && 0.11\textsuperscript{aa} 	&  && e-irradiated Ge \\
	&  \ce{V^{-/0}}	 &&  	& 0.02\textsuperscript{ab} && Anneal+quench Cz-Ge \\
	&  \ce{V^{--/-}}	 &&  	& 0.26\textsuperscript{ab} && Anneal+quench Cz-Ge \\
	&  \ce{V2}	 && 0.29\textsuperscript{ac,ad}~0.31\textsuperscript{ae,af} 	&  && \makecell[{{p{\linewidth}}}]{e-\textsuperscript{ac}, n-\textsuperscript{ac}, p-\textsuperscript{ad}irradiated, sputter-\textsuperscript{ae}, e-beam-deposited\textsuperscript{af} Ge:Sb} \\
	&  \ce{V2^{--/0}}	 &&  	& 0.19\textsuperscript{ag} && p-irradiated Cz-Ge \\
	&  \ce{V3^{-/0}}	 &&  	& 0.08\textsuperscript{ag} && e-, p-irradiated Cz-Ge\\
	&  Small V cluster	 &&  0.1\textsuperscript{ac}	&  && n-irradiated Ge:Sb \\

	\midrule
	
	\multirow{7}{*}{\makecell[c]{Linear\\defects\\ in Ge}} 
	&  Decorated \glspl{td}	 &&  	& 0.29\textsuperscript{ah,ai},~0.25\textsuperscript{ai}  && \gls{vge} \\
	&  Clean? D	 &&  	& 0.02\textsuperscript{aj},~0.10\textsuperscript{aj}  && D-rich pGe crystal \\
	&  Clean? D	 &&  0.09\textsuperscript{aj}	&  && D-rich nGe crystal \\
	&   D-related	 &&   0.3?\textsuperscript{ak}	& 0.16\textsuperscript{ak},~0.18\textsuperscript{ak} && Relaxed Ge:B on Si \\
	&   D-V-related	 &&  0.28\textsuperscript{al}  	&  0.18\textsuperscript{al} && Relaxed Ge on grSiGe/Si \\
	&   60\degree{}/90\degree{} partials	 &&   0.27\textsuperscript{am}	& 0.07\textsuperscript{am}, 0.19\textsuperscript{am}, 0.27\textsuperscript{am}  && Plastically deformed Ge:Ga  \\
	
	\midrule
	
	\multirow{3}{*}{\makecell[c]{DBs \\ in Ge}} 
	& DBs	 &&  \multicolumn{2}{c}{Below \acrshort{vb}\textsuperscript{an}}  && DFT calculations \\
	& DBs at GB	 &&  	& 0.05 -- 0.10\textsuperscript{ao}  && \acrshort{spc} poly-Ge* \\
	& GB-related	 &&  	& 0.32\textsuperscript{ah}  && poly-Ge \\

	\midrule
	\multirow{11}{*}{\makecell[c]{O \\ in Ge}} 	
	&  \ce{O_{Ge}}	 &&    0.017, 0.04, 0.2	&   && Unspecified, reported in Ref.~\cite{Madelung2004} \\
	&  \ce{O_{i}}	 &&  \multicolumn{2}{c}{Neutral\textsuperscript{ap}}   &&  \\
	&  \ce{O4}	 &&    0.017\textsuperscript{ap}	&   && Annealed O-rich Cz-Ge$^\ast$ \\
	&  \ce{VO^{--/-}}	 &&    0.21\textsuperscript{aq,ar},~0.27\textsuperscript{ad,as} 	&   && \textgamma-\textsuperscript{aq,ar}, p-\textsuperscript{ad,as}, e-\textsuperscript{ad,aq}irradiated Ge \\
	&  \ce{VO^{-/0}}	 &&    &   0.27\textsuperscript{aq,ar}	&& \textgamma-\textsuperscript{aq,ar}, e-\textsuperscript{aq}irradiated O-rich Ge \\
	&  \ce{VO_2^{--/-}}	 &&    0.195\textsuperscript{at}	&   && e-irradiated O-rich Ge\\
	&  \ce{VO_2^{-/0}}	 &&    0.365\textsuperscript{at}	&   && e-irradiated O-rich Ge\\
	&  O-related	 &&    0.14\textsuperscript{ad}, 0.19\textsuperscript{ad} 	&   && p-, e-irradiated O-rich n-Ge \\
	&  O-, H-related?	 &&   	&  0.15\textsuperscript{au} && p-irradiated Ge:Sb \\
	&  \ce{I_{Ge}}-O\textsubscript{2i}&&  0.06\textsuperscript{aa,as},~0.08\textsuperscript{aa}  	&  && e-\textsuperscript{aa}, p-\textsuperscript{as} irradiated O-rich Ge:Sb\textsuperscript{aa}, Ge:P\textsuperscript{as} \\

	\midrule
	\multirow{2}{*}{\makecell[c]{C \\ in Ge}}
	&  \ce{C_{Ge}}	 &&  \multicolumn{2}{c}{Neutral\textsuperscript{av,aw}}   &&  \\
	&  \ce{V2C}\,\textsuperscript{ax}	 &&  \multicolumn{2}{c}{Unknown}   &&   \\
	
	\midrule
	\multirow{4}{*}{\makecell[c]{H \\ in Ge}} &  H\textsubscript{i}	 &&    \multicolumn{2}{c}{Shallow acceptor/neutral\textsuperscript{ay} }&& p-irradiated Ge:Sb  \\
	&  \ce{V2H}	 &&    	&  0.07\textsuperscript{av} && \textgamma-irradiated Ge  \\
	&  H\ce{Si_{Ge}}, H\ce{C_{Ge}}	 &&    	&  Shallow\textsuperscript{az,av} && \\
	&  HO\textsubscript{i}	 &&  Shallow\textsuperscript{az,av}  	&   && \\
	
	\midrule
	\multirow{3}{*}{\makecell[c]{N \\ in Ge}} 	&  \ce{N_{Ge}}	 &&  Shallow\textsuperscript{ba}  	&   && \gls{dft} calculations  \\
	&  \ce{N_{2i}}	 &&  \multicolumn{2}{c}{Neutral\textsuperscript{ba}}   && \gls{dft} calculations  \\
	&  \ce{V_nN_m}\,\textsuperscript{bb}	 &&  \multicolumn{2}{c}{Unknown}   &&  \gls{dft} calculations \\
	
	\midrule
	\midrule
	
	\multirow{4}{*}{\makecell[c]{Known \\defects \\ in GeSn}} 	& Clean 60\degree{} ED	 &&  	& BL$\leq0.15$\textsuperscript{bc}  && Rlx. Ge$_{0.922}$Sn$_{0.078}$ on Ge \\
	&   \gls{td}-related	 &&  	& BL\,0.29\textsuperscript{bd}  && Str. Ge$_{0.93}$Sn$_{0.07}$ on vGe\\
	&   \ce{VSn^{--/-}}	 &&  	& 0.19\textsuperscript{be},~0.14\textsuperscript{as}  &&  e-\textsuperscript{be}, p-\textsuperscript{as}irradiated GeSn \\ 
	&   \ce{V2Sn}\textsuperscript{bf}	 &&  	\multicolumn{2}{c}{Unknown}  &&  e-irradiated Ge:Sn$^{\mathparagraph}$ \\
	
	\midrule
	
	\multirow{8}{*}{\makecell[c]{Uniden-\\tified\\defects \\ in GeSn}} 
	& \multirow{3}{*}{Unassigned} &&  	& 0.14\textsuperscript{bg},~0.075\textsuperscript{bg}  && Rlx. Ge$_{0.9994}$Sn$_{0.0006}$ on nSi$^\ast$ \\
	& &&  \mbox{0.12--~0.14}\textsuperscript{as}	&   && p-irr. rlx. GeSn($x_{Sn}<0.1$) on Si \\
	&   &&  	& 0.08\textsuperscript{bh}  && Rlx. Ge$_{0.95}$Sn$_{0.05}$ on vGe$^{\mathsection}$\\
	
	\cmidrule{2-7}
	
	& \multirow{2}{*}{D-related?}  &&  	& $\leq0.05$\textsuperscript{bi}  && Rlx. Ge$_{0.94}$Sn$_{0.06}$ on pGe$^+$\\
	&   &&  	& \mbox{0.085--0.090\textsuperscript{bj}}  && Rlx. GeSn($x_{Sn}\leq0.04$) on vGe$^+$\\
	
	\cmidrule{2-7}
	& \multirow{2}{*}{Sn-related PDs?} &&  0.23\textsuperscript{bi},~0.27\textsuperscript{bi}	&   && Str. GeSn($x_{Sn}\leq0.032$) on nGe \\
	&   &&  	& 0.14\textsuperscript{bj}, 0.16\textsuperscript{bj}  && Rlx. Ge$_{0.906}$Sn$_{0.094}$ on vGe$^\dagger$\\
	\cmidrule{2-7}
	& GeSn/Ge int.  &&  	& \mbox{0.20--0.25\textsuperscript{bk}}  && Rlx. Ge$_{0.962}$Sn$_{0.058}$:B on nGe$^\ddagger$\\
	
	\bottomrule
	\addlinespace
	\multicolumn{7}{l}{ \makecell[{{p{\linewidth}}}]{
			$^\ast$Measured by Hall effect, $^+$Measured by \gls{cv}, $^{\mathsection}$Extracted from conductivity, $^\dagger$Measured by \gls{pc}, $^\ddagger$Simulated from I-V curves, $^\mathparagraph$Observed with \gls{ftir} spectroscopy. \\ 
			\addlinespace
			References: \textsuperscript{aa}~\cite{Markevich2009},
			\textsuperscript{ab}~\cite{Vanhellemont2009},
			\textsuperscript{ac}~\cite{Capan2009},
			\textsuperscript{ad}~\cite{Fage-Pedersen2000},
			\textsuperscript{ae}~\cite{Auret2007},
			\textsuperscript{af}~\cite{Coelho2013},
			\textsuperscript{ag}~\cite{ChristianPetersen2010},
			\textsuperscript{ah}~\cite{Simoen2019},
			\textsuperscript{ai}~\cite{Simoen2013},
			\textsuperscript{aj}~\cite{Simoen1985},
			\textsuperscript{ak}~\cite{DiLello2012},
			\textsuperscript{al}~\cite{Grillot1996},
			\textsuperscript{am}~\cite{Baumann1983},
			\textsuperscript{an}~\cite{Weber2015},
			\textsuperscript{ao}~\cite{Imajo2022},
			\textsuperscript{ap}~\cite{Clauws1996},
			\textsuperscript{aq}~\cite{Markevich2003},
			\textsuperscript{ar}~\cite{Markevich2002},
			\textsuperscript{as}~\cite{Hogsed2020},
			\textsuperscript{at}~\cite{Carvalho2007},
			\textsuperscript{au}~\cite{Nyamhere2011},
			\textsuperscript{av}~\cite{Haller1981},
			\textsuperscript{aw}~\cite{Gulyas2021},
			\textsuperscript{ax}~\cite{Chroneos2007},
			\textsuperscript{ay}~\cite{Dobaczewski2004},
			\textsuperscript{az}~\cite{Stavola2008},
			\textsuperscript{ba}~\cite{BergRasmussen1994},
			\textsuperscript{bb}~\cite{Kuganathan2022},
			\textsuperscript{bc}~\cite{Gupta2018},
			\textsuperscript{bd}~\cite{Gupta2013a},
			\textsuperscript{be}~\cite{Markevich2011},
			\textsuperscript{bf}~\cite{Khirunenko2018},
			\textsuperscript{bg}~\cite{Ryu2012},
			\textsuperscript{bh}~\cite{Scajev2020},
			\textsuperscript{bi}~\cite{Takeuchi2016},
			\textsuperscript{bj}~\cite{Kondratenko2020},
			\textsuperscript{bk}~\cite{Baert2015}.}}\\
\end{longtable*}

\subsubsection*{Impurity-related traps in Ge}
Additional sources of trap states in the bandagap are spurious impurities introduced in the films during growth.
Impurities originating from background pressure in the growth chamber correspond to O, C, N, H.
In the following, we summarize the known trap states that they can induce in pure Ge growth:
\begin{itemize}
	\item \textbf{Oxygen impurity}: O occupies interstitial positions, but is electrically inactive~\cite{Clauws1996}. On the other hand, upon annealing of O-rich Ge, \ce{O4} clusters can form and lead to n-type doping of the material~\cite{Clauws1996}.
	O also acts as vacancy sink, and O-V complexes have been found to induce both acceptor and donor deep levels in Ge~\cite{Markevich2003,Fage-Pedersen2000}.
	
	\item \textbf{Carbon impurity}: C occupies neutral substitutional positions~\cite{Stavola2008} and, due to its small size, the \ce{C_{Ge}} defect has a structure similar to a Ge monovacancy~\cite{Gulyas2021}.
	It only weakly binds to vacancies, but it is stabilized by double vacancies~\cite{Chroneos2007}. It also tends to bind to and neutralize dopant-vacancy complexes, inactivating  dopants~\cite{Brotzmann2008}.
	Despite its large contamination levels during epitaxy~\cite{Carroll2006}, trap  levels induced by C complexes are seldom studied~\cite{Sgourou2022}, and, to the best of our knowledge, there exists no information on the resulting trap states. However, knowing that C is universally present in the base pressure, often due to the use of graphite heaters, it may be assumed that its contributions to the trap concentration in Ge are not dominant, since they have not been observed in experimental \gls{dlts} studies.
	
	\item \textbf{Hydrogen impurity}: Supported by experimental findings~\cite{Dobaczewski2004}, first-principle calculations predict interstitial \ce{H^-} impurities to behave as acceptors with shallow traps resonant or very close to the \gls{vb}~\cite{VandeWalle2008,Weber2007,Weber2015}.
	In Ge, atomic H has been observed to induce shallow acceptor levels by forming complexes with isoelectric Si and C, and donor levels with O~\cite{Stavola2008}.
	\ce{V2H} complexes instead induce shallow acceptor levels in pure Ge~\cite{Haller1981}.
	Contrary to Si, H does not passivate completely dangling bonds in Ge, with at best only 60\% of the surface dangling bonds passivated upon annealing in \ce{H2}~\cite{Stesmans2014b}.
	On the other hand, it has been experimentally verified that atomic H species introduced from a \ce{H2} plasma can passivate dislocations reducing their electrical activity in Ge~\cite{Pearton1983}. The same effect was not observed in He plasma or simple annealing in \ce{H2} atmosphere, demonstrating that the passivation effect is induced by atomic H species generated by the plasma process diffusing into in the material. 
	Dislocations act as sinks for vacancies and excess H~\cite{Haller1981}, and therefore in \gls{vge} we can expect H impurities to be attracted by dislocations and passivate defects at dislocation cores.
	\ce{VH_n} complexes with $n\le3$ have also been theoretically predicted, and were found to possess acceptor levels within 0.35\,eV of the \gls{vb}, and donor levels close to the \gls{vb}~\cite{Coomer1999}.
	The effect of atomic hydrogen in sputtered Ge should thus be accurately evaluated to see if it is negative or positive in terms of electrical properties.
	
	\item \textbf{Argon impurity}: The effect of Ar plasma or implantation was studied in Ge, and no Ar-specific trap levels were found~\cite{Auret2007a,Kusumandari2013,Wen2023}. All  trap levels were induced by non-Ar-specific ion bombardment.
	This indicates that, as expected, Ar does not electrically interact with the material.
	
	\item \textbf{Nitrogen impurity}: Being from group-V in the periodic table of elements, N is expected to be a n-type dopant for Ge. However, N is a poor dopant~\cite{Chambouleyron1998} because it  tends to form electrically inactive N interstitial pairs~\cite{BergRasmussen1994}.
	When in substitutional positions, N gives a shallow level close to the \gls{cb}~\cite{BergRasmussen1994}, although it has been suggested that it also gives rise to deep traps due to lattice distortions~\cite{Chambouleyron1998}.	
\end{itemize}


\subsubsection*{Trap states in GeSn}
In low-Sn-content GeSn alloys, we can typically find the same trap levels of pure Ge~\cite{Schulte-Braucks2017a}.
The lower processing temperatures of GeSn however lead to the presence of additional defects that are absent in Ge.
For example, intrinsic Ge point defects that are expected to annihilate during high-temperature annealing of the Ge buffer may instead be observed in GeSn due to its low growth temperature. This is the case of Ge divacancies, which have been reported to show mid-gap trap levels~\cite{Auret2007,ChristianPetersen2010,Capan2009}. These defects are expected to be stable up to 400\degree C~\cite{Kuitunen2008}, well above typical GeSn growth temperatures in \gls{pvd} methods. Hence, these divacancies forming in the film during growth are not expected to be annihilated, and  will likely affect the film electrical properties.
The growth temperatures should therefore be maintained as high as possible to limit the formation of vacancies.

In addition to the Ge-related traps, in GeSn alloys there will be trap states induced by the presence of Sn in the material, which we summarize in the following.
There exists several studies of defect levels arising in GeSn, but rarely the observed traps are unambiguously assigned to specific defects.
Tab.~\ref{tab:TrapLevels} reports the measured trap levels induced by the presence of Sn in the GeSn alloy. These are generally independent on the alloy composition~\cite{Gupta2018, Takeuchi2016,Hogsed2020}.
Concerning point defects in GeSn, Markevich\etal~\cite{Markevich2011} measured by \gls{dlts} hole trap levels 0.19\,eV in electron-irradiated GeSn and assigned them to Sn-V complexes. Similar traps were measured in proton-irradiated GeSn~\cite{Hogsed2020}.
As seen in Sec.~\ref{sec:Defects_GeSn}, monovacancy complexes were not observed in as-grown epitaxial GeSn; \gls{pas} characterisation rather evidenced  significant concentrations of divacancy complexes and larger vacancy clusters~\cite{Assali2019a,Slotte2014}. Hence, we can deduce Sn-V complexes are annihilated at the temperatures employed for epitaxial growth.
Sn-divacancies complexes have been observed in irradiated Sn-doped Ge~\cite{Khirunenko2014,Khirunenko2018}, but, to the best of our knowledge, their electrical levels have never been reported. In analogy with the \ce{SnV2} defect in Si~\cite{Kaukonen2001}, we can expect \ce{SnV2} complexes to introduce a deep level in Ge.

Dislocations present in GeSn will also affect its electrical properties. Besides the electronic levels known to be induced in pure Ge, there have been reports of electronic trap states induced by dislocations in GeSn.
Gupta\etal~\cite{Gupta2018} studied by \gls{dlts} the trap levels in a Ge$_{0.922}$Sn$_{0.078}$ film grown by \gls{cvd} on a n+Ge substrate. They observed band-like shallow acceptor defects with energies $\leq0.15$\,eV, and attributed them to clean \glspl{ed} observed in proximity of the GeSn/Ge interface. These defects, consisting in \glspl{sf} bound by Shockley partials, showed trap states similar to Shockley partials in pure Ge~\cite{Baumann1983}. Despite the low activation energy of these \glspl{ed}, the authors ruled out their role as acceptor dopants due to their small capture cross-sections, implying these defects act as donor-like repulsive centers.
They further demonstrated that the \glspl{ed} determined the \gls{srh} minority electron generation rate in GeSn, by analyzing the Arrhenius behavior of a pGeSn/nGe diode, where the GeSn was \emph{p}-doped unintentionally.
The presence of defects at the GeSn/Ge interface determining the dark currents of pGeSn/nGe diodes had already been supposed through simulations and fittings of the dark currents in an earlier work by the same authors, though the simulated trap levels yielded defects of 0.20--0.25\,eV above the \gls{vb}~\cite{Baert2015}.
Kondratenko\etal~\cite{Kondratenko2020} fitted \gls{pc} curves to find activation energies of 85--90\,meV above the \gls{vb}, which they attributed to the large dislocation density in their \gls{cvd}-grown relaxed GeSn ($x_{Sn}\leq0.04$) films on \gls{vge}. On the other hand, in GeSn films with better structural properties -- as measured by \gls{xrd} -- and larger Sn contents ($x_{Sn}>0.04$) they found that dominant traps were placed at 0.14--0.16\,meV above the \gls{vb}. They suggested these traps were not related to dislocations, and tentatively associated them to SnV complexes~\cite{Markevich2011,Hogsed2020}. The latter trap level was also observed by Ryu\etal~\cite{Ryu2012} with Hall measurements of \gls{cvd}-grown relaxed Ge$_{0.9994}$Sn$_{0.0006}$ on nSi substrates. Furthermore, the authors reported the appearance of a \emph{p}-type degenerate conductive layer at the GeSn/Si interface, attributing to the arrays of 90\degree~\glspl{md} generated due to epitaxial strain relaxation.
This result is in agreement with the decrease in lasing threshold observed when removing \glspl{md} in the active region of GeSn microdisks lasers~\cite{Elbaz2020}.

In conclusion, the few studies reporting trap states of GeSn are mostly speculative, with often ambiguous assignment of trap states to specific defects. The investigation of trap states in GeSn is still at its infancy, and more systematic studies are required to evaluate the influence of the epitaxial technique of choice, and of the employed growth parameters, which will ultimately determine the film electrical properties.



Lastly,  it is important to mention possible defects arising from the atomic impingement of species from the plasma in plasma-based growth techniques, such as \gls{ms} and \gls{pecvd}.
\gls{dlts} investigations of  bulk Ge:Sb exposed to Ar plasma during sputtering have evidenced the absence of hole traps~\cite{Auret2010}. On the other hand, several electron traps were observed down to the first 400\,nm of the nGe crystal, mostly associated to Sb and Ge interstitial induced by Ar impingement. They reported only one intrinsic trap level --- at 0.31\,eV below the \gls{cb} --- tentatively attributed to Ge divacancies. All observed sputtering-induced defects were annihilated above 250\degree C~\cite{Auret2007,Auret2010}, suggesting the growth temperature should be above this value to prevent plasma-induced defects.

\subsection{Unintentional Doping Concentration}
\label{sec:ElecProp_UnintDoping}
Epitaxial GeSn alloys, including pure Ge, always show carrier concentration levels higher than the intrinsic values at room temperature, despite being nominally intrinsic.
This charge carrier concentration is therefore termed \emph{unintentional}, and originates from the presence of defects and impurities in the material.
High levels of unintentional doping can be detrimental for the operation of diode devices. For example, they can increase the junction capacitance and decrease the frequency bandwidth of \glspl{pd}~\cite{Xu2019}, or induce breakdown in the GeSn absorber of eventual GeSn-on-Si \glspl{spad}~\cite{Chen2021}.
Representative studies reporting unintentional doping levels in Ge and GeSn thin films are listed in Tab.~\ref{tab:GeSn_UnintDoping}. Carrier concentrations refer to majority holes at 300\,K ($p_{300K}$) unless differently specified with the symbol ``$n=$'' inserted before the concentration values.
In most cases,  GeSn thin films show unintentional doping of type \emph{p}, while pure Ge is reported possessing both \emph{n-} and \emph{p-}type unintentional doping.

\subsubsection*{Potential sources of unintentional doping in Ge}

In pure Ge films, unintentional doping concentrations in the $10^{16}$\,cm$^3$ range can be achieved~\cite{Atalla2021,Hogsed2020,Wei2020,Oehme2014a,Yeh2014,Roucka2011a}. To the best of our knowledge, a concentration in the $10^{15}$\,cm$^3$ range was measured only by Roucka\etal~\cite{Roucka2011a}, who reported $p_{300K}=7\cdot10^{15}$\,cm$^3$ in Ge grown by \gls{gsmbe}.
Unintentional doping in pure Ge is often attributed to the presence of multi-level vacancy complexes~\cite{Zaima2015} or divacancies~\cite{Assali2019a}.
Considering the trap levels reported in the literature, summarized in Tab.~\ref{tab:TrapLevels}, divacancies and vacancy clusters may effectively yield both \emph{n-} and \emph{p-}type doping.
However, \gls{dlts} analysis of neutron- and proton-irradiated Ge~\cite{Kuitunen2008,Elsayed2015} showed that intrinsic vacancy defects in Ge annihilate completely at temperatures above 500\degree C.
Hence, intrinsic Ge vacancies -- or vacancy complexes -- are not expected to be present in annealed Ge films and are unlikely to be the source of unintentional doping in the material.
We thus discuss other possible sources of shallow traps that may lead to unintentional doping.

Clean dislocations have also been proposed as source of residual doping in Ge~\cite{Simoen2019}.
This may be in accordance with observations made in Ref.~\cite{Yeh2014}, by Yeh\etal{}, who measured the dependence of unintentional doping concentration on Ge film thickness, reported in Fig.~\ref{fig:GeHallmeasurements}(a).
They observed a decrease in $p$ with Ge film thickness, explained as a reduction in defect concentration.
We believe there are two possible explanations for this behavior:
(1) On the one hand, if we consider solely the film defects, the \gls{tdd} is well known to decrease with film thickness as a consequence of enhanced \gls{td} interaction due to the geometric effect~\cite{Loo2009}. On the other hand, there is no reason why the point defect concentration should change with film thickness if the growth conditions are kept constant. Hence, the results from Fig~\ref{fig:GeHallmeasurements}(a) would point at \glspl{td} contributing to unintentional \emph{p}-type doping at room temperature.
Tab.~\ref{tab:TrapLevels} shows various trap levels measured in Ge and associated to the presence of \glspl{td}.
(2) The decrease in apparent concentration with thickness can also be explained due to a fixed concentration of traps at the interface, e.g. \glspl{md} or impurities adsorbed on the substrate prior to film growth, whose contribution to the total amount of carriers in the film decreases with thickness, leading to an apparent decrease in film carrier concentration.

\begin{turnpage}
	\begin{table*}
		\renewcommand{\arraystretch}{1.5}
		\caption{\label{tab:GeSn_UnintDoping}Unintentional doping concentrations in Ge$_{1-x}$Sn$_x$ thin films reported in the literature. Concentrations refer to holes ($p$), unless differently specified ($n$). Acronyms: amrp.:~amorphous, \acrlong{spc} (\gls{spc}), \acrlong{cv} (\gls{cv}) and \acrlong{ecv} (\gls{ecv}) measurements,\acrlong{vdp} (\gls{vdp}), \acrlong{pda} (\gls{pda}) \acrlong{soi} (\gls{soi}), \acrlong{goi} (\gls{goi}), \acrlong{r} (\gls{r}).}
		\begin{tabular}{L{0.1\linewidth}cM{0.07\linewidth}cccM{0.17\linewidth}M{0.06\linewidth}m{0.35\linewidth}}
		
		\toprule
		
		\centering\arraybackslash Ref. 	& \makecell{Comp. \\ Sn\atp} 		& Substrate 	& \makecell{Growth\\ method}	& \makecell{Gr. \gls{t}\\ (\degree C)}& \makecell{Thick.\\ (nm)}& \makecell{$p_{300\,K}$ \\ (cm$^{-3}$)}  &  \makecell{Meas. \\ method}	 & \centering\arraybackslash Comments
		\\
		
		\midrule
		
		\rowcolor{lightgrey} 
		
		\makecell[{{L{\linewidth}}}]{Mizoguchi, 2021 \cite{Mizoguchi2021}}&
		1.2--4.0	&  SiO$_2$		& \makecell[c]{\gls{mbe} amrp. \\+ \acrshort{pda}\\ + \acrshort{spc}} & 	$\ast$	& 150 & $>5$E17  & \makecell[c]{Hall\\ \gls{vdp}}& \makecell[{{p{\linewidth}}}]{ $\ast$:\,500--700\degree C \gls{pda} for 2\,h. Then \acrshort{spc} in N$_2$ at 400--650\degree C for 40\,h gave poly-GeSn.}
		\\
		\addlinespace
		
		Atalla, 2021 \cite{Atalla2021}&
		0--17	&  Ge/Si(001)		& \gls{lpcvd} 			& 	-	& 120-160 & \makecell[c]{Pure Ge: 7E16 \\ $x_{Sn}\geq10.5\%$: 1--5E17 } &\gls{cv}  & \makecell[{{p{\linewidth}}}]{Increase in dark current in GeSn \gls{pd} with larger Sn\atp.}	\\
		
		\addlinespace
		\rowcolor{lightgrey}
		Hogsed, 2020 \cite{Hogsed2020}&
		0--9.4	&  Si(001)		& \gls{cvd} 			& 	-	& 300--800 & \makecell[c]{Pure Ge: $n=5$E16 \\ $x_{Sn}=2\%$: 2E16 \\$x_{Sn}=9.4\%$: 2E17} & \makecell[c]{\gls{cv}\\ \gls{dlts}}  & \makecell[{{p{\linewidth}}}]{	Measured in \emph{pin} devices. $p$ increases with \atp\,Sn.}\\

		\addlinespace
		Tao, 2020 \cite{Tao2020}&
		0--14	&  Ge(001)		& \gls{mbe} 			& 	200	& 30 &\makecell[c]{Pure Ge: $n=5.8$E17 \\ $x_{Sn}=4\%$: $n=3.9$E17 \\$x_{Sn}=8\%$: $n=5.2$E17 \\$x_{Sn}=14\%$: $n=5.4$E17}  & Hall & \makecell[{{p{\linewidth}}}]{Films reported partially relaxed, but  \gls{mbe} 30-nm-thick GeSn with $x_{Sn}\leq$8\atp{} is expected to be pseudomorphic on Ge~\cite{Wang2015}.	\textbf{Carriers are \emph{n}-type also in GeSn}.} \\
		
		\addlinespace
		\rowcolor{lightgrey}
		Wei, 2020 \cite{Wei2020}&
		0	&  Si(001)		& \gls{mbe} 			& 	$200$	& 500 & \makecell[c]{As-grown Ge: 9E16\\
			\gls{pda} $\geq500$\degree C: $\sim2$E16}  & Hall & \makecell[{{p{\linewidth}}}]{$p_{<300K}$ is higher with \gls{pda} of 800\degree C compared to 600\degree C.} \\
		
		\addlinespace
		\makecell[{{L{\linewidth}}}]{Julsgaard, 2020 \cite{Julsgaard2020}}&
		12.5	&  Ge/Si(001)		& \gls{cvd} 			& 	-	& 350 & 9E17  & \gls{ecv} & 	\\
		
		\addlinespace
		\rowcolor{lightgrey} 
		Scajev, 2020 \cite{Scajev2020}&
		5	&  \makecell[c]{Ge/SiGe/\\/Si(001)}		& \gls{mbe} 			& 	150	& 200 & \makecell[c]{1.1E17\\$p_{200\,K}=6$E16}  & \gls{r} & \makecell[{{p{\linewidth}}}]{\gls{tdd}$\sim3\cdot10^6$\,cm$^{-2}$. Hole activation energy of $80\pm10$\,meV.}	\\
		
		\addlinespace
		Gupta, 2018 \cite{Gupta2018}&
		7.8	&  Ge(001)		& \gls{cvd} 			& 	320	& 315 & 2E16  & \gls{cv} & GeSn is partially relaxed.	\\
		
		\addlinespace
		\rowcolor{lightgrey} 
		\makecell[{{L{\linewidth}}}]{Schulte-Braucks, 2017 \cite{Schulte-Braucks2017b}}&
		0--12.5	&  Ge/Si(001)		& \gls{cvd} 			& 	350	& \makecell[c]{2500\\768\\835\\414} & \makecell[c]{Pure Ge: 4E17 \\ $x_{Sn}=8.5\%$: 5E17 \\$x_{Sn}=10\%$: 1.5E18 \\$x_{Sn}=12.5\%$: 4E18}  & Hall & \makecell[{{p{\linewidth}}}]{Cooling down, $p_{200K}$ lower by 1  order of magnitude in GeSn, and by 2.5 in Ge.}	\\
		
		\addlinespace
		Asano, 2015 \cite{Asano2015}&
		5	&  Ge(001)		& \gls{mbe} 			& 	150	& 55--70 & \makecell[c]{8E17 no H$_2$\\1E17 with H$_2$} & \gls{cv}& \makecell[{{p{\linewidth}}}]{Growth in partial pressure of H$_2$. Pseudomorphic films.}	\\
		
	\end{tabular}
\end{table*}
\end{turnpage}

\begin{turnpage}
\begin{table*}\ContinuedFloat
\begin{tabular}{L{0.1\linewidth}cM{0.07\linewidth}cccM{0.17\linewidth}M{0.06\linewidth}m{0.35\linewidth}}

		\rowcolor{lightgrey} 
		\makecell[{{L{\linewidth}}}]{Kamiyama, 2014 \cite{Kamiyama2014}}&
		0--1.7	&  \gls{goi}(001)		& \gls{mbe} 			& 	170	& 200 & \makecell[c]{Pure Ge: 1.7E17 \\ $x_{Sn}=0.1\%$: 2.4E16\\ $x_{Sn}=1.7\%$: 5.5E17} & Hall & \makecell[{{p{\linewidth}}}]{\textbf{Carrier type unspecified}. \gls{pas} used to measured vacancy concentration in films, which matched $p$ ($n$).}	\\
		
		\addlinespace
		Oehme, 2014 \cite{Oehme2014a}&
		0--4.2	&  Ge/Si(001)		& \gls{mbe} 			& 	160	& 300 & \makecell[c]{Pure Ge: $\sim1$E16\\
			$x_{Sn}=2\%$: $\sim1$E16\\ $x_{Sn}=4.2\%$: $\sim1$E17}  & \gls{cv} & \makecell[{{p{\linewidth}}}]{\textbf{Carrier type unspecified}. Indirect measurements of conc. in \emph{pGe-iGeSn-nGe}	\gls{pd}.} \\
		

		\addlinespace
		\rowcolor{lightgrey} 
		Yeh, 2014 \cite{Yeh2014}&
		0	&  Si(001)		& \gls{ms} 			& 	$<360$	& 1000 & \makecell[c]{As-grown Ge: 5E16\\
			\gls{pda} 700\degree C: 1.3E16}  & Hall & \makecell[{{p{\linewidth}}}]{$p_{200K}$ is higher in film annealed at 700\degree C.} \\
		

 \addlinespace
		\makecell[{{L{\linewidth}}}]{Nakatsuka, 2013 \cite{Nakatsuka2013}}&
		0--2	&  \gls{soi}(001)		& \gls{mbe} 			& 	200	& - & \makecell[c]{Pure Ge:$\sim1$E18\\ $x_{Sn}=0.01,2\%$: $\sim1$E18\\$x_{Sn}=0.1\%$:  $\sim1$E17} & Hall & \makecell[{{p{\linewidth}}}]{Anneal in H$_2$ decreases $p$ by one order of magnitude.}\\

			\addlinespace	
		\rowcolor{lightgrey} 
		Ryu, 2012 \cite{Ryu2012}&
		0.06	&  Si(001)		& \gls{uhv} \gls{cvd} 			& 	\makecell[c]{390\\ + \gls{rta} \\ 680\degree C}	& 800 & 1.4E16  & \makecell[c]{Hall\\ \gls{vdp}}&
		\makecell[{{p{\linewidth}}}]{Degenerate \emph{p}-type conductive layer at GeSn/Si interface due to \glspl{md}.}	\\
		
		\addlinespace

		Roucka, 2011 \cite{Roucka2011a}&
		0	&  Si(001)		& \gls{gsmbe} 			& 	390	& \makecell[c]{~~~~~350~~~~~\\900} & \makecell[c]{Pure Ge\\$n=7$E15--5E16} & Hall &  \\
		
		\addlinespace
		\rowcolor{lightgrey} 
		\makecell[{{L{\linewidth}}}]{Nakatsuka, 2010 \cite{Nakatsuka2010}}&
		0--5.8	&  \gls{soi}(001)		& \gls{mbe} 			& 	200	& 160--170 & \makecell[c]{$>7$E17 as-grown\\3E17 after \gls{pda}} & \makecell[c]{Hall\\ \gls{vdp}}& \makecell[{{p{\linewidth}}}]{Anneal in N$_2$ at 500\degree C for 60\,min decreases \emph{p}. Slight increase in $p$ with Sn\atp.}	\\

\addlinespace
		Soref, 2006 \cite{Soref2006}&
		2	&  -		& \gls{cvd} 			& 	-	& - & 5.6E16 & Hall & \gls{pda} improved mobility and decreased $p$.	\\
		
		\bottomrule
			\end{tabular}
	\end{table*}
\end{turnpage}

\subsubsection*{Influence of Ge annealing on unintentional doping}
\begin{figure*}[htb]
	\centering
	\includegraphics[width=\textwidth]{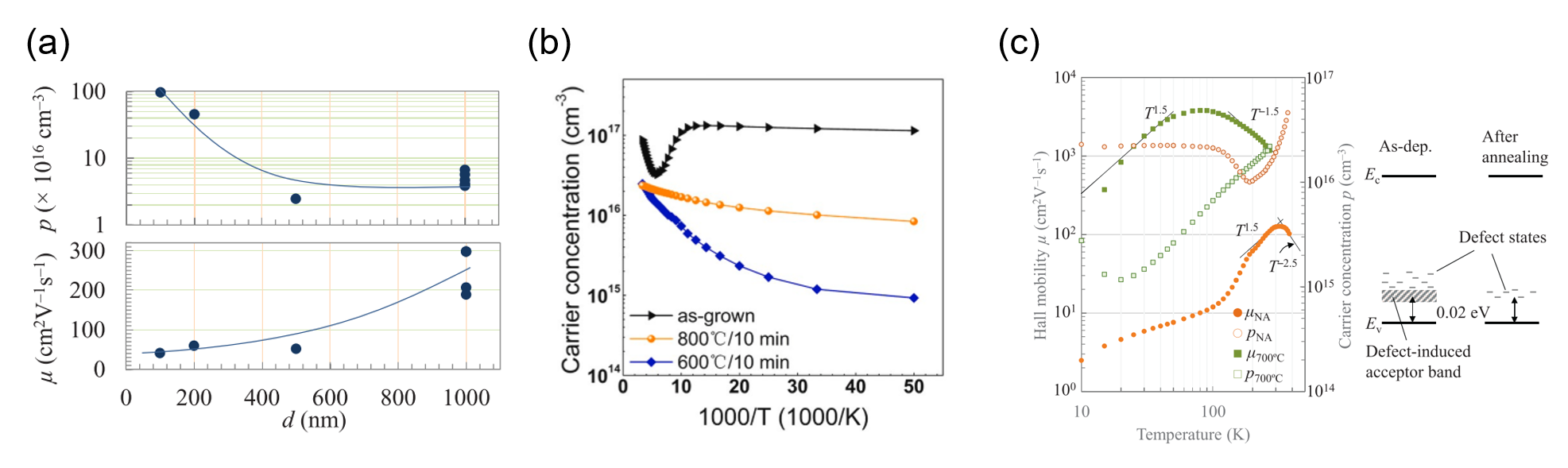}
	\caption{Hall measurements of Ge films from (a,c)\,Ref.~\cite{Yeh2014} and (b)\,\cite{Wei2020}. More details of these studies are reported in Tab.~\ref{tab:GeSn_UnintDoping}. In both cases Ge is unintentionally doped \emph{p}-type. Figures (a,c) reproduced from Ref.~\cite{Yeh2014} under terms of the CC-BY license. Figure (b) with permission from Wei\etal~\cite{Wei2020}, \copyright\,2020 \emph{Elsevier}. }
	\label{fig:GeHallmeasurements}
\end{figure*}

Annealing of Ge films has been reported to be beneficial for the material electrical properties. In particular, a reduction of unintentional doping and an improvement in mobility have been observed as a consequence of a decrease in defect concentration during annealing~\cite{Yeh2014,Wei2020}.
Nevertheless, Hall measurements from 300\,K to cryogenic temperatures revealed that the unintentional doping concentration does not decrease over the whole temperature range after annealing. Two examples are shown in Fig.~\ref{fig:GeHallmeasurements}(b-c).
After annealing in vacuum a \gls{mbe} Ge film, Wei\etal~\cite{Wei2020} measured a reduced carrier concentration across the entire temperature measurement range, shown in (b). However, a thermal treatment of 600\degree C allowed to obtain a lower unintentional doping levels compared to 800\degree C, except for room-temperature measurements, where $p$ was equal.
This suggests that shallow trap levels formed at higher processing temperatures, despite an overall reduction in defect concentration.
On the same line, Yeh\etal~\cite{Yeh2014}, in (c), compared Hall measurements of \gls{ms} Ge with and without thermal treatment (30~minutes at 700\degree C in \ce{N2}+3.8\%\ce{H2}).
They observed a higher unintentional doping level after Ge annealing at 700\degree C between 300\,K and $\sim150$\,K. On the other hand, at lower temperatures, the carrier concentration was strongly reduced with annealing.
Hence, in this study, it was observed that deeper trap levels, which contribute to unintentional doping at 300\,K, were created as a consequence of the annealing process. Conversely, shallower trap levels were eliminated during the thermal processing, leading to a reduction in unintentional doping at lower temperatures.
The formation of deep trap levels after annealing cannot be associated to intrinsic point defects, nor \glspl{td}, since their concentration decreases with thermal processing. On the other hand, \glspl{md} elongate during annealing, and may be a source of deep traps. This could explain also the  results in Fig.~\ref{fig:GeAnnealing}(b), where a higher density of \glspl{md} is expected following annealing at 800\degree C.
Alternatively, the increase in unintentional doping observed with annealing could also be associated to the evolution during annealing of defect complexes involving impurities.
It is important to note that the role of impurities is mostly overlooked when trying to individuate the origin of the material unintentional doping. However, intrinsic defects in Ge, being \ce{V}, \ce{V2}, or larger clusters, may form complexes with impurities that are stable at higher temperatures~\cite{Markevich2011,Markevich2011a} or may form during the annealing process through diffusion and recombination.
In particular, Tab.~\ref{tab:TrapLevels} shows that O- and H-related defects can be source of doping (\emph{n}-type: \ce{I_{Ge}-O_{2i}}, \ce{O_{Ge}}, \ce{O4}, \ce{HO_i}. \emph{p}-type: \ce{H_i}, \ce{V2H}, \ce{HSi_{Ge}}, \ce{HC_{Ge}}). Trap levels associated with C and N complexes are unknown, but cannot be discarded a priori as source of doping. In addition, precise data on the thermal stability of these complexes is often lacking, complicating unambiguous individuation of the source of shallow traps.
The different impurity levels, growth techniques and thermal processes of the grown Ge films can therefore explain the apparent inconsistencies found in the literature.

\subsubsection*{Unintentional doping behavior with temperature}
On a side note, it is interesting to observe the analogous temperature-dependence of $p$ in as-grown \gls{ms} and \gls{mbe} Ge films in Figs.~\ref{fig:GeAnnealing}(b-c);
as the films are cooled down from 300\,K, the carrier concentration first decreases and, round $150$\,K $p$ starts increasing again. It then saturates around 100\,K, remaining unchanged as the film is further cooled down to $T<20$\,K. Yeh\etal{} explain the saturation behavior at low temperatures with the presence of an impurity band formed due to defects at energies approximately 20\,meV above the \gls{vb}, schematized in Fig.~\ref{fig:GeAnnealing}(c). This impurity band disappears upon annealing as a consequence of the reduction in defect concentration.
From Tab.~\ref{tab:TrapLevels}, plausible defects with trap energy of 20\,meV are dislocations~\cite{Simoen1985}. This is in line with the strong decrease in \gls{tdd} occurring upon annealing.
Monovacancies ($V^{-/0}$) also show the same trap level, but they are not expected to be stable above 65\,K~\cite{Slotte2021}, and are thus to be excluded as possible source of this behavior\,\footnote{\,Vacancies cannot form upon cooling down in equilibrium conditions, as thermodynamics predict they are less stable compared to room temperature}. 

\subsubsection*{Potential sources of unintentional doping in GeSn}
In contrast to Ge, whose unintentional doping type is equally reported to be \emph{n-} or \emph{p-type}, GeSn alloys are predominantly reported to be \emph{p}-type, as shown in Tab.~\ref{tab:GeSn_UnintDoping}.
Typical room-temperature carrier concentrations are in the order of $10^{16}-10^{17}$\,cm$^{-3}$. Concentrations in the order of $10^{15}$\,cm$^{-3}$ have never been reported to the best of our knowledge.
The fact that GeSn is consistently reported to be \emph{p}-type, with generally increasing $p$ for higher Sn contents~\cite{Atalla2021,Hogsed2020,Tao2020,Schulte-Braucks2017b,Oehme2014a,Nakatsuka2010}, could suggest that there exists a Sn-related dominant acceptor trap that overshadows any other defect.
This is often associated to Sn-vacancy complexes~\cite{Atalla2021,Gupta2018,Oehme2014a}.
In agreement with this hypothesis, Kamiyama\etal~\cite{Kamiyama2014} have observed a proportionality between vacancy concentration measured by \gls{pas} and hole doping in Ge and GeSn.
However, the exact nature of this defect remains unknown: Sn-V pairs are expected to induce hole trap levels between 140\,meV and 190\,meV, as shown in Tab.~\ref{tab:TrapLevels}, but \gls{pas} characterization of epitaxial GeSn revealed a predominance of divacancies and vacancy clusters~\cite{Assali2019a, Slotte2014}. \ce{Sn2}-V~\cite{Khirunenko2018} and higher order \ce{Sn_nV_m} complexes~\cite{Chroneos2009} are thus more likely candidates, though their trap  energies remain unknown to date.
Alternatively, acceptor defects in GeSn may also arise from dislocations~\cite{Takeuchi2016,Kondratenko2020}, or their interaction with Sn, since most studies are performed on relaxed GeSn films.
A systematic comparison of pseudomorphic and relaxed GeSn would shed light on the role of dislocations, though Asano\etal~\cite{Asano2015} reported high concentrations of \emph{p}-type dopants in pseudomorphic GeSn, suggesting dislocations are not the source of \emph{p}-type doping.

Ultimately, the growth conditions and impurity levels govern the electronic trap states in the material and thus its unintentional doping concentration.
The consistency in \emph{p}-type doping in GeSn, as opposed to the inconsistency in majority carrier type observed in Ge, may also arise from the lower deposition temperature employed for the alloy.
Higher growth temperatures have been reported to be beneficial to reduce the concentration of point defects in the film~\cite{Ueno2000,Knights2001}, and thus in principle the unintentional doping. Nevertheless, a systematic study of the effect of temperature on $p$ has not been performed to the best of our knowledge.
It has been reported that the presence of \ce{H2} gas in the \gls{mbe} growth atmosphere has a positive effect in reducing $p$ by almost one order of magnitude, possibly due to point defect passivation from dissociated H species~\cite{Asano2015}.
\Acrfull{pda} of Ge at 500--700\degree C was also seen to be beneficial in reducing $p$~\cite{Nakatsuka2010,Soref2006}, especially when performed in \ce{H2} ambient~\cite{Nakatsuka2013}.
However, this approach is constrained when dealing with metastable GeSn alloys due to their limited thermal stability, as discussed in Sec.~\ref{sec:GeSn_Challenges_Segregation}.
To circumvent this limitation, subsecond annealing methods~\cite{Rebohle2016}, namely laser and flash annealing,  may prove beneficial. These processes have the potential to reduce defect density responsible for unintentional doping, all the while preventing excessive Sn diffusion and segregation thanks to the limited employed thermal budget.

\subsection{Carrier Lifetime}
\label{sec:ElecProp_CarrierLifetime}
The free carrier lifetime is a key figure of merit for the quality of semiconductor materials in view of their employment in optoelectronic devices.
It determines the device performance, affecting its noise and efficiency~\cite{Gallagher2015,Ghosh2020,Scajev2020,Marzban2021,Chen2021}.
The term \emph{lifetime} can refer to both carrier generation and recombination phenomena, and the relative importance depends on the type of optoelectronic device.
For example, in optical detectors, the \emph{generation} lifetime determines the rate at which free carriers are generated in the reverse-biased depletion region in dark conditions. These \emph{dark } carriers induce currents that affect the noise performance of the device~\cite{Ghosh2020,Chen2021}.
In this case, eq.~\eqref{eq:srhGen} can be used to calculate the device dark currents, assuming the carrier generation in the depletion region is governed by \gls{srh} processes.
In the presented form, eq.~\eqref{eq:srhGen} is expressed in function of the recombination lifetimes (i.e., $\tau_r$). It can however be expressed in function of the generation lifetime ($\tau_g$) in the form of $G\sim n_i/\tau_g$~\cite{Gonzalez2011}, considering that $n_i^2>>pn$ in the depletion region of the device.
The recombination and generation lifetimes are thus not equal, but are related via material's parameters~\cite{Schroder1982}.
For example, in the the depletion region of a \emph{p-n} junction in reverse bias, if the hole and electron capture cross-sections ($\sigma_p$, $\sigma_n$) do not differ by more than 2 orders of magnitude, the generation and recombination lifetimes are related via
\begin{equation}
	\tau_g \simeq 2\tau_r\sqrt{\sigma_n/\sigma_p}\cosh[(E_t-E_g/2)/kT]
\end{equation}
where $E_t$ and $E_g$ are respectively the trap energy and the bandgap energy~\cite{Schroder1982}.

The carrier lifetime is highly sensitive to the electronic trap states in a semiconductor, and can thus be employed to assess its crystal quality.
The recombination lifetime $\tau_r$ can be measured by injecting carriers in a material and monitoring their decay time with time-resolved measurements.
In this setup, $\tau_r$  depends non-linearly on the excess carrier density ($\Delta n$), and can be decomposed in three contributions~\cite{Schroder1982}:
\begin{equation}
	1/\tau_r=1/\tau_{SRH}+1/\tau_{rad}+1/\tau_{Auger}=A+B(\Delta n)+C(\Delta n)^2
\end{equation}  
where $A,B,C$ are material constants, and $\tau_{SRH}$, $\tau_{rad}$, $\tau_{Auger}$, are the \gls{srh}, radiative and Auger lifetimes, respectively. The latter becomes important only at high injection regimes, generally avoided when assessing carrier lifetime in thin films.
In a highly-defective material, $\tau_{SRH}<<\tau_{rad}$, and thus $\tau_r\sim\tau_{SRH}$.

\subsubsection*{Carrier lifetimes in Ge}
Tab.~\ref{tab:CarrierLifetime} summarizes carrier lifetimes reported for Ge and GeSn films in the literature.
In Ge crystals, bulk recombination lifetimes ($\tau_{r,B}$) have been measured to be between 100\,\textmu s and 5000\,\textmu s~\cite{Gaubas2006,Derhacobian1994}.
These values are however far from the reported carrier lifetimes in Ge films grown on Si substrates, which are of only a few ns~\cite{Saito2014,Sheng2013,Geiger2014}. This is attributed to the large density of dislocations generated by epitaxial relaxation phenomena, and to the presence of a surface and an interface in the vicinity of injected carriers~\cite{Grzybowski2011}. In thin films, carriers can in fact diffuse and recombine at surface traps with a non-negligible contribution to the overall carrier recombination time.
The measured effective carrier lifetime can thus be decomposed in two separate contributions~\cite{Gaubas2006}:
\begin{equation}
	\label{eq:recombLifetime}
	\tau_r=\left[\frac{1}{\tau_{r,B}}+\frac{1}{\tau_{r,S}+\tau_D} \right]^{-1}
\end{equation}
where $\tau_{r,S}=d_{eff}/v_{r,S}$ is the surface recombination lifetime, and $\tau_D=d_{eff}/(\pi^2D)$ is the carrier diffusion time. $v_{r,S}$ is the surface recombination velocity, $D$ the carrier diffusion coefficient, and $d_{eff}$ the effective probed depth of the material.
$\tau_D$ takes into account the loss of carriers due to diffusion out of the probed region, and can be mostly neglected with the low $\tau_{r,B}$ measured in epitaxial Ge and GeSn thin films.
$\tau_{r,B}$ is a proper figure of merit of material quality,  while $\tau_{r,S}$ is indicative of the presence of trap states at the film surface or at the film/substrate interface.
$\tau_{r,S}$ is also sensitive to surface roughness~\cite{Hudait2022}.
In configurations where  $\tau_{r,S}$ is comparable to $\tau_{r,B}$, the resulting measurements of $\tau_r$ are strongly affected by the film thickness, as carriers can diffuse and recombine at the interfaces. Thicker films will be less sensitive to $\tau_{r,S}$, yielding longer recombination lifetimes.
This situation is typical of few-\textmu m-thick Ge films grown and annealed on Si substrates, as dislocations remain confined near the interface after a \gls{pda} process~\cite{Loo2009}.
To properly account for the effect of this defective interface on $\tau_r$, eq.~\eqref{eq:recombLifetime} can be fitted over $\tau_r$ measured at different film thicknesses -- i.e., $d_{eff}$ --~\cite{Sheng2013}.
This yields an effective $\tau_{r,S}$ that describes the recombination rate at the defective epitaxial interface or film surface~\cite{Kako2015}.
Doing so, Ge bulk lifetimes ($\tau_{r,B}$) from few ns~\cite{Saito2014} up to 11\,ns~\cite{Sheng2013} were reported, with recombination velocities ($v_{r,S}$) of tens of m/s at the Si/Ge interface.
An exceptional $\tau_{r,B}=91$\,ns was measured by Kako\etal~\cite{Kako2015}, possibly due to a low \gls{tdd} in the film owed to its large thickness ($>3$\,\textmu m).

\begin{turnpage}
	\begin{table*}
		\renewcommand{\arraystretch}{1.5}
		\caption{\label{tab:CarrierLifetime}Carrier lifetimes reported in the literature. Introduced acronyms: surface recombination velocity ($v_{r,S}$), bulk and surface recombination lifetimes ($\tau_{r,B}$, $\tau_{r,S}$), \acrlong{pecvd} (\gls{pecvd}), \acrlong{ligt} (\gls{ligt}), \acrlong{el} (\gls{el}), \acrlong{mwpcd} (\gls{mwpcd}), \acrlong{trdr} (\gls{trdr}), \acrlong{trpls} (\gls{trpls}), \acrlong{ir} (\gls{ir}), broadband pump-probe differential transmission measurements ($\Delta T/T(t)$), \acrlong{icpcd} (\gls{icpcd}).}
		
	\begin{tabular}{L{0.12\linewidth}P{0.05\linewidth}P{0.075\linewidth}cP{0.04\linewidth}cP{0.15\linewidth}P{0.063\linewidth}m{0.3\linewidth}}
		\toprule
		\centering\arraybackslash Ref. 			& \makecell{Comp. \\ Sn\atp} 		& Substrate 	& \makecell{Growth\\ method}	& \makecell{Gr. \gls{t}\\ (\degree C)}& \makecell{Thick.\\ (nm)}& \makecell{$\tau_r$ \\ (ns)}  &  \makecell{$\tau$ meas. \\ method}	 & \centering\arraybackslash Comments \\
		
		\midrule
		
		
		\rowcolor{lightgrey} 
		Hudait, 2022 \cite{Hudait2022} &
		0--6	&  \makecell[c]{III-V buf./\\/GaAs}		& \gls{mbe} 			& 	\makecell[c]{175--\\--250}	& 50--600 & \makecell[c]{@300\,K:\\90--470} & \gls{mwpcd} & \makecell[{{p{\linewidth}}}]{GeSn lattice matched or pseudomorphic. $\tau_r=\tau_{SRH}$, increased with higher gr. \gls{t} or lower \gls{rrms}. $\tau_{r,S}$ not properly considered.}\\
		
		\addlinespace
		
		\makecell[{{L{\linewidth}}}]{Rogowicz, 2021 \cite{Rogowicz2021}}&
		6--12	&  Ge/Si(001)		& \gls{rpcvd} 			& 	-	& 36--60 & \makecell[c]{$\tau_{r,B}$@300\,K:\\$x_{Sn}=6\%$: 1.5  \\  $x_{Sn}=11\%$: 1.2 \\ $\tau_{r,S}$@300\,K: 0.7--0.5} & \makecell[c]{\gls{trpls}\\ \gls{trdr}} &  \makecell[{{p{\linewidth}}}]{Pseudomorphic GeSn. Slightly longer lifetimes with less Sn. Report T-dependence of $\tau_{r,B}$ and $\tau_{r,S}$.}	\\
		
		\addlinespace
		
		\rowcolor{lightgrey} 
		\makecell[{{L{\linewidth}}}]{Julsgaard, 2020 \cite{Julsgaard2020}}&
		12.5	&  Ge/Si(001)		& \gls{cvd} 			& 	-	& 350 & \makecell[c]{0.22 @20\,K \\Same @300\,K}  & \gls{trpls} & \makecell[{{p{\linewidth}}}]{Claim $\tau_r=\tau_{SRH}$. Modest variation with measurement T.} \\

		\addlinespace
		
		Scajev, 2020 \cite{Scajev2020}&
		5	&  \makecell[c]{Ge/SiGe/\\/Si(001)}		& \gls{mbe} 			& 	150	& 200 & \makecell[c]{0.05--0.15 \\ @ 50--350\,K}
		& \gls{ligt} & \makecell[{{p{\linewidth}}}]{Pseudomorphic GeSn, with \gls{tdd}$\sim3$E6\,cm$^{-2}$ inherited from buffers. At 300\,K, $\tau_h>>\tau_e$. Measured diffusion length of 450\,nm.}	\\
		
		\addlinespace
		
		\rowcolor{lightgrey} 
		Vitiello, 2020 \cite{Vitiello2020}&
		2--10	&  Ge/Si(001)		& \gls{rpcvd} 			& 	\makecell[c]{200--\\--300}	& \makecell[c]{\\~~~~~~150~~~~~~\\45\\35\\70\\ 40 \\40} & \makecell[c]{@10\,K:\\ $x_{Sn}=2\%$: 0.70 \\ $x_{Sn}=3\%$: 0.75 \\ $x_{Sn}=4\%$: 0.90 \\$x_{Sn}=6\%$: 2.5 \\$x_{Sn}=8\%$: 0.65 \\$x_{Sn}=10\%$: 0.4}
		& Hanle effect & \makecell[{{p{\linewidth}}}]{GeSn pseudomorphic on 600-nm-thick Ge buffer. Claim $\tau_r=\tau_{SRH}$. Not a clear trend on alloy composition, justified with different growth conditions. Observed a proportional decrease in $\tau_r$ with \gls{tdd}.} \\
		
		\addlinespace
		
		\makecell[{{L{\linewidth}}}]{De Cesari, 2019 \cite{DeCesari2019}}&
		5	&  Ge/Si(001)		& \gls{rpcvd} 			& 	280	& 70 & \makecell[c]{1.1 @300\,K \\ 0.9 @210\,K}
		& \gls{trpls} & \makecell[{{p{\linewidth}}}]{GeSn pseudomorphic on 650-nm-thick Ge buffer. Traps at 13\,meV and 17\,meV above Fermi level.}	\\
		
		\addlinespace
		
		\rowcolor{lightgrey} 
		Geiger, 2016 \cite{Geiger2016}&
		0--12.5	&  Ge/Si(001)		& \gls{rpcvd} 			& 	\makecell[c]{350--\\--390}	& \makecell[c]{\\~~~~~2700~~~~~\\800\\560} & \makecell[c]{@300\,K: \\ Pure Ge buffer: 2.6 \\8.5\%: 0.47 (0.43@20K)\\ 12.5\%: 0.26 (0.17@25K)}  & \makecell[c]{\gls{ir}\\$\Delta T/T(t)$} & \makecell[{{p{\linewidth}}}]{Ge buffer underwent \gls{pda}. $\tau_r$ slightly decreases at $\sim20$\,K compared to 300\,K. Pure Ge shows 10-fold longer $\tau_r$ compared to GeSn.}	\\
		
		\addlinespace
		
		Wirths, 2015 \cite{Wirths2015}&
		8--13	&  Ge/Si(001)		& \gls{rpcvd} 			& 	\makecell[c]{350--\\--390}	& 200--560 & \makecell[c]{0.35 @300\,K \\ 2.1 @15\,K}  & Indirect & \makecell[{{p{\linewidth}}}]{$\tau_r$ extracted from \gls{pl} measurements. Fit \gls{srh} temperature behavior of $\tau_r$ and obtain $E_t-E_F=19$\,meV.}	\\
		
		\addlinespace
		
		\rowcolor{lightgrey} 
		\makecell[{{L{\linewidth}}}]{Gallagher, 2015 \cite{Gallagher2015}}&
		0--11	&  Ge(001)		& \makecell[c]{~~~~~\gls{uhv}~~~~~ \\ \gls{cvd}} 		& \makecell[c]{285--\\--320}	& 400--700& \makecell[c]{@300\,K\\Rlxd. 400nm: 30\\Rlxd. 700nm: 95\\Psdmrp.: 100--300}	& Indirect & \makecell[{{p{\linewidth}}}]{$\tau_r$ extracted from \gls{el} measurements in \emph{pin} diodes. Strong reduction of $\tau_r$ in presence of interface relaxation defects.}	\\
		
		\end{tabular}
	\end{table*}
\end{turnpage}

\begin{turnpage}
	\renewcommand{\arraystretch}{1.5}
	\begin{table*}\ContinuedFloat
		\begin{tabular}{L{0.12\linewidth}P{0.05\linewidth}P{0.075\linewidth}cP{0.04\linewidth}cP{0.15\linewidth}P{0.063\linewidth}m{0.3\linewidth}}
		
		Kako, 2015 \cite{Kako2015}&
		0	&  Si(001)	& \gls{cvd} 			& 	-	& 230--3150 & \makecell[c]{Pure Ge @300\,K: \\ $\tau_{r,B}\sim91$\,ns \\ $v_{r,S}\sim130$\,m/s }  & \gls{trpls} & \makecell[{{p{\linewidth}}}]{Surface recombination dominated by defective Ge/Si interface. Demonstrate fitting \gls{pl} decay time with bulk and surface components.}	\\

		\addlinespace
		
		\rowcolor{lightgrey} 
		\makecell[{{L{\linewidth}}}]{Srinivasan, 2015 \cite{Srinivasan2015}}&
		0	&  \gls{soi}	& - 			& 	-	& 350 & \makecell[c]{Pure Ge @300\,K: 1.6 \\ $v_{r,S}<200$\,m/s }  & \gls{trpls} & \makecell[{{p{\linewidth}}}]{$\tau_r$ measured in Ge waveguides. $v_{r,S}$ at Ge/Si interface.}	\\
		
		\addlinespace
		
		Saito, 2014 \cite{Saito2014}&
		0	&  Si(001)	& \gls{cvd}			& 	\makecell[c]{-\\ \gls{pda}}	& 500 & \makecell[c]{Pure Ge @300\,K: 2 \\ $\tau_{r,B}=3.5$\,ns \\$v_{r,S}<55$\,m/s }  & \gls{trpls} & \makecell[{{p{\linewidth}}}]{They consider $\tau_r$ is equal to twice the \gls{pl} decay time.	Growth \gls{t} and \gls{pda} conditions unspecified.}\\
		
		\addlinespace
		
		\rowcolor{lightgrey} 	
		
		\makecell[{{L{\linewidth}}}]{Geiger, 2014 \cite{Geiger2014}}&
		0	&  \makecell[c]{\\Si(001)	\\ \gls{goi}}	& \gls{pecvd} 			& 	\makecell[c]{500+ \\ \gls{pda}}	& \makecell[c]{\\~~~~~1700~~~~~\\2600} & \makecell[c]{Pure Ge @300\,K: \\ 2.0 ($v_{r,S}\sim850$\,m/s) \\ 5.3 ($v_{r,S}\sim490$\,m/s)}  & \makecell[c]{\gls{ir}\\$\Delta T/T(t)$} & \makecell[{{p{\linewidth}}}]{6-cycle \gls{pda} between 800\degree C and 600\degree C. Report $\tau_r=\tau_{r,S}$, assuming recombination occurs only at defective interface with substrate.}	\\
		
		\addlinespace
		
		\makecell[{{L{\linewidth}}}]{Conley, 2014 \cite{Conley2014}}&
		10	&  Ge/Si(001)		& \gls{rpcvd} 			& $<450$	& 95 & 120--1000 @ 77K	& Indirect & \makecell[{{p{\linewidth}}}]{$\tau_r$ extracted from optical responsivity at 1.55\,\textmu m. They however used an unrealistic abs. coeff. of 19200\,cm$^{-1}$ for Ge$_{0.093}$Sn$_{0.07}$.
		Mesa diameter determines $\tau_r$, indicating an unaccounted contribution of $\tau_{r,S}$.}	\\
		
		\addlinespace
		
		\rowcolor{lightgrey} 
		Sheng, 2013 \cite{Sheng2013}&
		0	&  Si(001)	& \gls{mbe}			& 	-	& 3500 & \makecell[c]{Pure Ge @300\,K: 7 \\ \gls{tdd}$\sim3.3$E7\,cm$^{-2}$:\\$\tau_{r,B}=8.2$\,ns \\\gls{tdd}$\sim1.6$E7\,cm$^{-2}$:\\$\tau_{r,B}=11$\,ns }  & \gls{mwpcd} & \makecell[{{p{\linewidth}}}]{Strong decrease in $\tau_r$ near Ge/Si interface. $\tau_{r,B}$ reflects \gls{tdd}.}  \\
		
		\addlinespace
		
		\makecell[{{L{\linewidth}}}]{Gonzalez, 2011 \cite{Gonzalez2011}}&
		0	&  Si(001)	& \gls{cvd}		& 	450	& 330 & \makecell[c]{Bulk Ge @300\,K:\\ \gls{tdd}$\sim1.6$E10\,cm$^{-2}$:\\$\tau_g=0.5$\,ns\\ \gls{tdd}$\sim4.2$E8\,cm$^{-2}$:\\$\tau_g=6.8$\,ns}  & Indirect & \makecell[{{p{\linewidth}}}]{\gls{pda} (850\degree C, 3\,min, in \ce{N2}) decreased \gls{tdd} 40-fold, but $\tau_g$ increased only 14-fold. $\tau_g$ was extracted from dark currents in Ge p\gls{fet}, and found to be governed by \gls{srh} and \gls{tat} generation.} \\
		
		\addlinespace
		
		\rowcolor{lightgrey} 
		Gaubas, 2006 \cite{Gaubas2006}&
		0	&  None	& Cz, FZ	& 	-	& - & \makecell[c]{Bulk Ge @300\,K:\\ $1-5\cdot10^5$}  & \gls{mwpcd} & \makecell[{{p{\linewidth}}}]{Study on Ge Czochralski (Cz) or float-zone (FZ) bulk Ge crystals.} \\
		
		\addlinespace
		
		\makecell[{{L{\linewidth}}}]{Derhacobian, 1994 \cite{Derhacobian1994}}&
		0	&  None	& -			& 	-	& - & \makecell[c]{Bulk Ge @300\,K:\\ $\tau_{r,B}=5\cdot10^6$\,ns \\ $v_{r,S}=13$\,m/s}  & \gls{icpcd} & \makecell[{{p{\linewidth}}}]{Study on bulk Ge crystal, with nominal doping $<10^{10}$\,cm$^{-3}$.} \\
		
		\bottomrule
	\end{tabular}
	\end{table*}
\end{turnpage}

\subsubsection*{Carrier lifetimes in GeSn}
For GeSn thin films, recombination lifetimes are even lower than in Ge films.
Measurements of $\tau_r$ in GeSn have been reported in a handful of studies, summarized in Tab.~\ref{tab:CarrierLifetime}.
Direct measurements of carrier lifetime in pseudomorphic GeSn on \gls{vge} substrates yielded values of $\tau_{r,300K}$ between a few hundreds of ps~\cite{Julsgaard2020,Scajev2020,Vitiello2020,Geiger2016,Wirths2015} to 1--2\,ns~\cite{Vitiello2020,DeCesari2019, Rogowicz2021}.
Considerably larger $\tau_{r,300K}$ of tens of ns was measured  by \gls{mwpcd} in a study by Hudait\etal~\cite{Hudait2022} on GeSn grown  pseudomorphically or lattice-matched\,\footnote{\,\emph{Pseudomorphic} refers to a film that grows epitaxially without relaxing the strain induced by a lattice-mismatched substrate. \emph{Lattice-matched} refers to epitaxial growth in absence of lattice mismatch.}  on III-V-buffered GaAs substrates. The high values of $\tau_{r,300K}$ measured in this study were possibly due to the absence of dislocations in the GeSn films.
However, comparing $\tau_r$ in GeSn to pure Ge films, one must consider that in studies concerning pure Ge films the surface and bulk recombination rates are often discerned by fitting eq.~\eqref{eq:recombLifetime}  to measurements in films of varying thickness. This is not the case for GeSn, as typically only the effective $\tau_r$ is reported. This value is thus dependent on the film thickness in few-hundred-nm-thick GeSn films, as confirmed in Refs~\cite{Hudait2022,Gallagher2015}.
In addition, Rogowicz\etal~\cite{Rogowicz2021} showed that measured \gls{pl} lifetimes of a few hundred of ps are consistent with surface-recombination phenomena, rather than the film bulk properties.
With \gls{trdr} measurements of various GeSn films with different compositions, they consistently observed two different decay lifetimes, attributed to surface and \gls{srh} bulk recombination processes. \gls{trpls} measurements of the same samples yielded \gls{pl} lifetimes matching almost perfectly the fast \gls{trdr} decay owed to surface processes. Their work highlighted the importance to decompose the bulk and surface lifetimes to accurately evaluate material performance.

\subsubsection*{Takeaways from $\tau$ measurements in literature}
From the measurements of carrier lifetime in Ge and GeSn thin films we can draw a few conclusions.
A clear correlation between the \gls{tdd} and $\tau_r$ has been observed in multiple works~\cite{Vitiello2020,Gallagher2015,Sheng2013,Gonzalez2011,Hudait2022}.
Vitiello\etal~\cite{Vitiello2020} measured $\tau_r$ in Ge$_{0.92}$Sn$_{0.08}$ films grown on \gls{vge} substrates with constant composition and different \gls{tdd}. They observed a linearly proportional decrease in $\tau_r$, and could estimate a recombination velocity at \gls{td} lines of $(1.77\pm0.03)\cdot10^5$\,cm/s, which is 2- to 10-fold higher than $v_{r,S}$ at Si/Ge interfaces.
We highlight that the authors of this study tuned the \gls{tdd} in GeSn films solely by controlling  their thickness, and thus the film \gls{dsr}, which is directly correlated to the \gls{tdd}.
On the other hand, when the \gls{tdd} is tuned through thermal processes, the varying temperatures employed to obtain different \glspl{tdd} also result in differing point defect concentrations, which also affect measurements of $\tau_r$. 
For example, Gonzalez\etal~\cite{Gonzalez2011} found that a decrease in \gls{tdd} obtained through high-temperature \gls{pda} processing of Ge films led to a subproportional increase in lifetime. This suggests that at high-temperature processing of the material, besides the known reactions occurring among dislocations (see Sec.~\ref{sec:GeSn-Ge_StrainRelax_GeAnneal}), additional reactions involving different types of defects occurred -- e.g. formation of point defect complexes -- affecting the trap concentration in the material and thus its carrier lifetime. This is in line with the investigations of unintentional doping concentrations in annealed films discussed in Sec~\ref{sec:ElecProp_UnintDoping}.

The GeSn alloy composition has also been found to influence the carrier recombination lifetimes. A few studies reported that $\tau_r$ decreases with increasing Sn content in the alloy~\cite{Rogowicz2021,Vitiello2020,Geiger2016}.
In particular, a large difference in carrier lifetime was observed between pure Ge and GeSn alloy films. A 10-fold decrease in $\tau_r$ was measured between a Ge buffer layer and a pseudomorphic GeSn film grown on the buffer itself~\cite{Geiger2016}.
Since the \gls{tdd} should be the same in the two materials, one can speculate the resulting lower $\tau_r$ in GeSn is owed to the increased concentration of vacancies that induce mid-gap traps, enhancing \gls{srh} recombination processes.
As elucidated in the previous sections, vacancies are expected in considerably lower concentrations in pure Ge as a consequence of the higher growth temperature and \gls{pda} processes of Ge films.
Lowering the growth temperature has been demonstrated to decrease the lifetime in GeSn~\cite{Hudait2022}.
Furthermore, the concentration of vacancies is known to increase with increasing Sn content~\cite{Assali2019a,Kamiyama2014}, explaining the observed trends in $\tau_r$ with alloy composition.


To conclude, the measured recombination lifetimes in GeSn films are always attributed to \gls{srh} or surface recombination phenomena. In GeSn, $\tau_r$ measures a few ns in the best cases, which is considerably smaller than its computed radiative recombination lifetimes~\cite{Dominici2016}. From Ref.~\cite{Dominici2016}, $\tau_r$ in GeSn is computed to decrease with Sn content, but still remains above 100\,ns up to 18\atp\,Sn; this computed lifetime is two order of magnitudes higher than any experimental carrier lifetime measured in GeSn on (Ge-buffered) Si substrates.
Hence, the reviewed experimental results elucidate the difficulty in achieving efficient room-temperature emission and lasing.
More systematic studies are required to understand the material defects in depth, and find effective passivation techniques for linear and point defects in order to increase the non-radiative lifetimes in GeSn. For example, \ce{H} species have been observed to passivate point defects in GeSn~\cite{Nakatsuka2013}.

\subsection{Mobility}
\label{sec:ElecProp_Mobility}
The carrier mobility ($\mu$) of a semiconductor is affected by multiple scattering phenomena in the material.
For epitaxial Ge and GeSn  films, the main scattering mechanisms involve phonon scattering, neutral and ionized   point defect (ND, ID) scattering, dislocation scattering, and alloy scattering.
Their contributions can be discerned thanks to their different temperature dependences, via the use of Matthiesen's rule~\cite{Grundmann2010}:
\begin{equation}	\frac{1}{\mu}=\sum_{i}\frac{1}{\mu_i}\simeq\frac{1}{\mu_{ph}}+\frac{1}{\mu_{ND}}+\frac{1}{\mu_{ID}}+\frac{1}{\mu_{disl}}+\frac{1}{\mu_{alloy}}+\cdots
\end{equation}
Carrier mobility values reported in the literature for Ge and GeSn films are summarized in Tab.~\ref{tab:Mobility}.
In the following, we will focus mainly on discussing the hole mobility, as it is the majority carrier type most frequently observed as a result of unintentional doping in the GeSn alloy (see Sec.~\ref{sec:ElecProp_UnintDoping}). 

\subsubsection*{Mobility in Ge}
For pure bulk Ge, the theoretical room-temperature hole mobility ($\mu_h$) is 1800\mobun~\cite{Sau2007}.
Experimental values of $\mu_h$ in as-grown Ge films are however significantly lower, between 100\mobun{} and 500\mobun~\cite{Wei2020,Schulte-Braucks2017b,Yeh2014,Nakatsuka2010} due to carrier scattering from point defects and dislocations present in the material.
In a systematic study of Ge hole mobility in function of dislocation density, Ghosh\etal~\cite{Ghosh2014} have demonstrated a clear inverse relation between \gls{tdd} and $\mu_h$. In particular, they observed a drop in $\mu_h$ for \gls{tdd}\,$>2$E7\,cm$^{-2}$, as dislocation scattering became the predominant carrier scattering mechanism in the material.
In this regime of \gls{tdd}, the Ge film thickness is  another factor indirectly influencing $\mu_h$, as the average \gls{tdd} decreases with increasing Ge film thickness~\cite{Loo2009}.
As a consequence, the hole mobility has been found to increase with film thickness~\cite{Yeh2014,Ghosh2014}, as shown in Fig.~\ref{fig:GeHallmeasurements}a, reprinted from Ref.~\cite{Yeh2014}.
High-temperature processing of Ge films has also been proven to be beneficial in terms of hole mobility, thanks to the associated reduction of defect density in the material.
Upon annealing, a significant increase in Ge $\mu_h$ beyond 1000\mobun{} was observed~\cite{Wei2020, Yeh2014, Ghosh2014}.
A record $\mu_h=1252$\mobun{}, not far from the material bulk mobility, was measured at room temperature in annealed 500-nm-thick \gls{mbe} Ge by Wei\etal~\cite{Wei2020}.

The temperature dependence of $\mu_h$ in Ge has been investigated in a few studies~\cite{Wei2020, Yeh2014, Schulte-Braucks2017b}.
An example of such behavior is reported in Fig.~\ref{fig:GeHallmeasurements}c, reprinted from Ref.~\cite{Yeh2014}, where $\mu_h$ in as-grown Ge is plotted with filled, orange dots.
All Ge films investigated in the literature showed a similar temperature dependence. As the measurement temperature was decreased from 300\,K, $\mu_h$ initially increased due to the reduction of phonon scattering, but below a threshold \gls{t} $\mu_h$ started decreasing as a consequence of stronger ionized defect scattering~\cite{Yeh2014}.
The threshold \gls{t} reported in the cited studies varied between 80\,K and 300\,K and was dependent on the material quality. In particular, it was found to decrease  after \gls{pda} of Ge~\cite{Yeh2014,Wei2020}, as visible in Fig.~\ref{fig:GeHallmeasurements}c,  suggesting a decrease in ionized defects after annealing, possibly associated to the reduction in \gls{tdd}.

\subsubsection*{Mobility in GeSn}
For the GeSn alloy, the electron mobility ($\mu_e$) is predicted to be higher than that of pure Ge.
This is a consequence of an increase in carrier population of the $\Gamma$ valley in the \gls{cb} of GeSn.
Conversely,  when solely considering the band structures, no significant change in $\mu_h$ is anticipated~\cite{Sau2007}.
Nevertheless, alloy scattering is expected to contribute to a decrease in $\mu_h$ compared to Ge~\cite{Sau2007,Roucka2011i}.
In the literature, the reported values of room-temperature $\mu_h$ in GeSn span from 100\mobun{} to 600\mobun{}~\cite{Liu2022c,Scajev2020,Prucnal2018ba,Huang2017,Schulte-Braucks2017b,Ryu2012,Roucka2011i,Vincent2011,Nakatsuka2010}, similarly to as-grown Ge.
The influence of GeSn alloy composition on experimental values of $\mu_h$ is still unclear. Two studies, by Schulte-Braucks\etal~\cite{Schulte-Braucks2017b} and Nakatsuka\etal~\cite{Nakatsuka2010}, measured $\mu_h$ in function of alloy composition in GeSn films grown keeping all other growth parameters constant.
However, these reports present somewhat contrasting results.
In Ref.~\cite{Nakatsuka2010} the room-temperature $\mu_h$ decreased with higher Sn content (Sn\atp{}: 2.0 -- 5.8), but did not follow a clear trend at cryogenic temperatures. In this study, the Ge $\mu_h$ was lower than for GeSn from room \gls{t} down to cryogenic temperatures.
On the other hand, in Ref.~\cite{Schulte-Braucks2017b}, $\mu_h$ decreased with higher Sn content (Sn\atp{}: 8.5 -- 12.5) across a large temperature window from room \gls{t} down to 80\,K. The Ge $\mu_h$ was comparable to GeSn at room \gls{t}, but considerably higher at cryogenic temperatures.
These results may suggest that larger Sn contents lead to enhanced impurity and defect scattering, similarly to what is found with carrier lifetime measurements described in Sec~\ref{sec:ElecProp_CarrierLifetime}. 
Additional studies are required to support this hypothesis.
Furthermore, an accurate determination of the contribution of  alloy scattering to  $\mu_h$ in GeSn is yet to be achieved.

To improve the GeSn mobility, a few techniques have been reported in the literature. Nakatsuka\etal~\cite{Nakatsuka2010} obtained a noticeable increase in $\mu_h$ by a \gls{pda} process for GeSn with Sn content 0 -- 5.8\atp{}. An annealing of 500\degree{}C for 1 hour in \ce{N2} led to an almost 2-fold improvement in mobility across the range of studied compositions, likely due to a reduction in the material defect density.
For thin GeSn films, of the order of a few nm to a few tens of nm, surface passivation by a 1-nm-thick Ge layer has been proven effective in enhancing the film $\mu_h$~\cite{Huang2017}. The Ge layer confines charge carriers within the GeSn film, reducing carrier surface scattering. Alternatively, a Si passivation layer has also be employed for improved carrier confinement~\cite{Cheng2021} at the expenses of a more complex epitaxial growth.

Finally, concerning the dependency of GeSn $\mu_h$ on temperature, a behavior analogous to Ge has been reported in the literature~\cite{Schulte-Braucks2017b,Nakatsuka2010,Ryu2012}.
As the measurement \gls{t} is decreased from 300\,K down to a threshold \gls{t}, $\mu_h$ increases as a result of reduced phonon scattering. The threshold \gls{t} has been most frequently measured to be between 100\,K and 200\,K, and is dependent on the growth conditions and composition of the GeSn film.
Below this \gls{t}, ionized defect scattering starts dominating, causing a decrease in $\mu_h$ with decreasing \gls{t}, as shown in the Hall measurements of a Ge film shown in Fig.~\ref{fig:GeHallmeasurements}c.

\begin{turnpage}
	\begin{table*}
		\renewcommand{\arraystretch}{1.5}
		\caption{\label{tab:Mobility}Carrier mobility in intrinsic Ge$_{1-x}$Sn$_x$ thin films reported in the literature. We report the hole mobility ($\mu_h$), unless differently specified. Acronyms: \acrlong{optp} (\gls{optp}), \acrlong{spc} (\gls{spc}), \acrlong{pda} (\gls{pda}), \acrlong{fla} (\gls{fla}), \acrlong{cv} (\gls{cv}), \acrlong{iv} (\gls{iv}), \acrlong{mosfet} (\gls{mosfet}).}
		\begin{tabular}{L{0.12\linewidth}P{0.05\linewidth}P{0.075\linewidth}cP{0.05\linewidth}P{0.045\linewidth}P{0.18\linewidth}P{0.063\linewidth}m{0.28\linewidth}}
			\toprule
			\centering\arraybackslash Ref. 			& \makecell{Comp. \\ Sn\atp} 		& Substrate 	& \makecell{Growth\\ method}	& \makecell{Gr. \gls{t}\\ (\degree C)}& \makecell{Thick.\\ (nm)}& \makecell{$\mu_h$ @\,300\,K \\ (\mobun)}  &  \makecell{$\mu$ meas. \\ method}	 & \centering\arraybackslash Comments \\
			
			\midrule
			\rowcolor{lightgrey} 
			\addlinespace
			Liu, 2022 \cite{Liu2022c}&
			1.5	&  Ge(001)		& \gls{ms} 			& 	300	& 350 &$\mu=80-210$  & \gls{optp} & \makecell[{{p{\linewidth}}}]{\textbf{Carrier type unspecified.} Mobility decreases with large pump fluences due to carrier-carrier scattering phenomena.} \\
			
			\addlinespace
			Tao, 2020 \cite{Tao2020}&
			0--14	&  Ge(001)		& \gls{mbe} 			& 	200	& 30 &\makecell[c]{Pure Ge: $\mu_e=1113$ \\ $x_{Sn}=4\%$: $\mu_e=1210$ \\$x_{Sn}=8\%$: $\mu_e=1123$ \\$x_{Sn}=14\%$: $\mu_e=1007$}  & Hall & \makecell[{{p{\linewidth}}}]{Films reported partially relaxed, but  \gls{mbe} 30-nm-thick GeSn with $x_{Sn}\leq$8\atp{} is expected to be pseudomorphic on Ge~\cite{Wang2015}.	\textbf{Carriers are \emph{n}-type also in GeSn}.} \\
			
			\rowcolor{lightgrey} 
			\addlinespace
			Wei, 2020 \cite{Wei2020}&
			0	&  Si(001)		& \gls{mbe} 			& 	$200$	& 500 & \makecell[c]{As-grown Ge: 343\\
				\gls{pda} 900\degree C: 1252}  & Hall & \makecell[{{p{\linewidth}}}]{Monotonic increase in $\mu_h$ with increasing annealing \gls{t}. Improvements with higher annealing \gls{t} across the entire \gls{t} range of 20 -- 300\,K.  } \\

			\addlinespace
			
			Scajev, 2020 \cite{Scajev2020}&
			5	&  \makecell[c]{Ge/SiGe/\\/Si(001)}		& \gls{mbe} 			& 	150	& 200 & 490  & - & \makecell[{{p{\linewidth}}}]{Carrier scattering from ionized and neutral impurities dominant  across the 70 -- 300\,K temperature range.}	\\
			
			\addlinespace
			
			\rowcolor{lightgrey} 
			Imajo, 2019 \cite{Imajo2019l}&
			0	& \ce{GeO2}/\ce{SiO2}		& \makecell[c]{aGeSn \gls{mbe}  \\ + \gls{spc}}			& $\ast$	& 500 & \makecell[c]{Pure Ge: 440\\After \gls{pda}: 620}  & \makecell[c]{Hall\\ \gls{vdp}} & \makecell[{{p{\linewidth}}}]{$\ast$: \gls{spc} at 450\degree C for 5\,h + 375\degree C for 150\,h (+ \gls{pda} at 500\degree C for 5\,h), resulting in \textbf{polycrystalline Ge}. Improvement in $\mu_h$ with increase in grain size.}	\\
			
			\addlinespace
			Prucnal, 2018 \cite{Prucnal2018ba}&
			5	&  Si(001)		& \gls{mbe} 			& 	\makecell[c]{160 \\ + \gls{pda}}	& 300 & 155  & \makecell[c]{Hall\\ \gls{vdp}} & \makecell[{{p{\linewidth}}}]{\gls{pda}: \gls{fla} with no pre-heating and a pulse of 3\,ms and 65\,J/cm$^2$.} \\
			
			
			\rowcolor{lightgrey} 
			\addlinespace
			Huang, 2017 \cite{Huang2017}&
			10	&  Ge/Si(001)		& \gls{cvd} 			& 	-	& 9 & 509  & \makecell[c]{\gls{cv} + \\ \gls{iv} in \\ \gls{mosfet}} & \makecell[{{p{\linewidth}}}]{GeSn capped with 1\,nm Ge to reduce surface scattering. Material underwent thermal processing at 400\degree C prior $\mu_h$ measurements. GeSn \gls{mosfet} outperforms Ge \gls{mosfet} in terms of $\mu_h$.} \\
			
			\addlinespace
			\makecell[{{L{\linewidth}}}]{Schulte-Braucks, 2017 \cite{Schulte-Braucks2017b}}&
			0--12.5	&  Ge/Si(001)		& \gls{cvd} 			& 	350	& \makecell[c]{2500\\768\\835\\414} & \makecell[c]{Pure Ge: $\sim350$ \\ $x_{Sn}=8.5\%$: $\sim400$ \\$x_{Sn}=10\%$: $\sim350$ \\$x_{Sn}=12.5\%$: $\sim300$}  & Hall & \makecell[{{p{\linewidth}}}]{While at room \gls{t} $\mu_h$ is comparable in GeSn and Ge, for $T<250$\,K, $\mu_h,Ge>>\mu_h,GeSn$.}	\\
			
			
			\rowcolor{lightgrey} 
			\addlinespace
			Yeh, 2014 \cite{Yeh2014}&
			0	&  Si(001)		& \gls{ms} 			& 	$<360$	& \makecell[c]{\\200\\1000\\1000} & \makecell[c]{Pure Ge:\\$\sim60$\\$\sim200$--$300$\\ After \gls{pda}: 1180}  & Hall & \makecell[{{p{\linewidth}}}]{Progressive increase in $\mu_h$ with film thickness. In 1\textmu m Ge, they observed a 4-fold improvement in $\mu_h$ after \gls{pda} at 700\degree C.} \\

		\end{tabular}
	\end{table*}
\end{turnpage}

\begin{turnpage}	
\begin{table*}\ContinuedFloat
\begin{tabular}{L{0.12\linewidth}P{0.05\linewidth}P{0.075\linewidth}cP{0.05\linewidth}P{0.045\linewidth}P{0.18\linewidth}P{0.063\linewidth}m{0.28\linewidth}}
			
			\rowcolor{lightgrey} 
			\addlinespace
			Ghosh, 2014 \cite{Ghosh2014}&
			0	&  Si(001)		& \gls{mbe} 			& 	\makecell[c]{600\\ + \gls{pda} \\ 800\degree C}	& \makecell[c]{\\100\\1000\\1000\\1000} & \makecell[c]{Pure Ge:\\$\sim950$\\ \gls{tdd}\,$=2$E7\,cm$^{-2}$: 1020\\ \gls{tdd}\,$=5$E7\,cm$^{-2}$: $\sim600$\\ \gls{tdd}\,$=7$E7\,cm$^{-2}$: $\sim360$}  & \makecell[c]{Hall\\ \gls{vdp}} & \makecell[{{p{\linewidth}}}]{Found two scattering regimes, dominated by (1) impurities and (2) dislocations. Clear drop in $\mu_h$ for \gls{tdd}\,$>2$E$7$\,cm$^{-2}$. $\mu_h$ slightly increases with Ge film thickness.} \\

			\addlinespace	
			Ryu, 2012 \cite{Ryu2012}&
			0.06	&  Si(001)		& \gls{uhv} \gls{cvd} 			& 	\makecell[c]{390\\ + \gls{rta} \\ 680\degree C}	& 800 & $\sim370$  & \makecell[c]{Hall\\ \gls{vdp}}&
			\makecell[{{p{\linewidth}}}]{Degenerate \emph{p}-type conductive layer at GeSn/Si interface due to \glspl{md}. $\mu_h$ measured down to 4\,K. $\mu_h,max=638$\mobun at $\sim130$\,K.}	\\
			
			\rowcolor{lightgrey} 
			\addlinespace	
			Roucka, 2011 \cite{Roucka2011i}&
			0--0.1	&  Si(001)		& \gls{cvd} 			& 	\makecell[c]{390\\ + \gls{rta} \\ 680\degree C}	& \makecell[c]{500\\800\\1100} & \makecell[c]{Pure Ge: 700--800\\$x_{Sn}=0.06\%$: 345--390\\$x_{Sn}=0.1\%$:270--360}  & Hall&
			\makecell[{{p{\linewidth}}}]{They show alloy scattering causes halvening of Hall mobility with respect to pure Ge.}	\\

			\addlinespace
			Vincent, 2011 \cite{Vincent2011}&
			8	&  Ge/Si(001)		& \gls{apcvd} 			& 	320	& 40 & $\sim600$  & \makecell[c]{micro-\\Hall} & \makecell[{{p{\linewidth}}}]{GeSn \emph{in-situ} doped with B to $p=2E17$\,cm$^{-3}$.} \\
			
			\rowcolor{lightgrey} 
			\addlinespace
			\makecell[{{L{\linewidth}}}]{Nakatsuka, 2010 \cite{Nakatsuka2010}}&
			0--5.8	&  \gls{soi}		& \gls{mbe} 			& 	200	& \makecell[{{P{\linewidth}}}]{160 -- 170} & \makecell[c]{Before/After \gls{pda}:\\Pure Ge: $\sim90$/$\sim150$\\$x_{Sn}=2.0\%$: $\sim170$/$\sim290$\\$x_{Sn}=4.0\%$: $\sim120$/$\sim260$\\$x_{Sn}=5.8\%$: $\sim110$/$\sim150$}  & \makecell[c]{Hall\\ \gls{vdp}} & \makecell[{{p{\linewidth}}}]{\gls{pda}: 500\degree C for 1\,h in \ce{N2}. Clear improvement in $\mu_h$ after \gls{pda}.  Impurity band induces parallel conduction. Authors present also higher $\mu_h$ values removing the contribution of the parallel conduction layer.} \\
			
			\bottomrule
		\end{tabular}
	\end{table*}
\end{turnpage}

\section{Conclusion}
\label{sec:Conclusion}
Over the last 15 years, GeSn has showcased its potential across a range of optoelectronic devices including photodetectors, \glspl{led} and lasers, showing its promise for applications in the \gls{swir} and \gls{mwir} domains.
However, despite significant performance improvements, no GeSn-based device has yet made it to commercialization.
The current bottleneck in device performance lies in the incomplete understanding of material defects and their associated trap states. This deficiency ultimately results in a lack of precise control over device properties.
With this review, we provided a perspective and critical view on how the material defects impact the optoelectronic properties, ranging from light absorption to electronic transport and unintentional doping.

The electrical properties of  GeSn depend on the microstructure and order at the atomic scale, and, in particular, on the material defects.
Typical defects in the GeSn crystal structure are dislocations and point defects.
Dislocations originate from the lattice mismatch with the substrate, while point defects form due to impurities and low-temperature, out-of-equilibrium growth. 
The defect density is determined by the epitaxial growth conditions and post-deposition thermal treatments.
Dislocations and point defects in Ge and GeSn are predominantly electrically active, as they are source of shallow and deep traps in the bandgap. Additional traps are induced by impurities.
These traps affect the material carrier lifetime, mobility, and unintentional doping concentration. Consequently, pinpointing the origin of each trap is pivotal for controlling electrical properties. However, this task is complex due to the simultaneous presence of multiple defects in the material.
For instance, to the best of our knowledge, there has been no investigation differentiating the trap states associated with threading or misfit dislocations. This distinction is crucial for managing the electrical properties of annealed Ge, where the density of threading dislocations is reduced while that of misfit dislocations is increased.
To precisely identify the defect responsible for the most common traps, it is imperative to utilize the purest materials available, and work in a synergy between computations and experiments.

The literature evidenced a general lack of agreement regarding the source of shallow trap states responsible for the high unintentional doping concentrations in Ge and GeSn.
Ge films have been reported to show both \emph{n-type} and \emph{p-type} unintentional doping, while GeSn is predominantly \emph{p-type}. Given the latter, shallow traps responsible for \emph{p-type} unintentional doping in GeSn are likely to be associated with the presence of Sn or to the lower growth temperatures employed with respect to Ge.
While shallow traps are generally attributed to vacancy complexes, in fact, \acrlongpl{td} have also been reported to induce shallow trap levels.
A systematic study of unintentional doping concentrations in pseudomorphic GeSn -- i.e., free of dislocations -- grown on Ge(001) may allow to rule out the latter.
On a different line, thermal treatment of Ge films has evidenced a non-linear variation in unintentional doping, which led in some cases to an increase in unintentional doping upon annealing. This phenomenon is presently not understood and may be owed to point defect evolution and/or formation of misfit dislocations.

Threading dislocations have been demonstrated to affect negatively the Ge/GeSn carrier lifetime, through \gls{srh} recombination mechanisms associated with deep traps.
Similarly, increasing Sn composition has also been associated with reduced carrier lifetimes. We speculate this is owed to an increase in vacancy complexes associated with the larger Sn contents. However, this hypothesis remains to be proven.

In conclusion, it is important to emphasize that both point defects and dislocations have been linked to the formation of shallow and deep traps, consequently impacting the material's carrier lifetime and unintentional doping concentrations. While trap levels in Ge have been relatively well examined in the past, defects and traps specific to GeSn remain largely unexplored.
A deeper understanding of these aspects could facilitate a greater degree of control of the material's electrical properties, thereby enabling the integration of GeSn into commercial devices.

Overall, this review provides the base to understand the state-of-the-art and challenges of GeSn as an optoelectronically grade semiconductor. It also provides a critical view on where the field may evolve so that the semiconductor reaches the optoelectronics market.

\begin{acknowledgments}
	This work was supported by Innosuisse, SNSF NCCR QSIT, Max Planck Institut für Festkörperforschung, and Max Planck Graduate Center for Quantum Materials. The authors wish to thank Prof. Paul McIntyre and Prof. Oussama Moutanabbir for the fruitful discussions.
	
	\textbf{Author contributions}: A.G. conceived the structure of the review and wrote the manuscript, with inputs from A.F.M. 
\end{acknowledgments}

\bibliography{bibliography}

\end{document}